\documentclass[10pt,a4paper,final]{iopart}
\usepackage{graphicx,epstopdf}
\usepackage{amssymb}
\usepackage{subfig}
\usepackage{caption}
\usepackage{appendix}
\def\pslash{p\!\!\!\slash }
\def\qslash{q\!\!\!\slash }
\def\xslash{x\!\!\!\slash }

\begin{document}

\title{Isovector Axial Vector Form Factors of Octet-Decuplet Hyperon Transition in QCD}
\author{A. Kucukarslan}%
\address{Physics Department, Canakkale  Onsekiz Mart University, 17100 Canakkale, Turkey}
\ead{akucukarslan@comu.edu.tr}
\author{U. Ozdem}%
\address{Physics Department, Canakkale  Onsekiz Mart University, 17100 Canakkale, Turkey}
\ead{ulasozdem@gmail.com}
\author{A. Ozpineci}%
\address{Physics Department, Middle East Technical University, 06531 Ankara, Turkey}
\ead{ozpineci@metu.edu.tr}

\begin{abstract}
We calculate the isovector axial vector form factors of the octet-decuplet hyperon transitions
within the framework of the light-cone QCD sum rules (LCSR) to leading in QCD and including higher-twist corrections.
 In particular, we motivate the most recent version of the $\Sigma$ and $\Lambda$ baryons
 distribution amplitudes which are examined up to twist-6 based on conformal symmetry of the massless QCD Lagrangian.
 \end{abstract}

\noindent{\it Keywords\/}: Hyperon axial  form factors, Octet-decuplet transition, light-cone QCD sum rules

\maketitle

\section{Introduction}

Form factors are extremely important in the studies of the hadron physics.
They give an information about the internal structure of composite particles.
The isovector axial vector form factors are important to measure the axial charge of the hadrons.
The axial charge $g_A$ is also viewed as an indicator of the phenomenon of spontaneous breaking of chiral symmetry
of non-perturbative QCD \cite{Choi:2010ty}.

The axial form factors of N $\rightarrow \Delta$ have been studied using  Lattice QCD \cite{Alexandrou:2006mc, Alexandrou:2007xj},
quark model \cite{BarquillaCano:2007yk}, light cone
QCD sum rules \cite{Aliev:2007pi},
chiral perturbation theory ($\chi$PT) \cite{Geng:2008bm, Procura:2008ze}
and weak pion production \cite{Amaro:2008hd, Hernandez:2010bx}. The experimental
information on the weak axial form factors come from parity-violating
electron scattering experiments at JLAB \cite {Androic:2012doa}.
For  $Q^2 $ = 0.34 $GeV^2$, they found that the value of axial form factors determined from the hydrogen asymmetry
was  $G_A^{N\Delta}$ = - $ 0.05\pm(0.35)_{stat} \pm(0.34)_{sys} \pm(0.06)_{theory}$
($G_A^{N\Delta}$ linear combination of $C^A_{4,5}$ form factors).
However recently there have been attempts to extract the octet-octet and decuplet-decuplet hyperon
axial charges using lattice QCD, quark model, $\chi$PT and QCD sum rules~\cite{ Choi:2010ty, Erkol:2010lat,
Lin:2008mr,Sasaki:2008ha, Erkol:2011qh, Jiang:2008aqa}.
The hyperon sector is interesting because they provide for an ideal system in which to study
$SU(3)$  flavor symmetry breaking by replacement of up or down quarks in nucleons by strange ones \cite{Lin:2008rb}.
For hyperon the axial charge is a major parameter for low-energy effective description of baryon sector as they
enter in the loop graphs of chiral perturbation theory \cite{Erkol:2010lat}.
Our information about the axial charge of the hyperons from experiment is also limited since their experimental measurements are difficult due
to their unstable nature.

In the present work, we calculate the isovector axial vector transition form factors of the
 $\Xi - \Xi^*$, $\Sigma - \Sigma^*$ and  $\Lambda - \Sigma^*$.
The considered processes take place in low energies far from
the perturbative region, hence  to calculate the form factors as the main ingredients, we need a non-perturbative method.
One of the most powerful non-perturbative methods is traditional QCD sum rules (QCDSR),
which is also a powerful tool to extract hadron properties
~\cite{Shifman:1978bx, Shifman:1978by, Reinders:1984sr, Ioffe:1983ju}.
An alternative to the traditional QCD sum rules is the light cone QCD sum rules (LCSR)
 ~\cite{Braun:1988qv, Balitsky:1989ry, Chernyak:1990ag}.
In this method, the hadronic properties are expressed in terms of the properties of
the vacuum and the light cone distribution amplitudes of one of the hadrons in the process. Since the form factors are
expressed in terms of the properties of the QCD vacuum and the distribution amplitudes, any uncertainty in these
parameters reflects in the uncertainty of the predictions of the form factors.
This method has been rather successful in determining hadron form factors at high $Q^2$ (see e.g. \cite{Erkol:2011qh, Aliev:2004ju, Aliev:2007qu, Wang:2006su,
Braun:2006hz, Erkol:2011iw, Aliev:2011ku, Aliev:2013jda}).

This work is organized as follows: In the next section, we obtain the formulation of the baryon axial
form factors in LCSR. In Section III, we present our numerical analysis and discussion.

\section{Hyperon axial form factors}

The baryon (hyperon) matrix element expressed in terms of four invariant form factors
can be written as \cite{Adler:1968tw, Adler:1975mt, LlewellynSmith1972261};

\begin{eqnarray}\label{fizz}
\hspace*{-2.5cm}
 \langle H^*(p',s')| A_\nu^3 (x)| H(p,s)\rangle & =
         i \overline{\upsilon}^{\lambda}(p',s')\bigg[\left\{\frac{C_3^{H^*}(q^2)}{M_H}\gamma_\mu \nonumber
         + \frac{C_4^{H^*}(q^2)}{M_H^{2}} p'_\mu\right\} (g_{\lambda\nu} g_{\rho\mu} \nonumber \\
    & -g_{\lambda\rho} g_{\mu\nu})q^\rho+ C_5^{H^*}(q^2) g_{\lambda\nu}
         +\frac{C_6^{H^*}(q^2)}{M_H^{2}}q_\lambda q_\nu \bigg]H(p,s)
\end{eqnarray}
where $H$ and $H^*$  denote the $\Xi$, $\Sigma$ or $\Lambda$ and  $\Xi^*$ or $\Sigma^*$, respectively,
and $A^3_\nu$ is an axial vector-isovector current defined as
\begin{eqnarray}
A^3_\nu(x) = \frac12\bigg( \bar u(x) \gamma_\nu \gamma_5 u(x) - \bar d(x) \gamma_\nu \gamma_5 d(x)\bigg).
\end{eqnarray}
Besides, $\upsilon^\lambda$ is a Rarita-Schwinger spinor describing spin 3/2 baryons.
In order to obtain the correlation function, phenomenological summation over spin
of Rarita-Schwinger spinor have been performed, and the following formula have been used;
\begin{eqnarray}\label{rar}
\hspace*{-2.5cm}
\upsilon_{\mu}(p',s)\overline{\upsilon_{\nu}}(p',s)&=
-(\pslash'+m_{H^*})\{g_{\mu\nu}
-\frac{1}{3}\gamma_{\mu}\gamma_{\nu}-\frac{2p'_{\mu}p'_{\nu}}{3m_{H^*}^{2}}
+\frac{p'_{\mu}\gamma_{\nu}-p'_{\nu}\gamma_{\mu}}{3m_{H^*}}\}
\end{eqnarray}

In deriving  Light cone QCD sum rules  for the axial baryon transitions form factors,
we start our analysis with the following two-point correlation function:
\begin{equation}\label{corrf}
	\Pi_{\mu\nu}(p,q)=i\int d^4 x e^{iqx} \langle 0 |T[j_{\mu}^{H^*}(0)A_\nu^3(x)]|H(p)\rangle,
\end{equation}
where $J^{H^*}_\mu(0)$ denotes the interpolating current $\Xi^*$ or $\Sigma^*$ baryon and T denotes the time ordering product.
In order to calculate this correlation function either phenomenologically in which we insert a complete
set of hadronic states into the correlator to represent in terms of the hadronic parameters, or theoretically in which
we calculate correlation function via the Operator Product Expansion (OPE) in Euclidean
region $p^2\rightarrow - \infty$ and $(p+q)^2 \rightarrow - \infty$ in terms of QCD parameters.
QCD sum rules for the considered form factors are obtained by mathching these two correlation function expressions and
applying Borel transformation in order to suppress contribution of the higher states and continuum.
In this work, we choose  the general form of the interpolating currents for $\Sigma^*$ and  $\Xi^*$
baryons as \cite{Colangelo:2000dp}:
 \begin{eqnarray}\label{cur}
 J_\mu^{\Sigma^*} (0) =\frac{\varepsilon^{abc}}{\sqrt{3}}[2(u^{aT}(0) C\gamma_\mu s^b(0))u^c(0)+(u^{aT}(0)C\gamma_\mu u^b(0))s^c(0)]\nonumber
\\
  J_\mu^{\Xi^*} (0) =\frac{\varepsilon^{abc}}{\sqrt{3}}[2(s^{aT}(0) C\gamma_\mu u^b(0))s^c(0)+(s^{aT}(0)C\gamma_\mu s^b(0))u^c(0)]
\end{eqnarray}
where  $a$, $b$, $c$ are the color indices and $C$ is the  charge conjugation operator.
Inserting a complete set of intermediate hadronic states with the same quantum numbers as the corresponding
interpolating currents, we determine the phenomenological part of the correlation function as follows
  \begin{eqnarray}\label{fiz1}
   \Pi_{\mu\nu}(p,q) = \sum_{s'}\frac{\langle 0 |j_{\mu}^{H^*}|H^*(p',s')\rangle \langle H^*(p',s')| A_\nu^3(x)|H(p,s)\rangle }{M_{H^*}^2-p'^2} +...
  \end{eqnarray}
where $M_{H^*}$ is the $\Xi^*$ or  $\Sigma^*$ mass and ... represent the contributions of the higher states and
continuum. The matrix element of the interpolating current between the vacuum and baryon state is determined as
  \begin{eqnarray}\label{amp}
\langle 0|j^{H^*}_\mu|H^*(p',s')\rangle=\lambda_{H^*}\upsilon_\mu(p',s')
\end{eqnarray}
where $\lambda_{H^*}$ is the baryon overlap amplitude and $\upsilon_\mu(p',s')$ is the baryon spinor.
Using the Eqs.(\ref{fizz}), (\ref{rar}) and (\ref{amp}) in Eq.(\ref{fiz1}) one determines the correlation function
in terms of the hadronic parameters as
\begin{eqnarray}\label{phys}
 \Pi^{H^*}_{\mu\nu}(p,q)&=-i\frac{\lambda_{H^*}}{M_{H^*}^2-p'^{2}}\bigg[C_3^{H^*}(q^2)\left\{1-\frac{M_{H^*}}{M_H}\right\}(q_\mu\gamma_\nu - g_{\mu\nu}  \qslash)\nonumber\\
         &   +\left\{C_5^{H^*}(q^2)+C_4^{H^*}(q^2)\frac{p'.q}{M^2_H}\right\} g_{\mu\nu}\qslash- \frac{C_4^{H^*}(q^2)}{M_H^2}q_\mu p'_\nu \qslash  \nonumber  \\
		&+C_3^{H^*}(q^2) (q_\mu\gamma_\nu-g_{\mu\nu}\qslash) \qslash +\left\{C_5^{H^*}(q^2)(M_{H^*}+M_H)\right\}g_{\mu\nu}\nonumber\\
		&+ \left\{-\frac{2C_3^{H^*}(q^2)}{M_H}-\frac{C_4^{H^*}(q^2)}{M_H}(1+\frac{M_{H^*}}{M_H})\right\}(q_\mu p'_\nu-g_{\mu\nu}p'.q)\nonumber\\
		&+\left\{C_6^{H^*}(q^2)(\frac{M_{H^*}+M_H}{M_H^2})\right\}q_\mu q_\nu
 		 + \frac{C_6^{H^*}(q^2)}{M_H^2}q_\mu q_\nu \qslash \bigg]    		
\end{eqnarray}
In this expression, the interpolating current $J_\mu$ couples to the  $J^P =3/2^+$ states.
However, the interpolating current  $J_\mu$ couples not only to  $J^P =3/2^+$ states, but also to the
 $J^P =1/2^-$ states.
Actually, the current $J_\mu$ has nonzero overlap with spin -$1/2$ states. Therefore, the matrix element of the current
$J_\mu$ can be written as the following relation
\begin{equation} \label{eli}
\langle 0 \vert j^{H^*}_\mu \vert 1/2(p') \rangle = \left(A p'_\mu + B \gamma_\mu \right) u(p').
\end{equation}
The condition $\gamma^\mu p_\mu = 0$ has been used to determine the relation in Eq.(\ref{eli}).
Hence spin-$1/2$ states contribute only the structures which include a $\gamma^\mu$ at the beginning
or which are proportional to $p'_{\mu}$. By choosing the appropriate structures,
the contributions from these states are eliminated  \cite{Belyaev:1993ss,  Belyaev:1982cd}.
In this work we will consider the ordering $\gamma_\mu\gamma_\nu\qslash\pslash'$.

The QCD side of the correlation function is obtained in terms of quark and gluon degrees of freedom.
For calculation of the theoretical part of the correlation function  from the QCD side the interpolating
fields in Eq. (\ref{cur}) are inserted into the correlation function in Eq. (\ref{corrf}), we obtained

 \begin{eqnarray} \label{eq:Qcds}
 \hspace*{-2.5cm}
 \Pi_{\mu\nu} &= \frac{i}{16\sqrt3}\int d^4 x e^{iqx}~(C\gamma_\mu)_{\alpha\beta}(\gamma_\nu \gamma_5)_{\rho\sigma}
 \left.\ \bigg[2\delta_\alpha^{\eta}\delta_\sigma^{\theta} \delta_\beta^{\phi}S(-x)_{\lambda\rho}\right.\nonumber\\
 &\left.+2\delta_\lambda^{\eta} \delta_\sigma^{\theta} \delta_\beta^{\phi}S(-x)_{\alpha\rho}\right.
\left.+ \delta_\alpha^{\eta} \delta_{\sigma\theta} \delta_\lambda^{\phi} S(-x)_{\beta\rho}\right.
\left.+\delta_\beta^{\eta} \delta_\sigma^{\theta} \delta_\phi^{\lambda} S(-x)_{\alpha\rho}\bigg] \right.\nonumber\\
&\left. 4\epsilon^{abc}\langle0|{q_1}_{\sigma}^a(0) {q_2}_{\theta}^b(x) {q_3}_{\phi}^c(0)|H(p,s) \rangle \right.\nonumber  \\
\end{eqnarray}
where $q_{1,2,3}$ denote the quark fields and  $S(x)$ represents the light-quark propagator as

\begin{equation}\label{pro}
	S(x)=\frac{i\xslash}{2\pi^2x^4}-\frac{\langle q\bar{q}\rangle}{12}\left(1+\frac{m_0^2 x^2}{16}\right).
\end{equation}

In this expressions, the first term describes the hard-quark propagator.
The other term represents the contribution from non-perturbative structure of the QCD vacuum.
Using Borel transformations, these contributions are removed.
In the background field the hard-quark propagator has corrections,
which are expected to give negligible contributions as they are related to four-
and five-particle baryon distribution amplitudes~\cite{Diehl:1998kh}.
In this work, we will not take into account such contributions, hence only the first term in Eq. (\ref{pro})
leaves for our discussion.
The matrix elements of the local three-quark operator is
\begin{eqnarray*}
4\epsilon^{abc}\langle 0|q_{1\alpha}^a(a_1 x) q_{2\beta}^b(a_2 x) q_{3\gamma}^c(a_3 x)|H(p,s)\rangle.
\end{eqnarray*}
This expression can be written in terms of
DAs using the Lorentz covariance, the spin and the parity of the baryon. Based on a conformal expansion using
the approximate conformal invariance of the QCD Lagrangian up to one-loop order, the DAs are then decomposed into
local non-perturbative parameters, which can be estimated using QCD sum rules or fitted so as to reproduce
experimental data. We refer the reader to Refs.~\cite{Liu:2009uc, Liu:2008yg, Liu:2013bxa, Liu:2014uha} for
a detailed analysis on DAs of hyperons, which we employ in our work to extract the axial-vector  form factors.

The QCD sum rules for axial form factors of the hyperon transitions are determined
by equating both representations of correlation function.
To do this, we choose the structures proportional to $ q_\mu\gamma_\nu\qslash$,
$q_\mu{p'}_\nu\qslash$, $(g_{\mu\nu}\qslash - q_\mu\gamma_\nu\qslash)$ and $q_\mu q_\nu\qslash$ for the form factors
 $C_3^{HH^*}$, $C_4^{HH^*}$, $C_5^{HH^*}$ and $C_6^{HH^*}$, respectively. After that, applying Borel transformation,
 the QCD sum rules for the axial form factors are obtained as
          \begin{eqnarray*}
          \hspace*{-2.5cm}
C_3^{\Xi\Xi^*}(q^2)  \frac{\lambda_{\Xi^*}}{M_{\Xi^*}^2-p'^{2}}&=\frac{1}{\sqrt{3}}\bigg\{\frac{M_\Xi^2}{8} \int_0 ^{1}d\alpha
         \frac{1-\alpha}{(q-p\alpha)^4} F_1
         -\frac{M_\Xi^2}{4} \int_0 ^{1}d\beta
         \frac{1-\beta}{(q-p\beta)^4} F_2\\
         & +\frac{M_\Xi^2}{2}\int_0 ^{1}d\beta \frac{1}{(q-p\beta)^4}F_3
         +\frac{1}{4}\int_0 ^{1}dx_2 \frac{1}{(q-px_2)^2}F_4 \bigg\}\\
          \end{eqnarray*}

         \begin{eqnarray*}
         \hspace*{-2.5cm}
C_4^{\Xi\Xi^*}(q^2)  \frac{\lambda_{\Xi^*}}{M_{\Xi^*}^2-p'^{2}}&=\frac{ M_\Xi}{\sqrt{3}}\bigg\{-\frac{M_\Xi^2}{2} \int_0 ^{1}d\beta
         \frac{1-\beta}{(q-p\beta)^6} F_5
         +\frac{M_\Xi}{2} \int_0 ^{1}d\alpha\frac{\alpha}{(q-p\alpha)^4} F_6 \\
         &+\frac{M_\Xi}{2} \int_0 ^{1}d\alpha\frac{1-\alpha}{(q-p\alpha)^4} F_7\bigg\}\\
          \end{eqnarray*}

         \begin{eqnarray*}
         \hspace*{-2.5cm}
C_5^{\Xi\Xi^*}(q^2)  \frac{\lambda_{\Xi^*}}{M_{\Xi^*}^2-p'^{2}}&=\frac{ M_\Xi}{\sqrt{3}}\bigg\{\frac{M_\Xi^2}{4} \int_0 ^{1}d\beta
         \frac{1}{(q-p\beta)^4} F_8
         +\frac{1}{8}\int_0 ^{1}dx_2 \frac{1}{(q-px_2)^2}F_9
         \\
         & +\frac{M_\Xi^2}{8} \int_0 ^{1}d\beta \frac{1}{(q-p\beta)^4} F_{10}
         +\frac{1}{8}\int_0 ^{1}dx_2 \frac{1-x_2}{(q-px_2)^2}F_{11}\\
         \\
         &  + \frac{M_\Xi^2}{4} \int_0 ^{1}d\beta \frac{\beta(1-\beta)}{(q-p\beta)^4} F_{12}
         +\frac{M_\Xi^2}{4} \int_0 ^{1}d\beta
         \frac{1-\beta}{(q-p\beta)^4} F_{13}\\
         \\
         &+\frac{M_\Xi^2}{8} \int_0 ^{1}d\alpha \frac{\alpha(1-\alpha)}{(q-p\alpha)^4} F_{14}
          -\frac{M_\Xi}{2}\int_0 ^{1}dx_2 \frac{x_2}{(q-px_2)^2}F_{15}\\
         \\
        & +\frac{M_\Xi^2}{2} \int_0 ^{1}d\beta \frac{\beta}{(q-p\beta)^4} F_{16}
                 + \frac{M_\Xi^2}{8} \int_0 ^{1}d\beta \frac{1-\beta}{(q-p\beta)^4} F_{17}\\
         &-\frac{1}{4} \int_0 ^{1}d\alpha \frac{1}{(q-p\alpha)^2} F_{18}\bigg\}\\
          \end{eqnarray*}

         \begin{eqnarray}\label{xi}
         \hspace*{-2.5cm}
 C_6^{\Xi\Xi^*}(q^2)  \frac{\lambda_{\Xi^*}}{M_{\Xi^*}^2-p'^{2}}&=-\frac{ M_\Xi^2}{2 \sqrt{3}}\bigg\{M_\Xi \int_0 ^{1}d\beta
         \frac{(1-\beta)^2}{(q-p\beta)^6} F_{19}
         + \int_0 ^{1}d\alpha\frac{1-\alpha}{(q-p\alpha)^4} F_{20}\bigg\}
        \end{eqnarray}

for the $\Xi$-$\Xi^*$ transition,

\begin{eqnarray*}
\hspace*{-2.5cm}
C_3^{\Sigma\Sigma^*}(q^2)\frac{\lambda_{\Sigma^*}}{M_{\Sigma^*}^2-p'^{2}}& =\bigg\{\frac{M_\Sigma^2}{\sqrt{3}} \int_0 ^{1}d\alpha
         \frac{1-\alpha}{(q-p\alpha)^4} F_{21}+\frac{2}{\sqrt{3}} \int_0^{1} dx_2 \frac{1}{(q-px_2)^2}F_{22}\\
         &- \frac{2 M_\Sigma^{2}}{\sqrt{3}} \int_0 ^{1}d\beta \frac{1}{(q-p\beta)^4} F_{23} \bigg\}\\
        \end{eqnarray*}
         \begin{eqnarray*}
         \hspace*{-2.5cm}
C_4^{\Sigma\Sigma^*}(q^2)\frac{\lambda_{\Sigma^*}}{M_{\Sigma^*}^2-p'^{2}}& =\frac{M_\Sigma}{\sqrt{3}}\bigg\{M_\Sigma^2 \int_0 ^{1}d\beta
         \frac{\beta(1-\beta)}{(q-p\beta)^6} F_{24}-\int_0^{1} d\alpha  \frac{\alpha}{(q-p\alpha)^4}F_{25}\\
         \\
        &-2 \int_0 ^{1}d\alpha \frac{1-\alpha}{(q-p\alpha)^4} F_{26}\bigg\}   \\
         \end{eqnarray*}
         \begin{eqnarray*}
         \hspace*{-2.5cm}
C_5^{\Sigma\Sigma^*}(q^2)\frac{\lambda_{\Sigma^*}}{M_{\Sigma^*}^2-p'^{2}}& =\frac{M_\Sigma}{\sqrt{3}}\bigg\{M_\Sigma^2 \int_0 ^{1}d\beta
         \frac{1}{(q-p\beta)^4} F_{27}+ \int_0^{1} dx_2 \frac{1}{(q-px_2)^2}F_{28}
         \\
         &  -2 M_\Sigma^{2} \int_0 ^{1}d\beta \frac{1}{(q-p\beta)^4} F_{29}+ \int_0^{1} dx_2 \frac{1-x_2}{(q-px_2)^2}F_{30}(x_2)\\
         \\
         &+M_\Sigma^2 \int_0 ^{1}d\beta \frac{1-\beta}{(q-p\beta)^4} F_{31}+M_\Sigma^2 \int_0^{1} d\alpha \frac{\alpha(1-\alpha)}{(q-p\alpha)^4}F_{32}\\
         \\
        &-2\int_0^{1} dx_2 \frac{x_2}{(q-px_2)^2}F_{33}+2 M_\Sigma^{2} \int_0 ^{1}d\beta \frac{\beta}{(q-p\beta)^4} F_{34}
      \\
       \\
       & +2 M_\Sigma^{2} \int_0 ^{1}d\beta \frac{1-\beta}{(q-p\beta)^4} F_{35}-\int_0^{1} d\alpha \frac{1}{(q-p\alpha)^4}F_{36}\bigg\}\\
        \end{eqnarray*}

         \begin{eqnarray}\label{sig}
         \hspace*{-2.5cm}
C_6^{\Sigma\Sigma^*}(q^2)\frac{\lambda_{\Sigma^*}}{M_{\Sigma^*}^2-p'^{2}}& =-\frac{4M_\Sigma}{\sqrt{3}}\bigg\{M_\Sigma^2 \int_0 ^{1}d\beta
         \frac{(1-\beta)^2}{(q-p\beta)^6} F_{37}- \int_0^{1} d\alpha \frac{1-\alpha}{(q-p\alpha)^4}F_{38} \bigg\}
        \end{eqnarray}
for the $\Sigma$-$\Sigma^*$ transition,

\begin{eqnarray*}
\hspace*{-2.5cm}
C_3^{\Lambda\Sigma^*}(q^2)\frac{\lambda_{\Sigma^*}}{M_{\Sigma^*}^2-p'^{2}}& =\bigg\{\frac{M_\Lambda^2}{\sqrt{3}} \int_0 ^{1}d\alpha
         \frac{1-\alpha}{(q-p\alpha)^4} F_{21}+ \frac{2}{\sqrt{3}} \int_0^{1} dx_2 \frac{1}{(q-px_2)^2}F_{22}\\
        & -  \frac{2 M_\Lambda^{2}}{\sqrt{3}} \int_0 ^{1}d\beta \frac{1}{(q-p\beta)^4} F_{23} \bigg\}\\
         \end{eqnarray*}
         \begin{eqnarray*}
         \hspace*{-2.5cm}
C_4^{\Lambda\Sigma^*}(q^2)\frac{\lambda_{\Sigma^*}}{M_{\Sigma^*}^2-p'^{2}}& =\frac{M_\Lambda}{\sqrt{3}}\bigg\{M_\Lambda^2 \int_0 ^{1}d\beta
         \frac{\beta(1-\beta)}{(q-p\beta)^6} F_{24}-\int_0^{1} d\alpha \frac{\alpha}{(q-p\alpha)^4}F_{25}\\
         &-2 \int_0 ^{1}d\alpha \frac{1-\alpha}{(q-p\alpha)^4} F_{26}\bigg\}   \\
         \end{eqnarray*}
         \begin{eqnarray*}
         \hspace*{-2.5cm}
C_5^{\Lambda\Sigma^*}(q^2)\frac{\lambda_{\Sigma^*}}{M_{\Sigma^*}^2-p'^{2}}& =\frac{M_\Lambda}{\sqrt{3}}\bigg\{M_\Lambda^2 \int_0 ^{1}d\beta
         \frac{1}{(q-p\beta)^4} F_{27}+ \int_0^{1} dx_2 \frac{1}{(q-px_2)^2}F_{28}
         \\
         \\
         &-2 M_\Lambda^{2} \int_0 ^{1}d\beta \frac{1}{(q-p\beta)^4} F_{29}+ \int_0^{1} dx_2 \frac{1-x_2}{(q-px_2)^2}F_{30}(x_2)\\
         \\
         &+M_\Lambda^2 \int_0 ^{1}d\beta \frac{1-\beta}{(q-p\beta)^4} F_{31}+M_\Lambda^2 \int_0^{1} d\alpha \frac{\alpha(1-\alpha)}{(q-p\alpha)^4}F_{32}\\
         \\
        &-2\int_0^{1} dx_2 \frac{x_2}{(q-px_2)^2}F_{33}+2 M_\Lambda^{2} \int_0 ^{1}d\beta \frac{\beta}{(q-p\beta)^4} F_{34} \\
       \\
       & +2 M_\Lambda^{2} \int_0 ^{1}d\beta \frac{1-\beta}{(q-p\beta)^4} F_{35}-\int_0^{1} d\alpha \frac{1}{(q-p\alpha)^4}F_{36}(\alpha)\bigg\}\\
        \end{eqnarray*}

         \begin{eqnarray}\label{lam}
         \hspace*{-2.5cm}
C_6^{\Lambda\Sigma^*}(q^2)\frac{\lambda_{\Sigma^*}}{M_{\Sigma^*}^2-p'^{2}}& =-\frac{4M_\Lambda}{\sqrt{3}}\bigg\{M_\Lambda^2 \int_0 ^{1}d\beta
         \frac{(1-\beta)^2}{(q-p\beta)^6} F_{37}- \int_0^{1} d\alpha \frac{1-\alpha}{(q-p\alpha)^4}F_{38}(\alpha) \bigg\}
        \end{eqnarray}
for the $\Lambda$-$\Sigma^*$ transition. The explicit form of the functions that appear
in Eqs. (\ref{xi}), (\ref{sig}) and (\ref{lam}) are given in Appendix A.

In order to eliminate the subtraction terms in the spectral representation of the correlation function, the Borel transformation is performed.
After the transformation, contributions from excited and continuum states are also exponentially suppressed. Then, using the quark-hadron duality
and subtracted, the contributions of the higher states and the continuum can be modelled. Both of the Borel transformation and the subtraction of higher states are applied by using following substitution rules (see e.g. \cite{Braun:2006hz}):

\begin{eqnarray*}
		&\int dx \frac{\rho(x)}{(q-xp)^2}\rightarrow -\int_{x_0}^1\frac{dx}{x}\rho(x) e^{-s(x)/M^2},\\
		&\int dx \frac{\rho(x)}{(q-xp)^4}\rightarrow \frac{1}{M^2} \int_{x_0}^1\frac{dx}{x^2}\rho(x) e^{-s(x)/M^2}+\frac{\rho(x)}{Q^2+x_0^2 m_B^2} e^{-s_0/M^2},\\
        &\int dt \frac{\rho(t)}{(q-xp)^6}\rightarrow -\frac{1}{2M^4}\int_{x_0}^1\frac{dx}{x^3}\rho(x) e^{-s(x)/M^2}\\
        &-\frac{1}{2M^2}\frac{\rho(x)}{x_0(Q^2+x_0^2m_B^2)}
		+\frac{1}{2}\frac{x_0^2}{Q^2+x_0^2m_B^2}\bigg[\frac{d}{dx_0}\frac{\rho(x_0)}{x_0(Q^2+x_0^2m_B^2)}\bigg],
\end{eqnarray*}
where
\[s(x)=(1-x)m_B^2+\frac{1-x}{x}Q^2,\]
$M$ is the Borel mass, $m_B = M_H$ and $x_0$ is the solution of the quadratic equation for $s=s_0$:
\[x_0=\left[\sqrt{(Q^2+s_0-m_B^2)^2+4m_B^2 Q^2}-(Q^2+s_0-m_B^2)\right]/(2m_B^2),\] where $s_0$ is the continuum threshold.

\section{Numerical Analysis and Conclusion}

In this section, we present LCSR results for the octet-decuplet hyperon transition form factors.
To obtain our numerical results the main input parameters are the baryon DAs.
The DAs of the hyperon depending on various non-perturbative 
parameters are studied  in \cite{Liu:2009uc, Liu:2008yg, Liu:2013bxa, Liu:2014uha}.
In Table.\ref{parameter_table} we present the values of the input parameters using the DAs of each hyperon.
In this section, we will only consider the central values of these parameters.

\begin{table}[t]
	\addtolength{\tabcolsep}{2pt}
	\begin{center}

\begin{tabular}{ccccccccc}
		\hline\hline
		Parameter  & $\Sigma$ & $\Xi$& $\Lambda$& \\[0.5ex]
		\hline
		 $f_B$~(GeV$^2$)         & 0.0094 & 0.0099&  0.0060 \\
		 $\lambda_1$~(GeV$^2$)   & -0.025 & -0.028&  0.0083  \\
		 $\lambda_2$~(GeV$^2$)   & 0.044  & 0.052 &  0.0083 \\
		 $\lambda_3$~(GeV$^2$)   & 0.02   & 0.017 &  0.010\\[0.8ex]
          \hline\hline\\
           & $\Sigma$ &\\
          \hline\hline
		  $V_1^s$ & $A_1^u$ & $f_1^s$ & $f_2^s$ \\
		
		     0.39 & 0.29    & -0.15   &  9.9 &\\
		\hline
		   $f_3^s$ & $f_1^u$ & $P_2^0$& $S_1^u$\\[0.8ex]
		
		   1.6 & -0.11 & 0.004 & -0.0014 &  \\[0.5ex]
		   \hline\hline\\
                      & $\Lambda$ &\\
                      \hline\hline
                       $A_1^s$ & $A_1^q$ & $f_1^s$  \\
		
		        0.31   & 0.032    & 0.23  & \\
		\hline
		   $f_1^q$ & $f_3^q$ & $f_4^q$\\[0.8ex]
		
		   -0.23 & 0.43 & 1.07 &   \\[0.5ex]

          \hline\hline
          \end{tabular}
          \caption{The values of the parameters are used in the DAs of  $\Sigma$, $\Lambda$  and $\Xi$.
          The first column includes the dimensionful parameters for each baryon.
          In the other four columns we represent the list of the values of parameters
          that determine the shape of the DAs,
          which have been extracted for $\Sigma$ and $\Lambda$. For $\Xi$ these parameters are taken as zero.}
\label{parameter_table}
	\end{center}
\end{table}

We use the values of some  parameters as follow; $M_\Lambda = 1.11~GeV$, $M_\Sigma = 1.18~GeV$,
$M_\Sigma* = 1.38~GeV$, $M_\Xi = 1.31 GeV$, $M_\Xi* = 1.53~GeV$ \cite{Beringer:1900zz}.
To evaluate a numerical prediction for the form factors, we need also specify the values of the residue
of $\Sigma^*$ and $\Xi^*$.
The residues can be determined from the mass sum rules as $\lambda_\Sigma*= 0.043~GeV^3$ and
$\lambda_\Xi*= 0.053~GeV^3$ \cite{Lee:1997ix} which are used in our calculations.
%
The predictions for  the form factors depend on two auxiliary parameters:
the Borel mass $M^2$, and the continuum threshold $s_0$.
The continuum threshold signals the scale at which, the excited states and
continuum start to contribute to the correlation function. Hence, it is choosen as
$s_0\simeq(m_{\Sigma^*}+0.3~GeV)^2=2.82~GeV^2$ for $\Sigma^*$ and $s_0\simeq(m_{\Xi^*}+0.3~GeV)^2=3.34~GeV^2$ for $\Xi^*$.
One approach to determine the continuum threshold and the working region of the Borel
parameter $M^2$ is to plot the dependence of the predictions on $M^2$ for a range of values
of the continuum threshold and determine the values of $s_0$ for which there is a stable
region with respect to variations of the Borel parameter $M^2$. For this reason,
in Figs.(1), (2) and (3) we plot the dependence of
the form factors on $M^{2}$
for  two fixed values of $Q^2$ and for various values of $s_0$ in the range $2.5 ~ GeV^2 < s_0 < 4,5~GeV^2$.
As can be seen from these figures, for $s_0=3.0\pm0.5~GeV^2$ for $\Sigma-\Sigma^*$, $\Lambda-\Sigma^*$ and
$s_0=3.5\pm0.5~GeV^2$ for $\Xi-\Xi^*$ the predictions are practically independent of the value of $M^2$ for the related range.
The uncertainty due to variation of $s_0$ in this range is much larger than the uncertainty due to variations
with respect to $M^2$.
Note that the determined range of $s_0$ is in the range that one would expect from the physical interpretation of $s_0$.
%
%
In Figs.(4), (5) and (6) we also represent twist contributions of DAs.
In these figures, the form factors of $\Sigma-\Sigma^*$ and $\Xi-\Xi^*$  are represent very good asymptotic behaviour.
Twist-3 and twist-4  give dominant contribution for these two transitions, respectively, but
 twist-5 and 6  give very small contribution.
In the form factors of $\Lambda-\Sigma^*$ transition we expect dominant contribution coming from twist-3
but it comes from higher twists.
In Figs. (7), (8) and (9) we plot the form factors of  hyperons, $C_3^{HH^*}$, $C_4^{HH^*}$, $C_5^{HH^*}$ and $C_6^{HH^*}$,
as a function of $Q^2$ in the region $1~GeV^2\leq Q^2 \leq 10$ $GeV^2$.
We see that the behavior of the form factors  agree well with our expectations except $C_5^{\Lambda \Sigma^*}$.
While the form factor $C_5^{\Lambda \Sigma^*}(Q^2)$ in the region of $1~GeV^2\leq Q^2 \leq 2$ $GeV^2$ has positive sign,
 in the region of  $2~GeV^2\leq Q^2 \leq 10$ $GeV^2$ the form factor changes it's sign.
For this reason we have not any prediction about the form factor $C_5^{\Lambda \Sigma^*}(Q^2)$.
In Ref.\cite{Liu:2008yg} DAs of $\Sigma$ and $\Lambda$ baryons have been calculated by employed QCD sum rules
but this DAs do not contain higher order terms. In Refs.\cite{Liu:2013bxa, Liu:2014uha} higher order corrections have been calculated for these baryons.
As seen second and third column of the Table.I, calculated higher order corrections of $\Sigma$ and $\Lambda$ give dominant contributions.
For that reason the form factors of $\Sigma$ and $\Lambda$ give higher values by comparison with the $\Xi$ form factors.

The form factor $C_5^{HH^*}$ is the dominant axial-vector form factor
and the only one that can be extracted directly from the matrix element at $Q^2 = 0$, determining the axial charge of
the transitions $\Sigma-\Sigma^*$, $\Xi-\Xi^*$ and $\Lambda-\Sigma^*$. In order to extrapolate,
we have tried to fit  LCSR results. To do this, we use an exponential form for $C_5^{HH^*}$ as
\begin{equation}
	C_5^{HH^*}(Q^2) = C_5^{HH^*}(0) \exp[-Q^2/m_{A}^2]
\end{equation}
with which we can make reasonable description of data with a two-parameter fit. Using this from, we have studied three
fit region $ Q^{2}$ $\geq $ $1~ GeV^2 $, $ Q^{2}$ $\geq $ $1.5~ GeV^2 $ and $ Q^{2}$ $\geq $ $2~ GeV^2 $.

Our results for axial charges and axial masses are presented in Table \ref{fit_table}.
Obtained values could not be compared any results that available recently.
We observe that the axial masses are very close to $\Sigma$ and $\Xi$.
At this point, it will be very interesting to compare our results to those from
different method in the near future.
The mass of the lightest axial vector meson is $m_A =1.23~ GeV$ \cite{Beringer:1900zz};
therefore the axial masses of the $\Sigma$ and $\Xi$ have close values with the predictions of the
VMD model.

\begin{table}[t]
	\addtolength{\tabcolsep}{2pt}
	\begin{center}
\begin{tabular}{ccccccc}
				\hline\hline
		Baryon & Fit Region~(GeV$^2$) & $C_5^{HH^*}(0)$ & $m_{A}$~(GeV)& \\[0.5ex]
		\hline

		                 & [1.0-10] & -41.80 & 1.22  &  \\
		$\Sigma-\Sigma^*$ & [1.5-10] & -28.19 & 1.38  &  \\
		                 & [2.0-10] & -21.96 & 1.49  &  \\[1ex]
		  \hline
			   & [1.0-10] & 2.73 & 1.20  &  \\
		$\Xi-\Xi^*$ & [1.5-10] & 2.13 & 1.29  &  \\
		           & [2.0-10] & 1.72 & 1.37  &  \\
		\hline\hline
	\end{tabular}
\caption{The values of exponential fit parameters, $C_5^{HH^*}(0)$ and $m_{A}$ for axial  form factors.
The results include the fits from three region. }
	\label{fit_table}
		\end{center}
\end{table}

To summarize, we have extracted the isovector axial vector form factors of octet-decuplet hyperons by applying the LCSR.
Studied the form factors in this work have been calculated for the first time in the literature.
These form factors bring information about the shape, size and axial charge of the baryons.
We also obtain the axial charges $C_5(0)$ for all transitions by using the fit to an exponential form.
Our axial charge results are $C_5^{\Sigma \Sigma^*}(0)$ = - $30.65 \pm 10$
and $C_5^{\Xi \Xi^*}(0)$ =  $2.19 \pm 0.5$.
We could not find any prediction for axial charge of $\Lambda-\Sigma^*$ transition, because of unstable behaviour of
$C_5^{\Lambda \Sigma^*}(Q^2)$.
Unfortunately, there is no sufficient experimental data yet to compare our results within this region.
Maybe we compare our axial charge result with $N-\Delta$ transition results.
We observed that the prediction of quark model results change from  $C_5^{N\Delta}(0)$ = $0.81$ to $1.53$ \cite{BarquillaCano:2007yk},
in the case of chiral perturbation theory is $C_5^{N\Delta}(0)$ = $1.16$ \cite{Geng:2008bm}, in the case of lattice QCD is $C_5^{N\Delta}(0)$ = $0.9 \pm 0.02$ \cite{Alexandrou:2007xj} and the results from weak pion production $C_5^{N\Delta}(0)$ = $1.08 \pm 0.1$ \cite{Hernandez:2010bx}.
The experimental result is $G_A^{N\Delta}$ = - $ 0.05\pm(0.35)_{stat} \pm(0.34)_{sys} \pm(0.06)_{theory}$ \cite{Androic:2012doa}.
We see that our results different from other theoretical approaches and experimental results.

\ack
  This work has been supported by The Scientific and Technological Research Council of Turkey (TUBITAK) under
project number 110T245. The work of A. O. and A.K. is also partially supported by the European Union (HadronPhysics2 project Study
  of strongly interacting matter).

  \section{References}
\bibliography{refs}

\newpage
\begin{figure}[htp]
\centering
 \subfloat[]{\label{fig:SigC3Msq.eps}\includegraphics[width=0.4\textwidth]{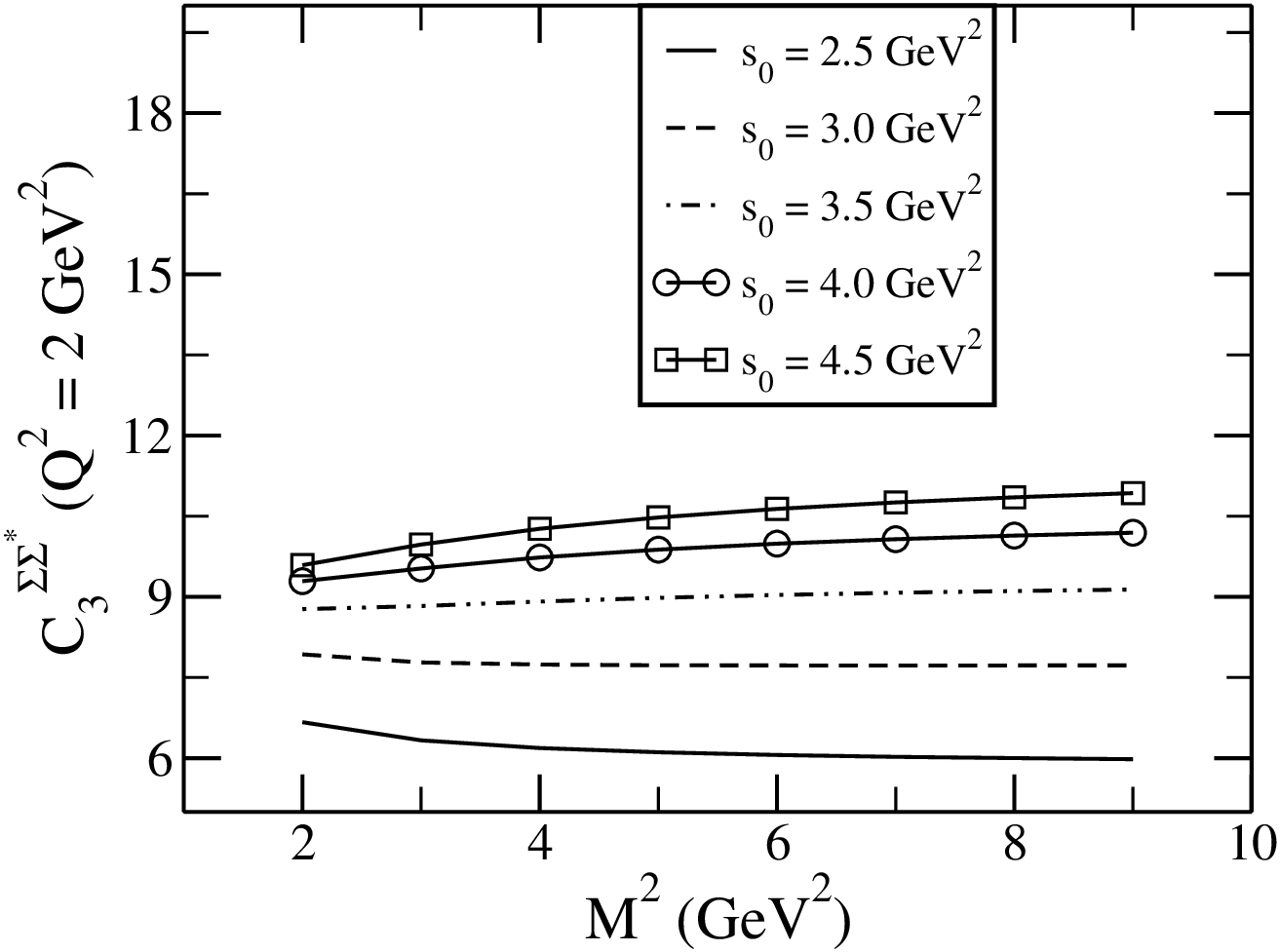}}
 \subfloat[]{\label{fig:SigC3Msq1.eps}\includegraphics[width=0.4\textwidth]{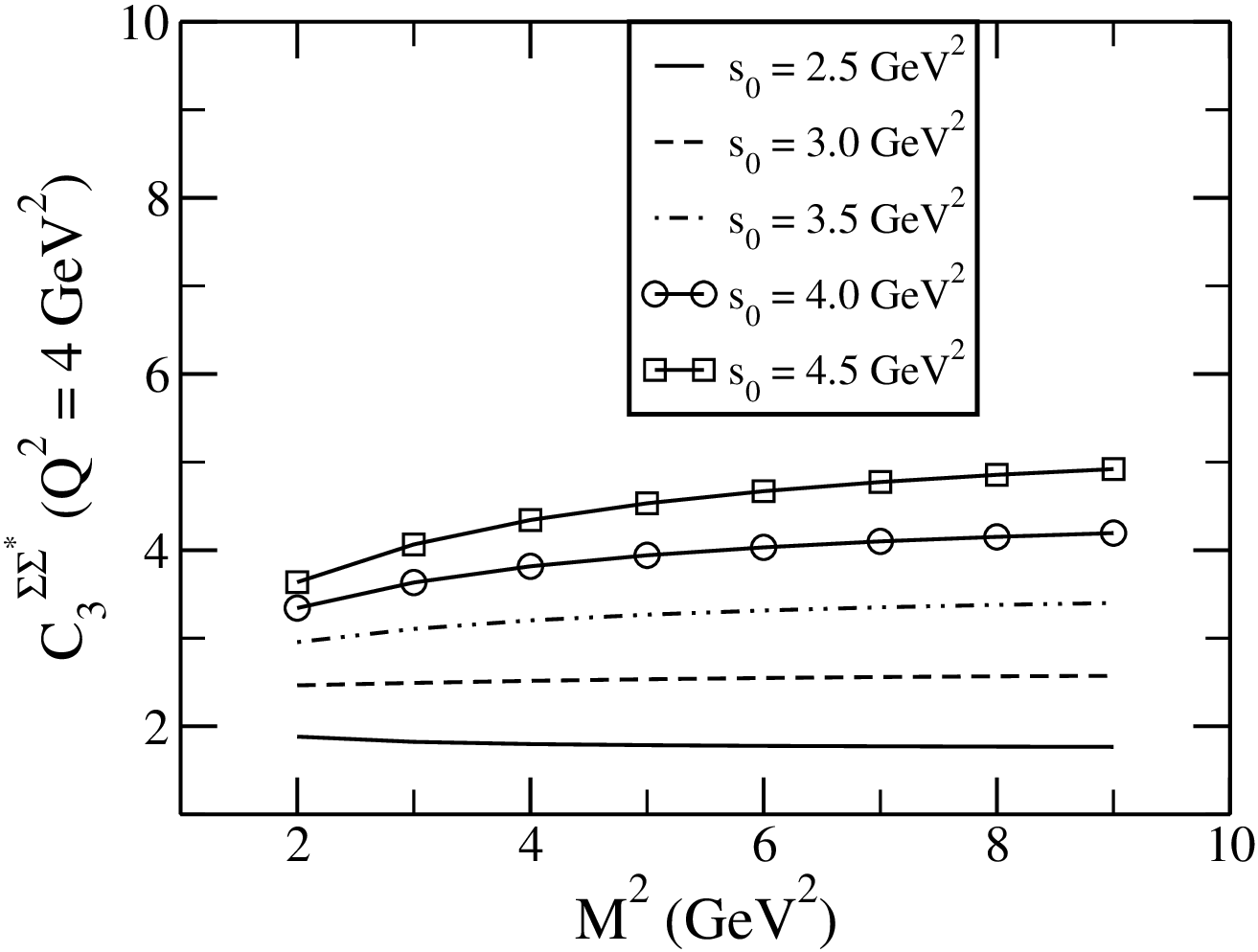}}\\
  \subfloat[]{\label{fig:SigC4Msq.eps}\includegraphics[width=0.4\textwidth]{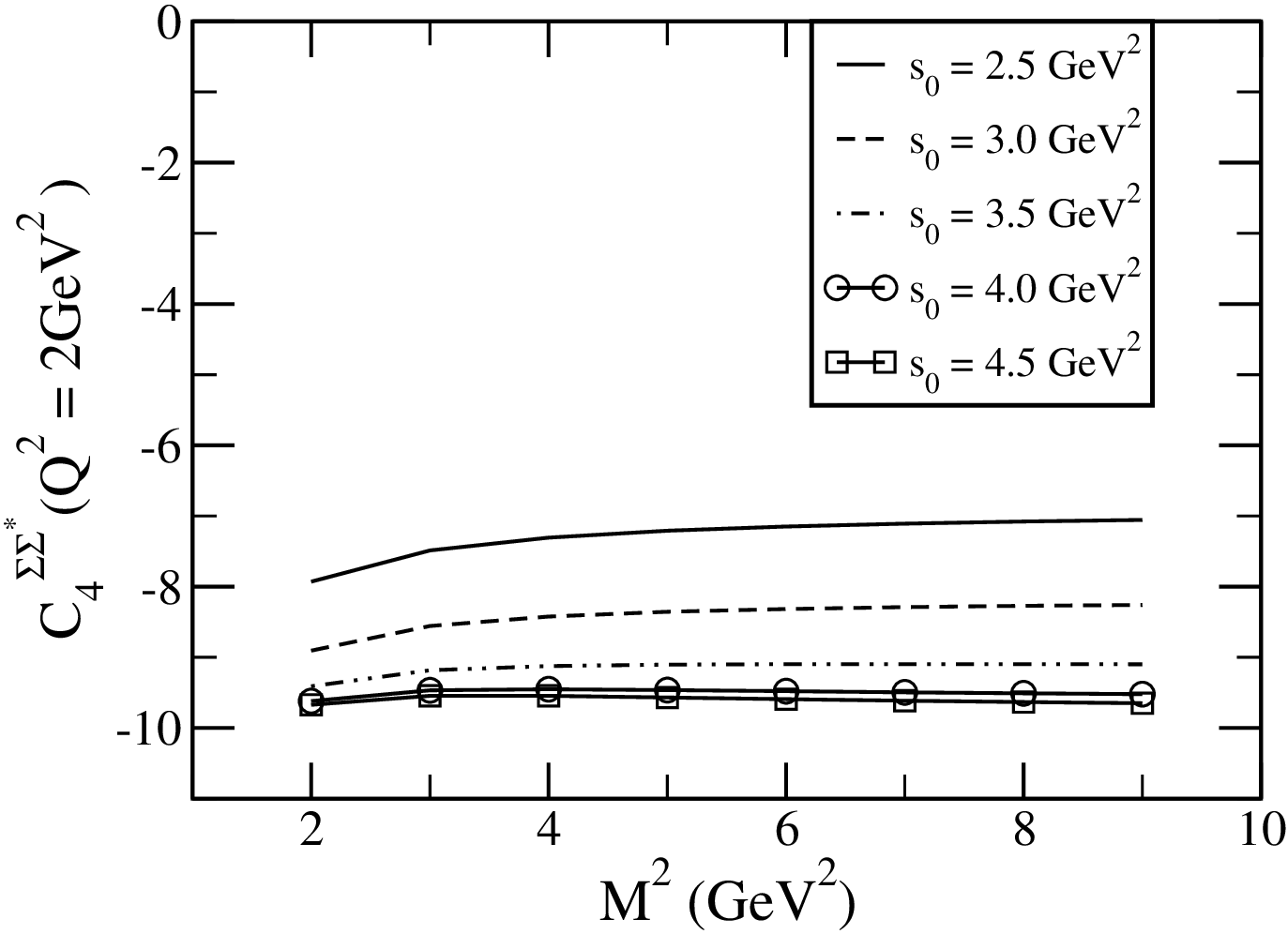}}
  \subfloat[]{\label{fig:SigC4Msq1.eps}\includegraphics[width=0.4\textwidth]{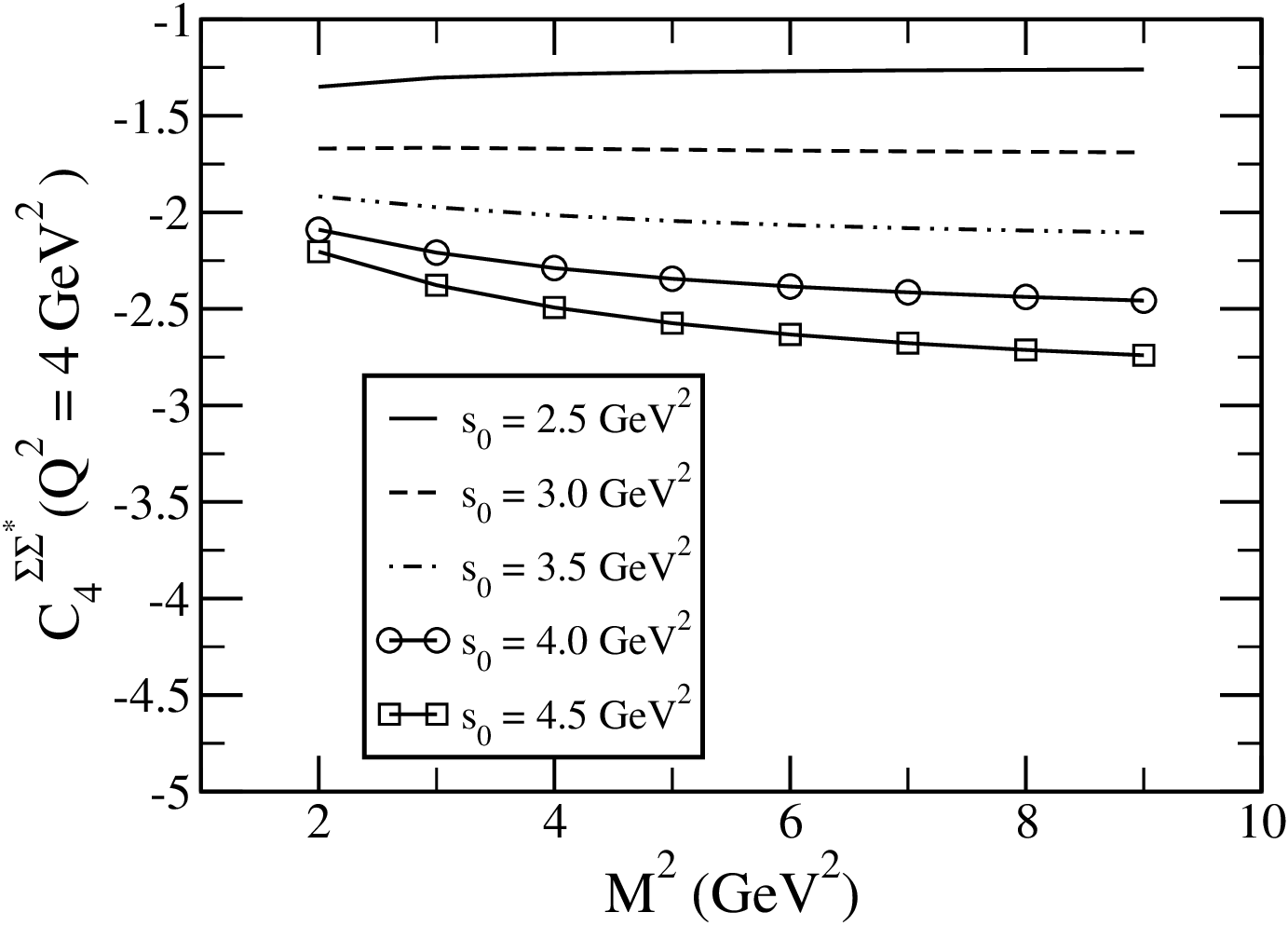}}
  \\
   \subfloat[]{\label{fig:SigC5Msq.eps}\includegraphics[width=0.4\textwidth]{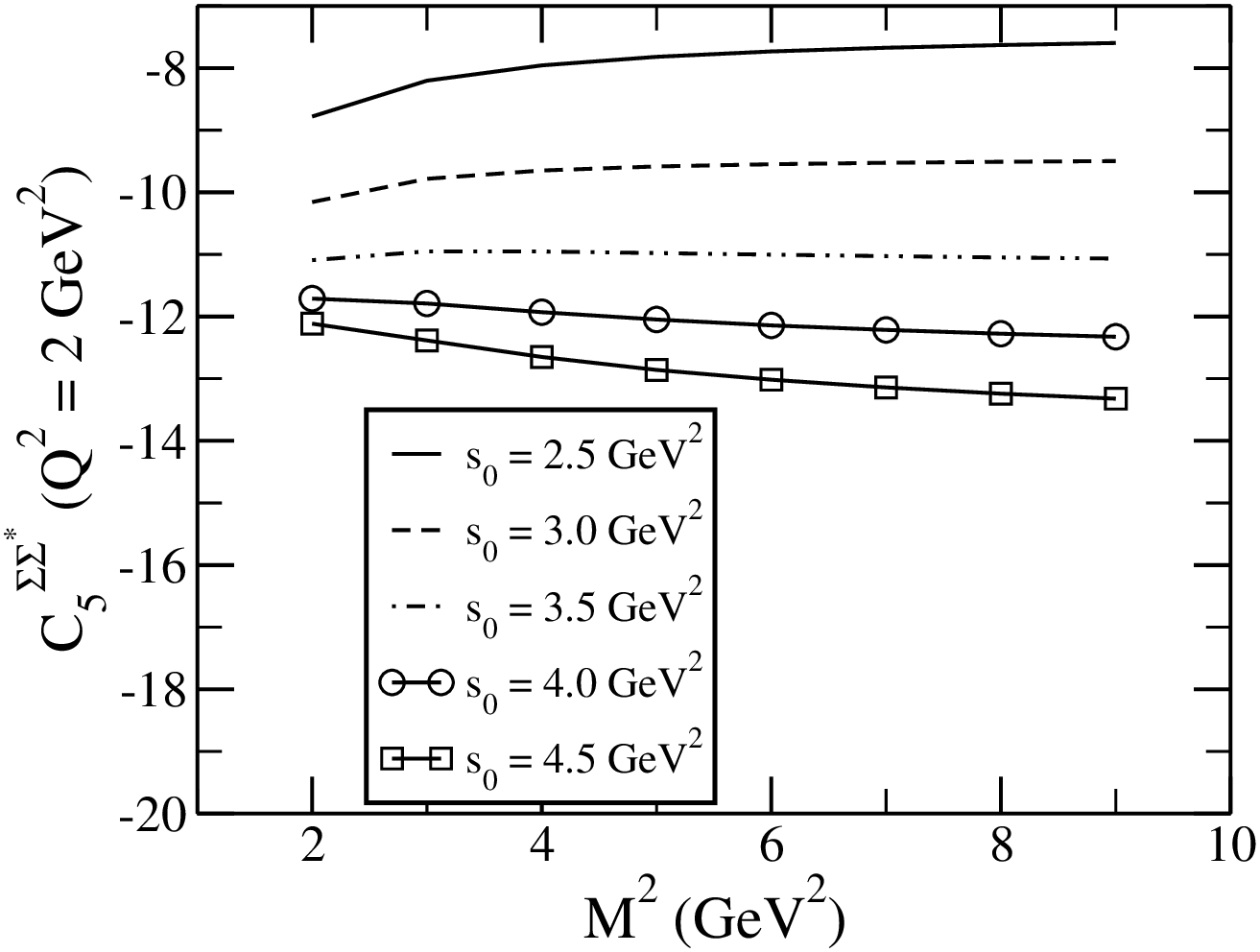}}
   \subfloat[]{\label{fig:SigC5Msq1.eps}\includegraphics[width=0.4\textwidth]{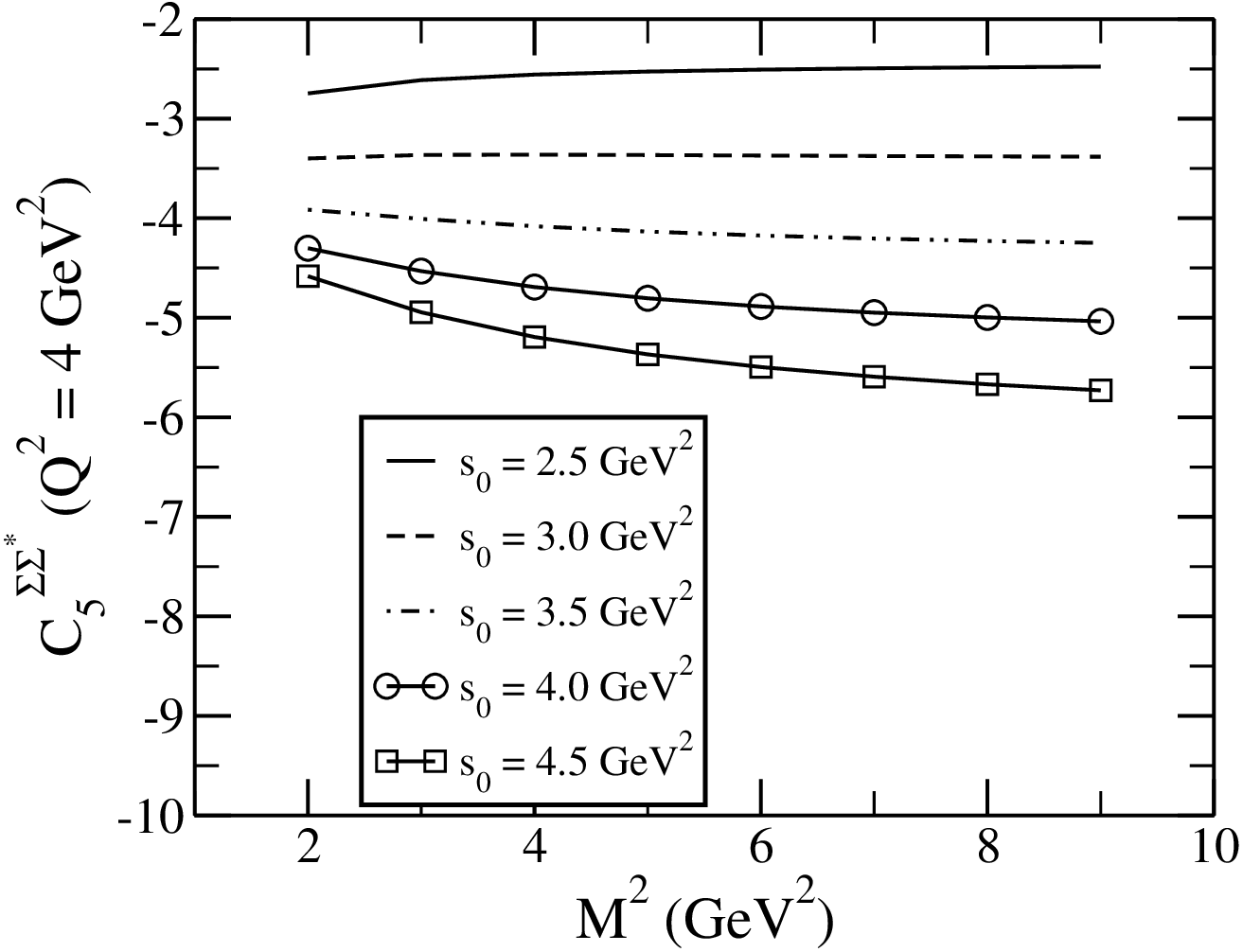}}\\
    \subfloat[]{\label{fig:SigC6Msq.eps}\includegraphics[width=0.4\textwidth]{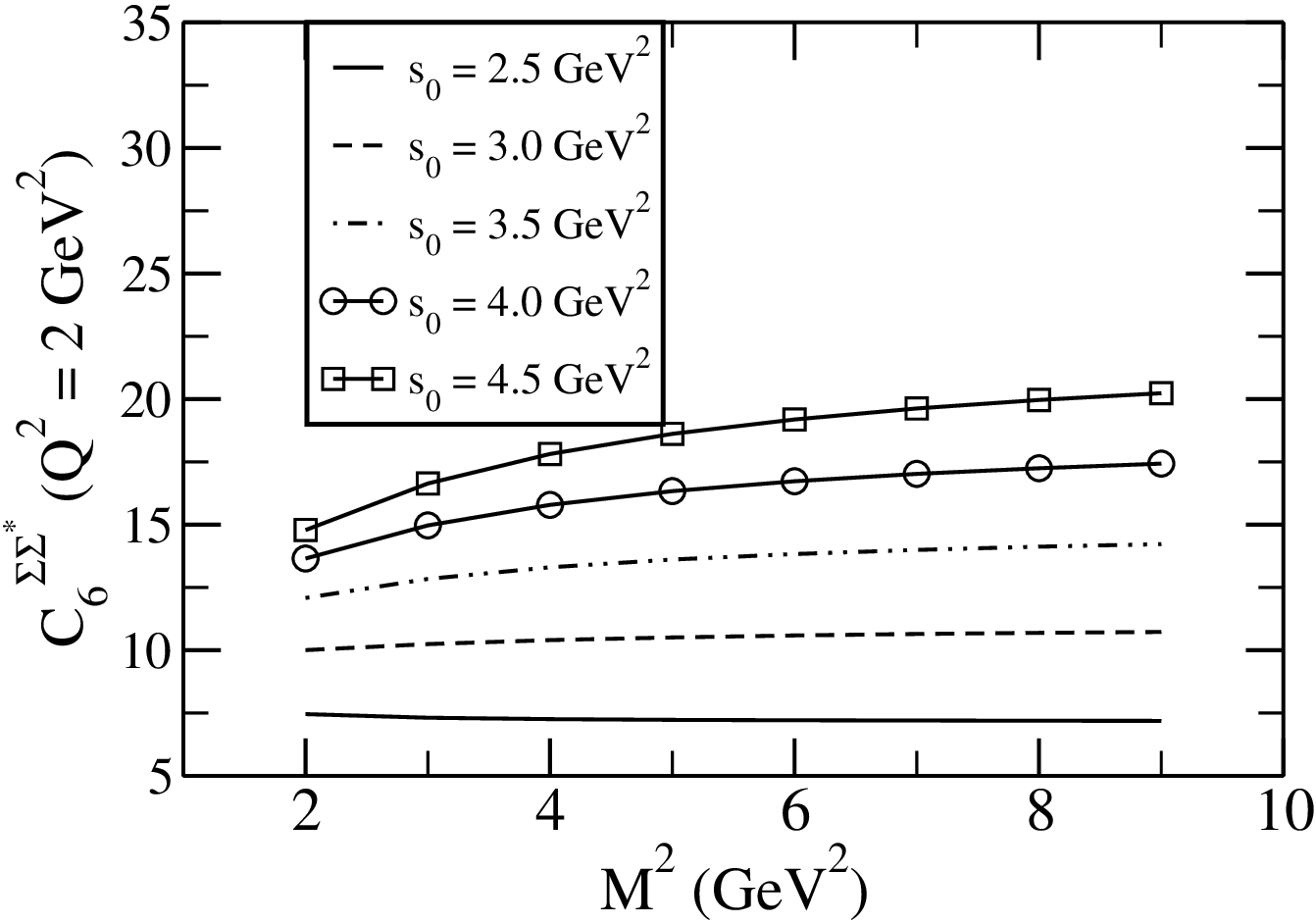}}
    \subfloat[]{\label{fig:SigC6Msq1.eps}\includegraphics[width=0.4\textwidth]{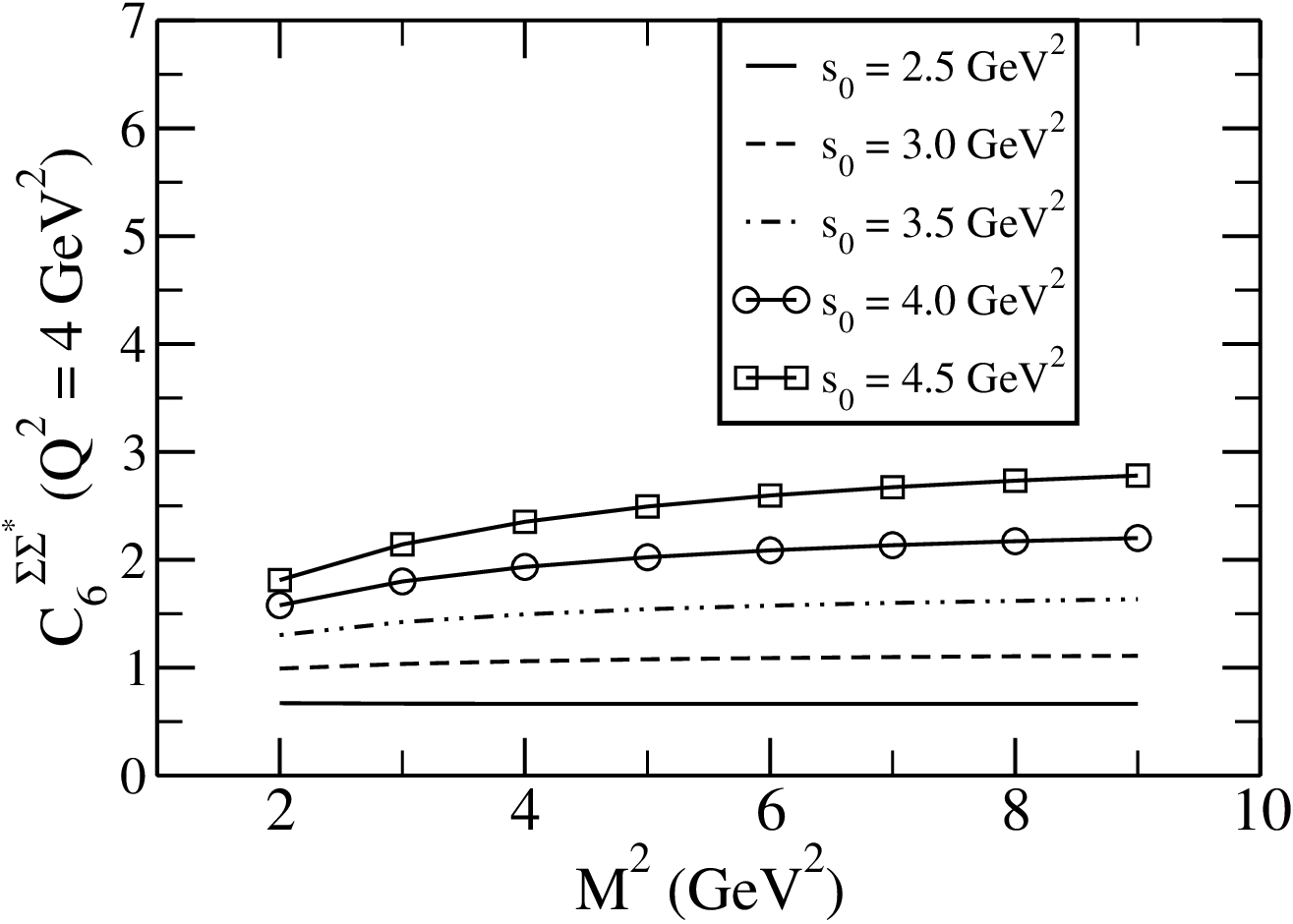}}
\caption{The dependence of the form factors; on the Borel parameter squared $M^{2}$
  for the values of the continuum threshold  $s_0 = 2.5 ~GeV^2$, $s_0 = 3.0~GeV^2$, $s_0 = 3.5~GeV^2$, $s_0 = 4.0~GeV^2$, $s_0 = 4.5~GeV^2$
  and $Q^2 = 2~ GeV^2$ and $4~ GeV^2$
(a) and (b) for $C_3^{\Sigma\Sigma^*}$ form factor,
(c) and (d) for $C_4^{\Sigma\Sigma^*}$ form factor,
(e) and (f) for $C_5^{\Sigma\Sigma^*}$ form factor and
(g) and (h) for $C_6^{\Sigma\Sigma^*}$ form factor.}
\end{figure}

\newpage
\begin{figure}[htp]
\centering
 \subfloat[]{\label{fig:C3Msq.eps}\includegraphics[width=0.4\textwidth]{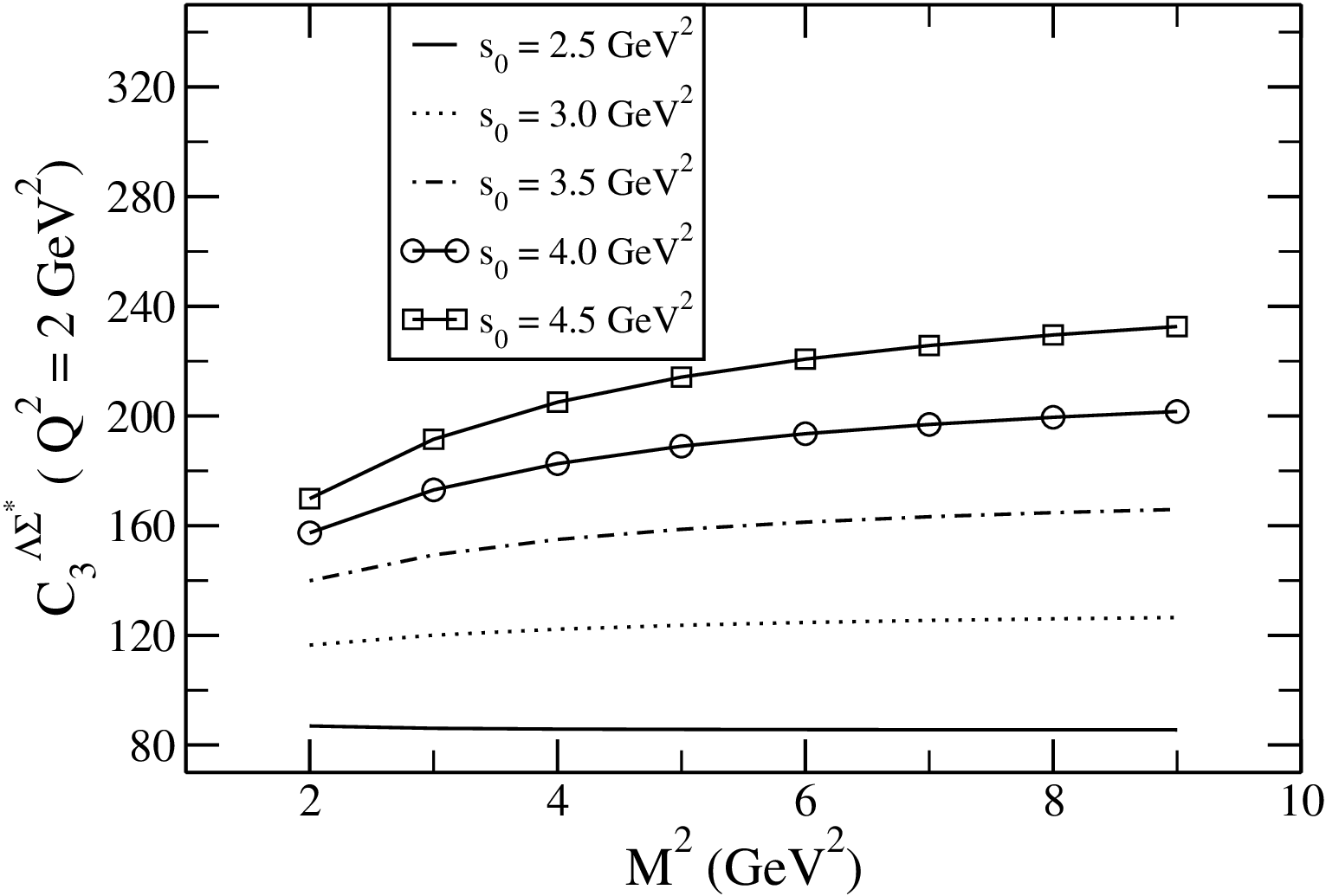}}
 \subfloat[]{\label{fig:C3Msq1.eps}\includegraphics[width=0.4\textwidth]{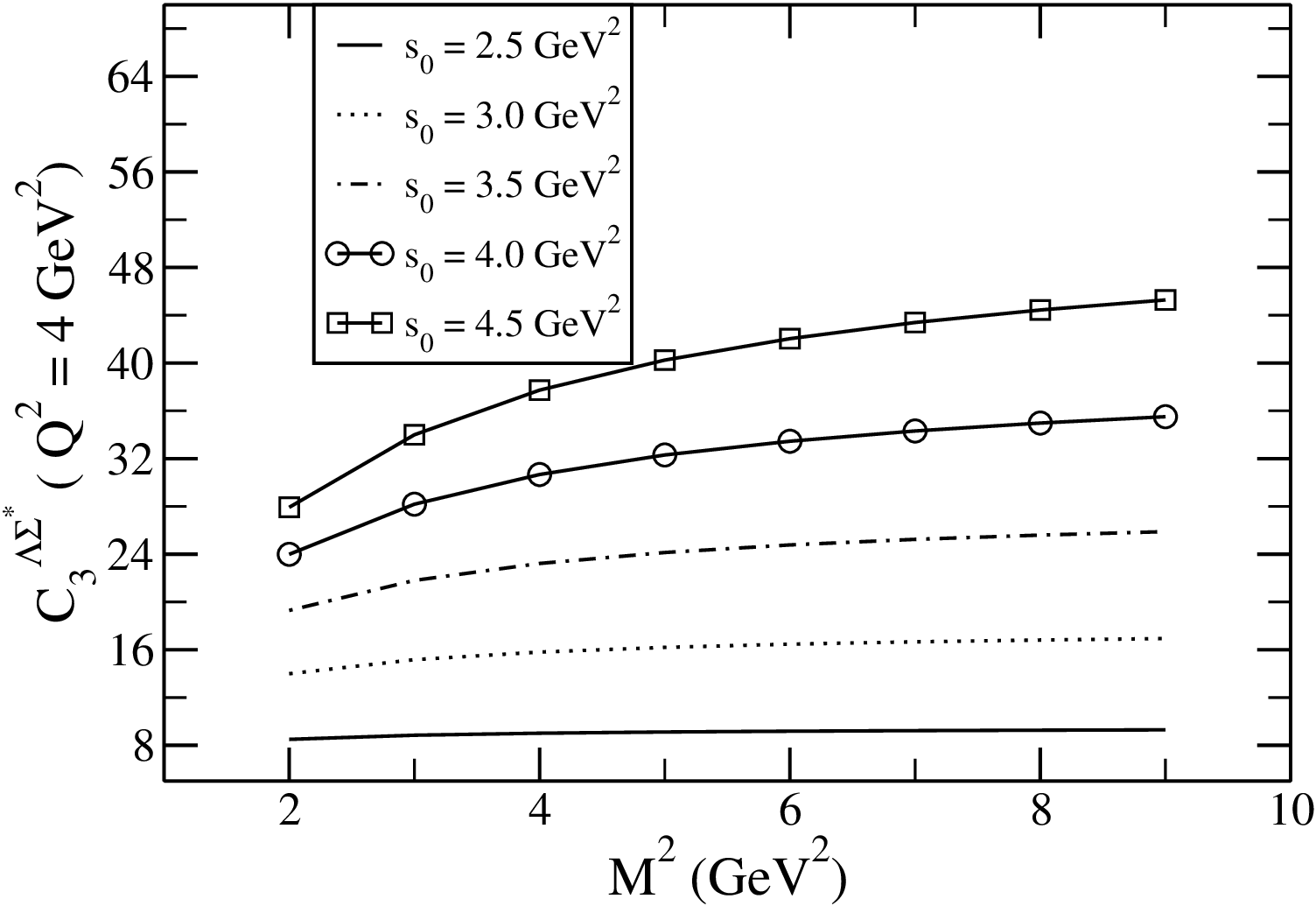}}\\
  \subfloat[]{\label{fig:C4Msq.eps}\includegraphics[width=0.4\textwidth]{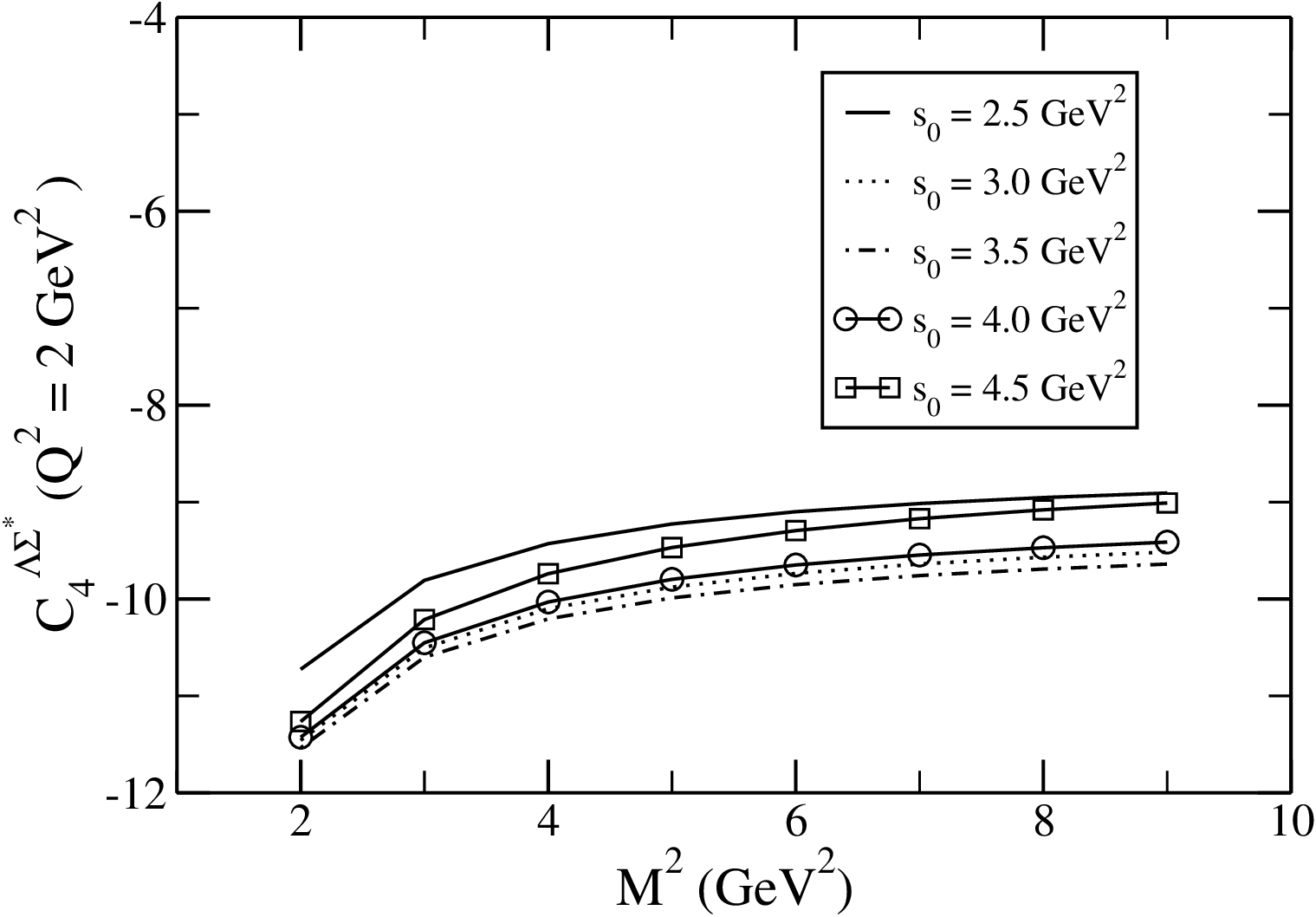}}
  \subfloat[]{\label{fig:C4Msq1.eps}\includegraphics[width=0.4\textwidth]{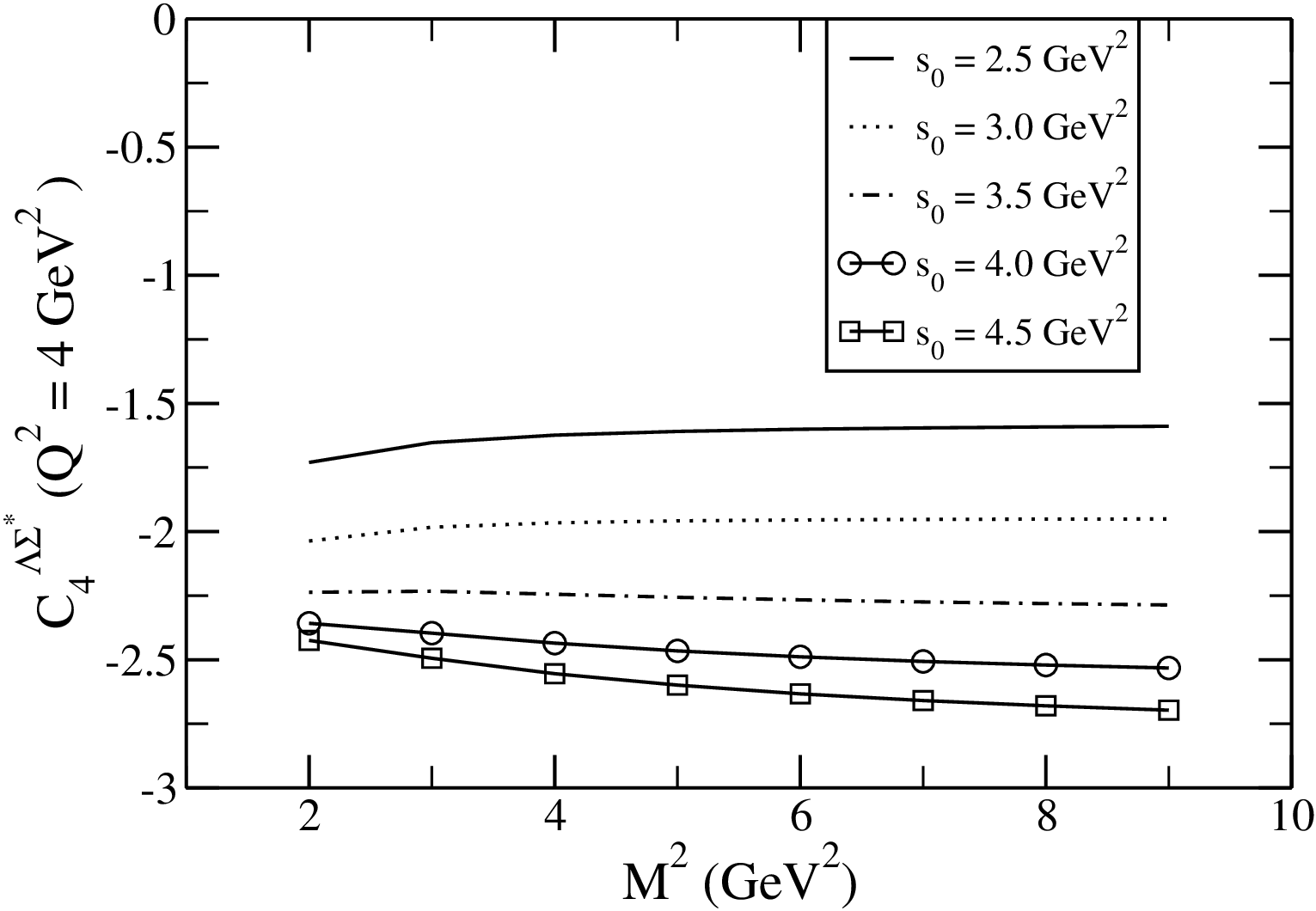}}
  \\
   \subfloat[]{\label{fig:C5Msq.eps}\includegraphics[width=0.4\textwidth]{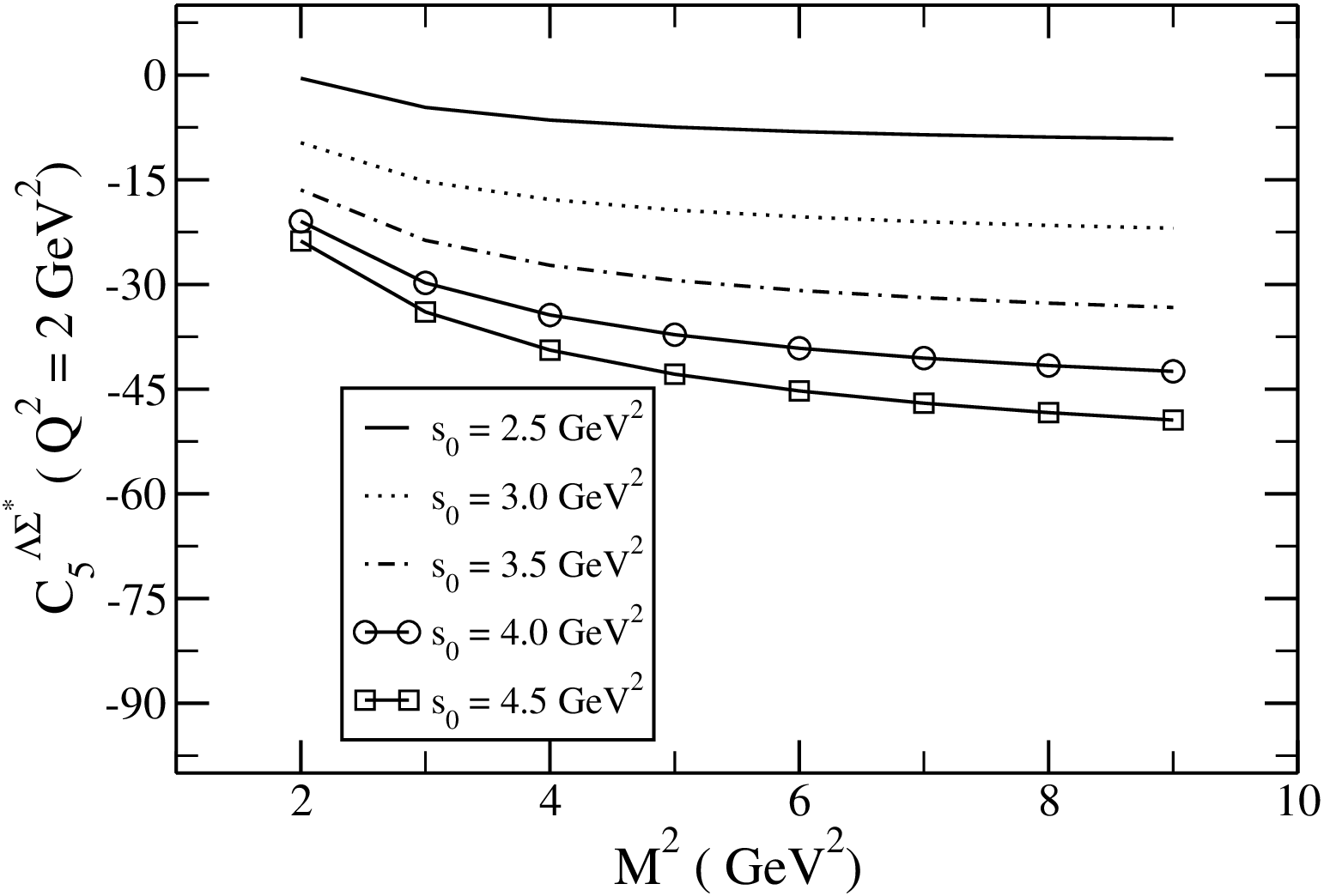}}
   \subfloat[]{\label{fig:C5Msq1.eps}\includegraphics[width=0.4\textwidth]{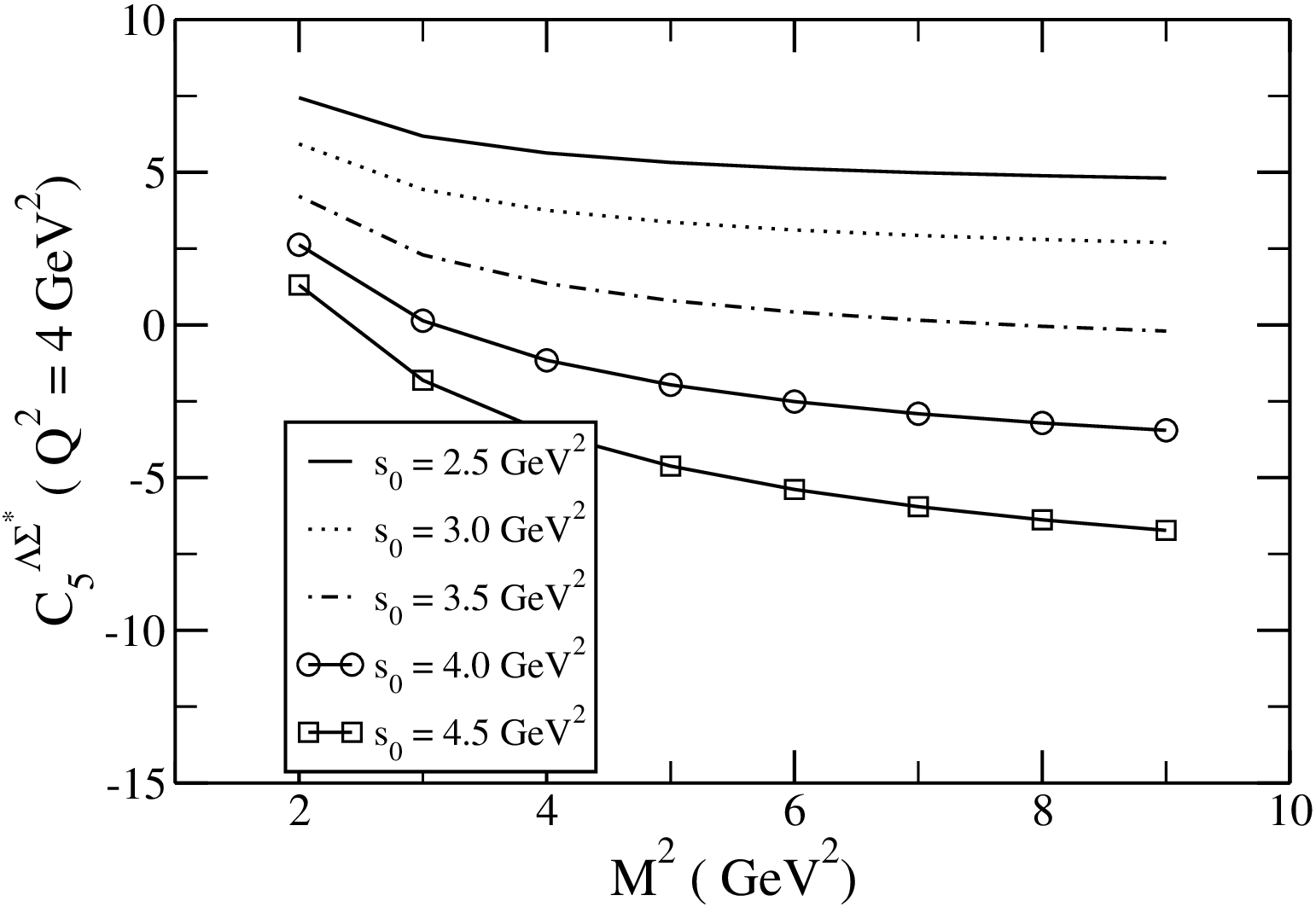}}\\
    \subfloat[]{\label{fig:C6Msq.eps}\includegraphics[width=0.4\textwidth]{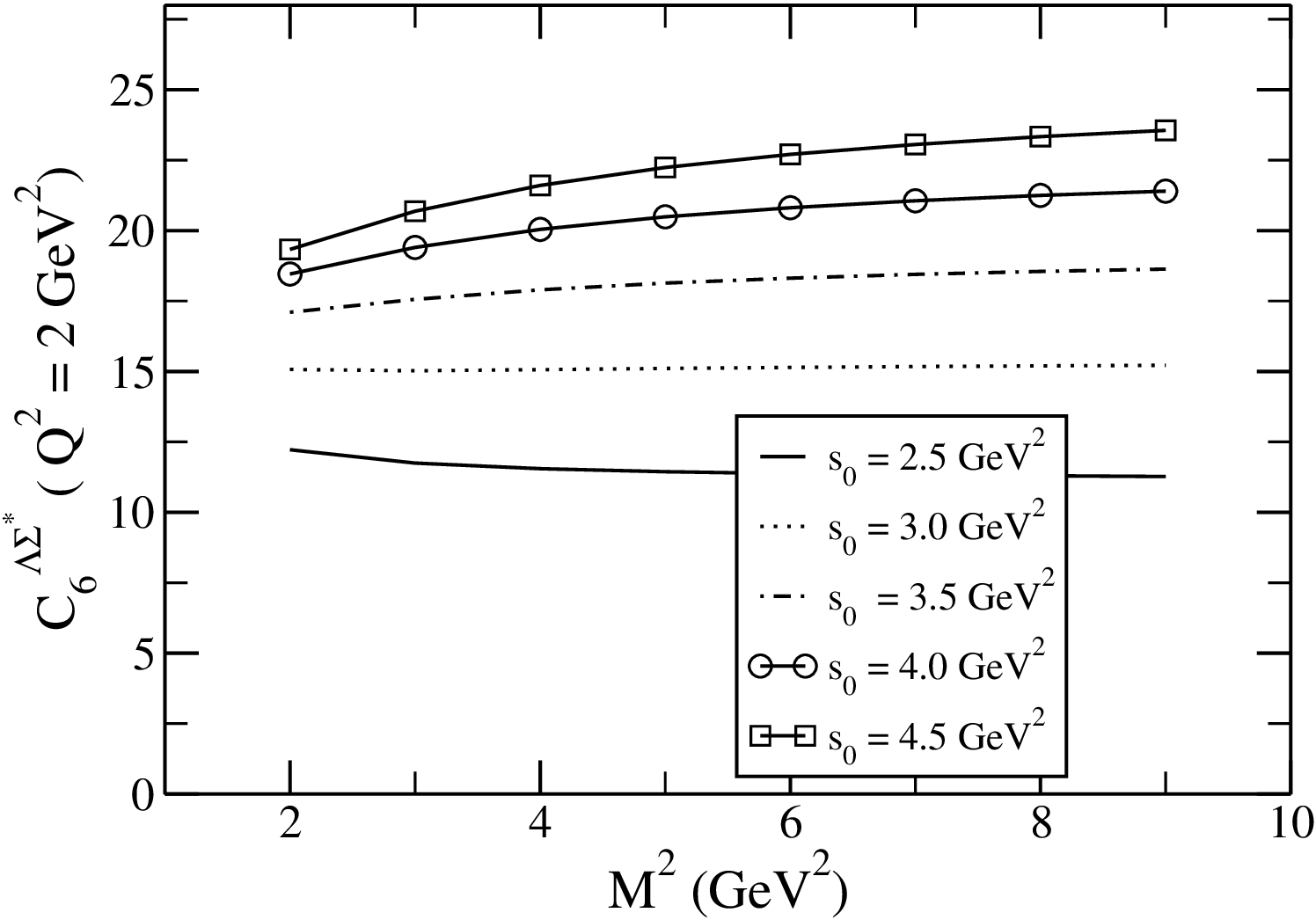}}
    \subfloat[]{\label{fig:C6Msq1.eps}\includegraphics[width=0.4\textwidth]{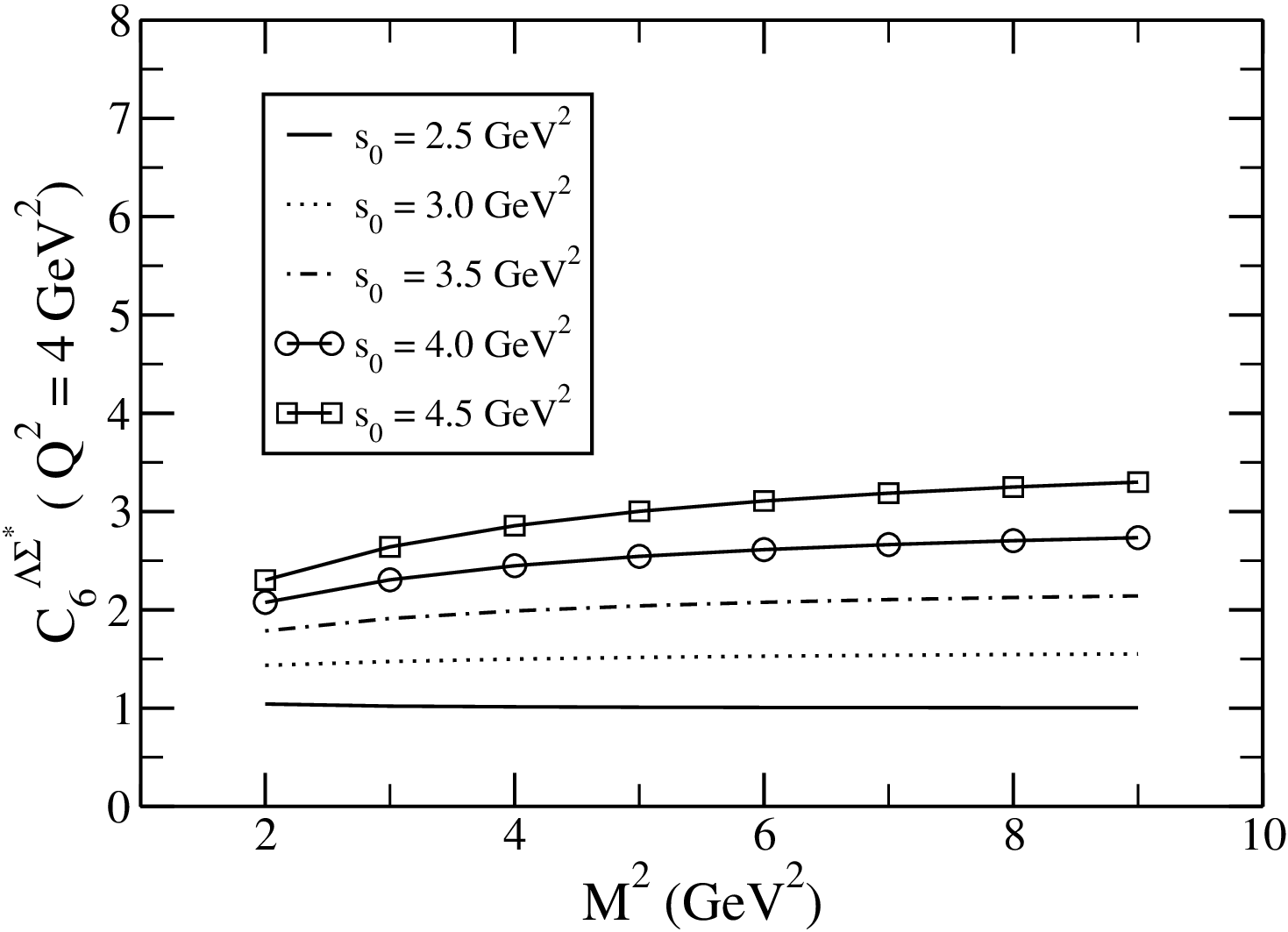}}
\caption{The dependence of the form factors; on the Borel parameter squared $M^{2}$
  for the values of the continuum threshold  $s_0 = 2.5 ~GeV^2$, $s_0 = 3.0~GeV^2$, $s_0 = 3.5~GeV^2$, $s_0 = 4.0~GeV^2$, $s_0 = 4.5~GeV^2$
  and $Q^2 = 2~ GeV^2$, $4~ GeV^2$
(a) and (b) for $C_3^{\Lambda\Sigma^*}$ form factor,
(c) and (d) for $C_4^{\Lambda\Sigma^*}$ form factor,
(e) and (f) for $C_5^{\Lambda\Sigma^*}$ form factor and
(g) and (h) for $C_6^{\Lambda\Sigma^*}$ form factor.}
\end{figure}

\newpage

\begin{figure}[htp]
\centering
 \subfloat[]{\label{fig:ChiC3Msq.eps}\includegraphics[width=0.4\textwidth]{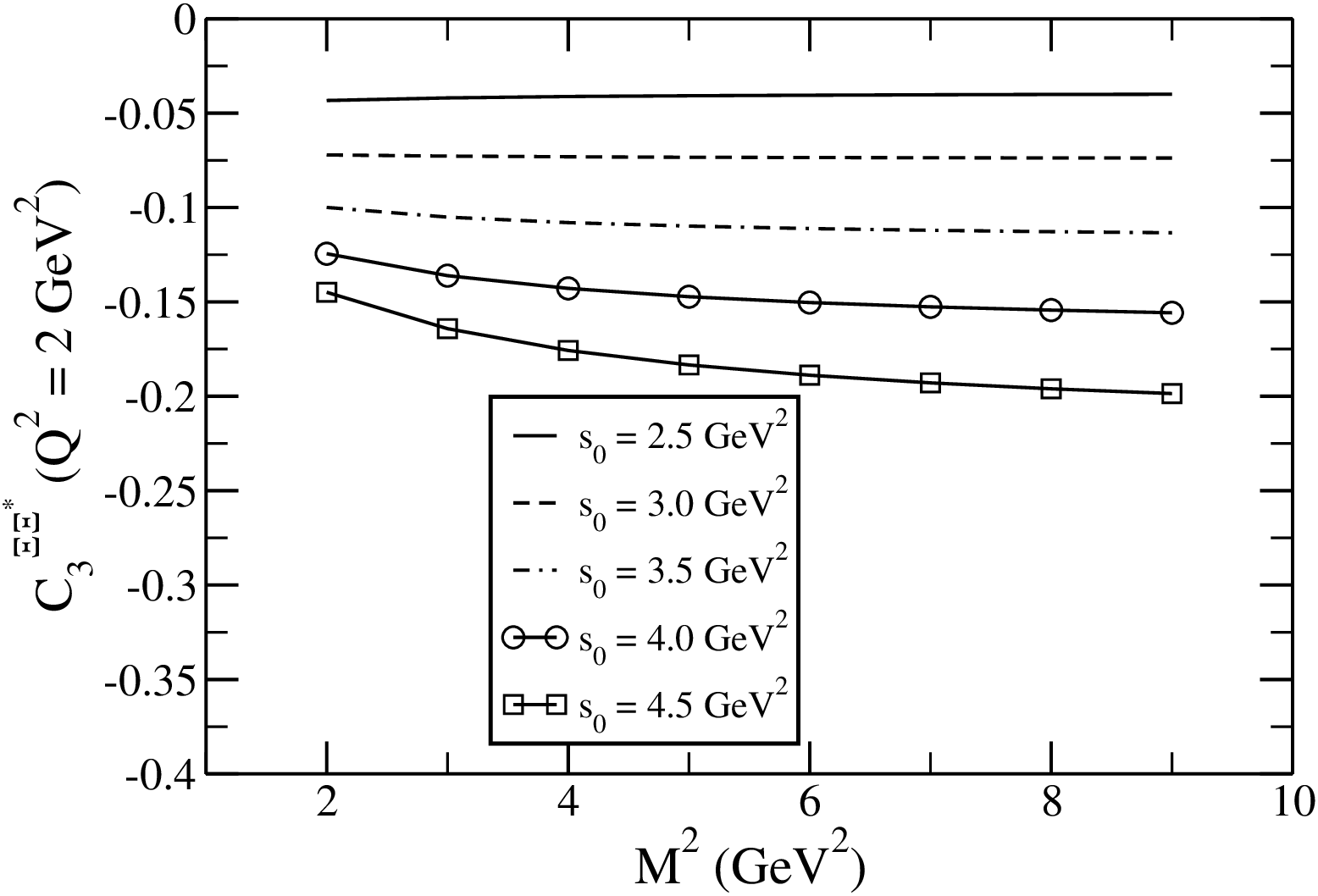}}
 \subfloat[]{\label{fig:ChiC3Msq1.eps}\includegraphics[width=0.4\textwidth]{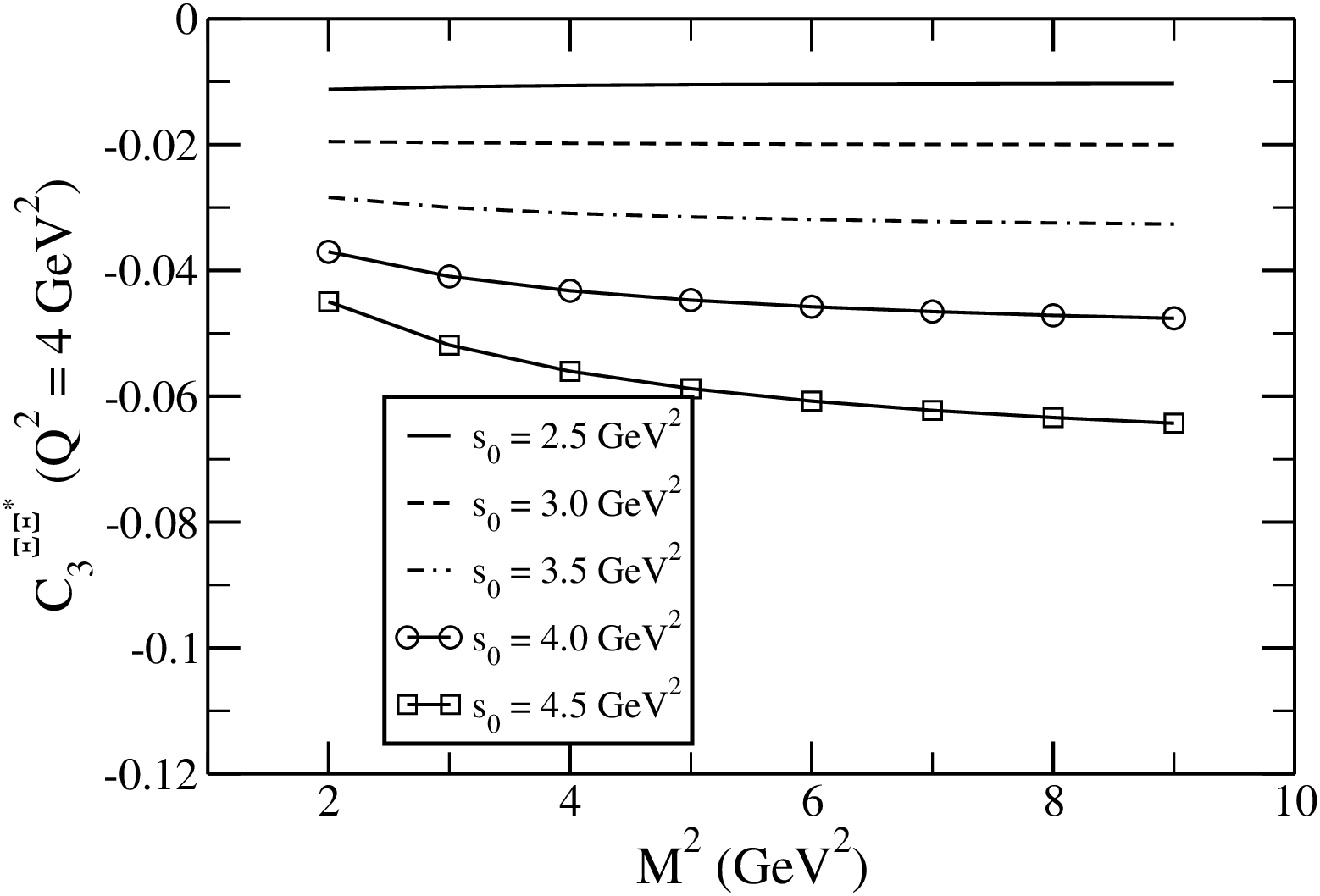}}\\
  \subfloat[]{\label{fig:ChiC4Msq.eps}\includegraphics[width=0.4\textwidth]{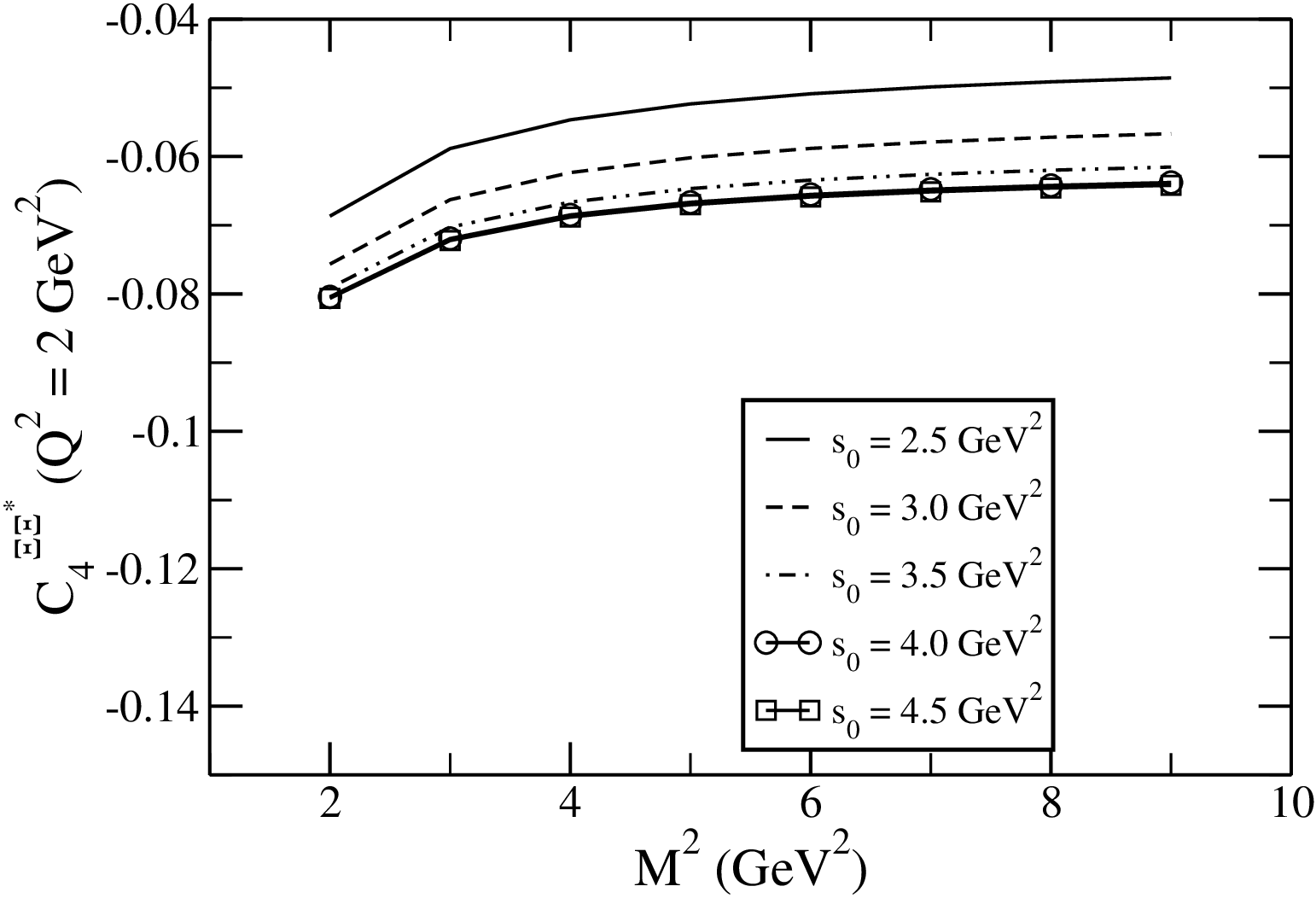}}
  \subfloat[]{\label{fig:ChiC4Msq1.eps}\includegraphics[width=0.4\textwidth]{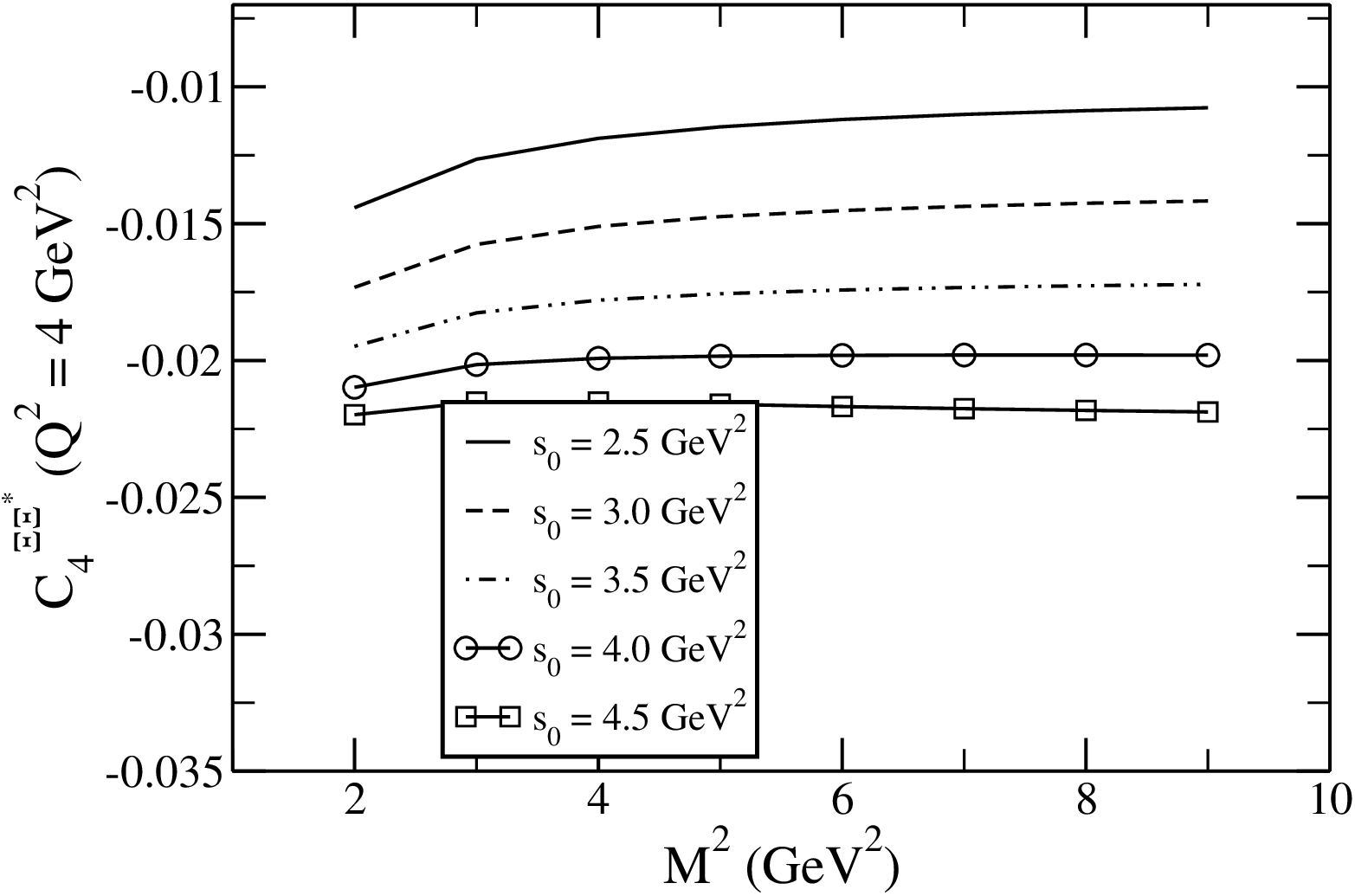}}
  \\
   \subfloat[]{\label{fig:ChiC5Msq.eps}\includegraphics[width=0.4\textwidth]{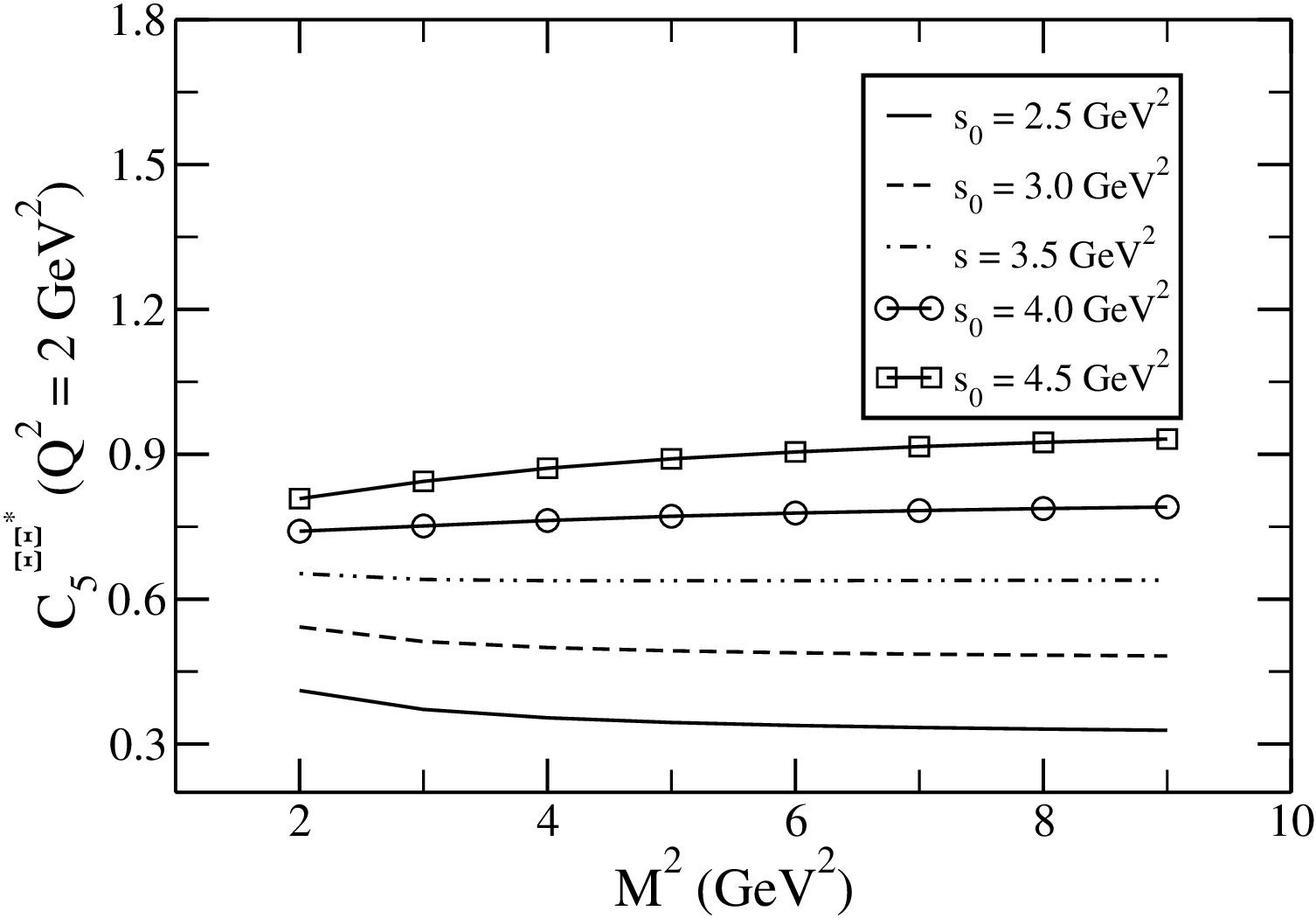}}
   \subfloat[]{\label{fig:ChiC5Msq1.eps}\includegraphics[width=0.4\textwidth]{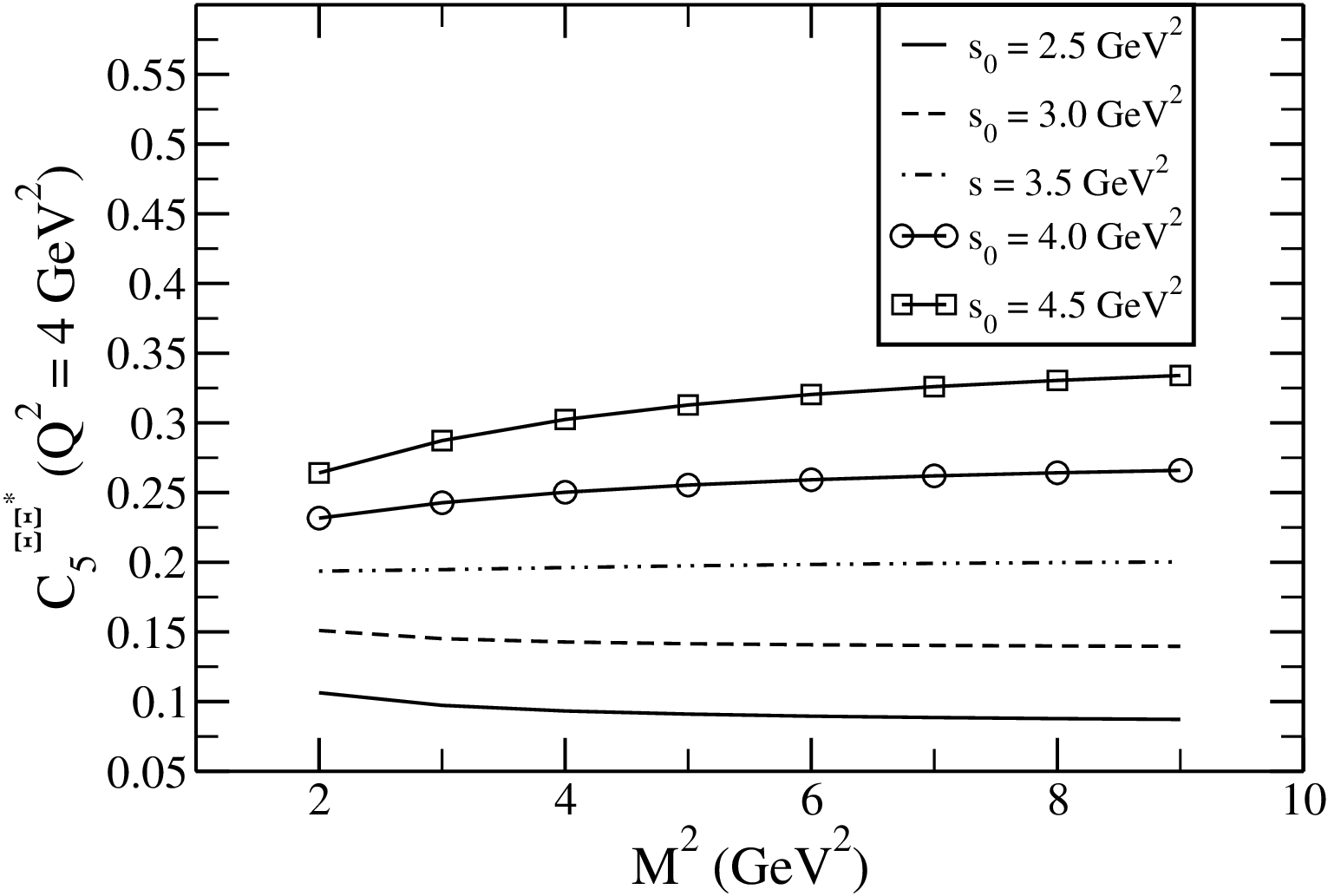}}\\
    \subfloat[]{\label{fig:ChiC6Msq.eps}\includegraphics[width=0.4\textwidth]{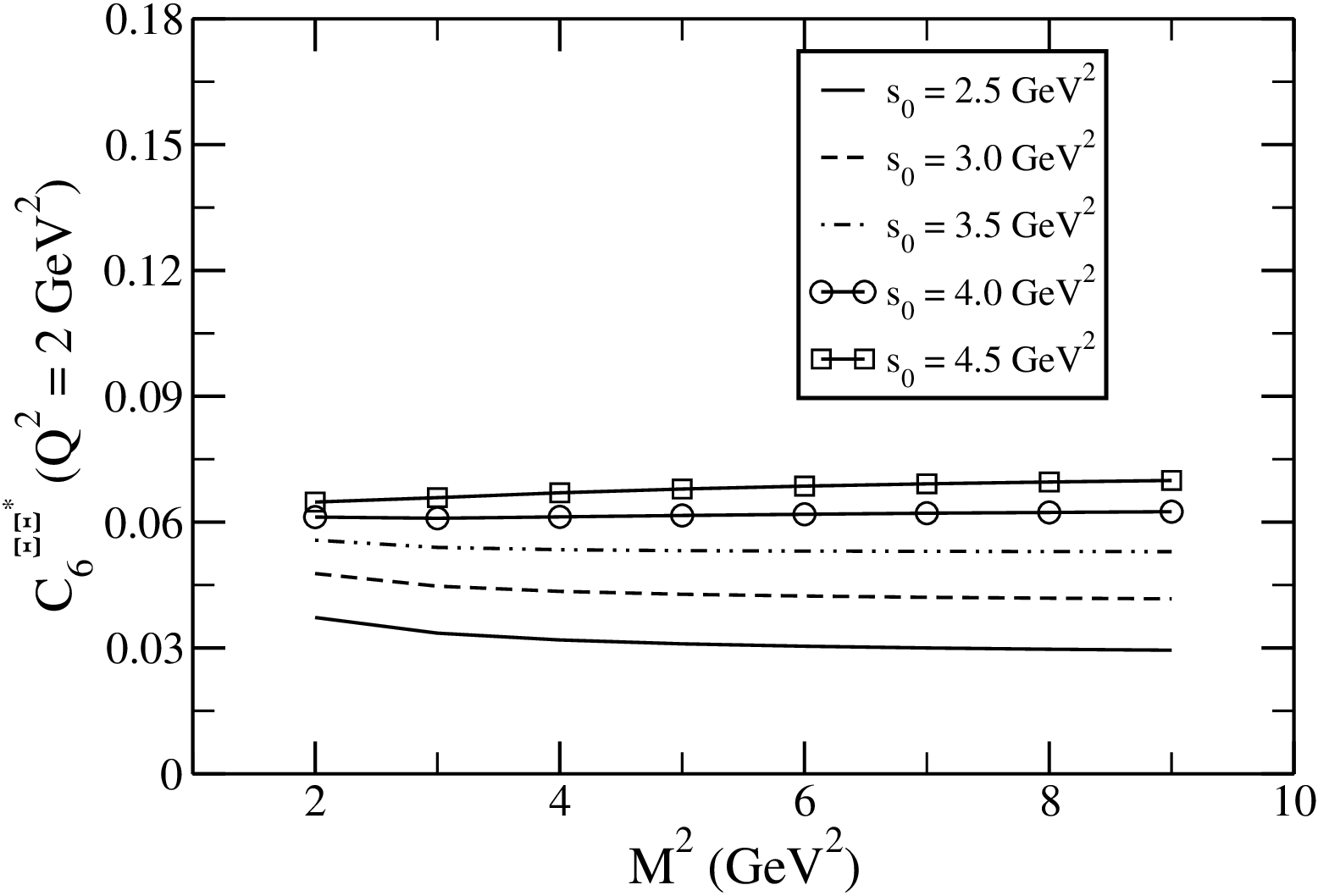}}
    \subfloat[]{\label{fig:ChiC6Msq1.eps}\includegraphics[width=0.4\textwidth]{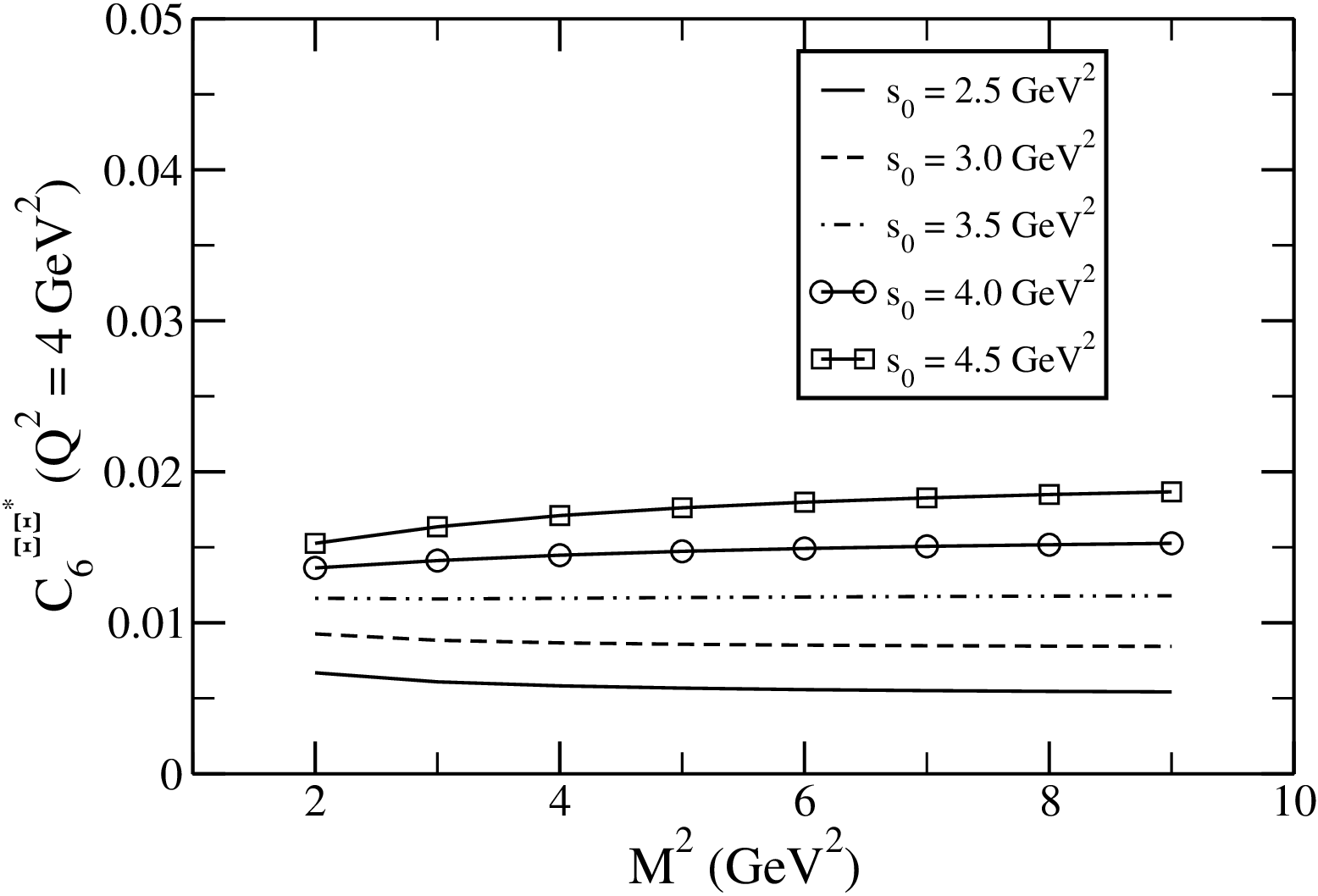}}
\caption{The dependence of the form factors; on the Borel parameter squared $M^{2}$
  for the values of the continuum threshold $s_0 = 2.5 ~GeV^2$, $s_0 = 3.0~GeV^2$, $s_0 = 3.5~GeV^2$, $s_0 = 4.0~GeV^2$, $s_0 = 4.5~GeV^2$
  and $Q^2 = 2~ GeV^2$,  $4~ GeV^2$
(a) and (b) for $C_3^{\Xi\Xi^*}$ form factor
(c) and (d) for $C_4^{\Xi\Xi^*}$ form factor,
(e) and (f) for $C_5^{\Xi\Xi^*}$ form facto and
(g) and (h) for $C_6^{\Xi\Xi^*}$ form factor.}
\end{figure}

\newpage
\begin{figure}[htp]
\centering
 \subfloat[]{\label{fig:SigC3con.eps}\includegraphics[width=0.4\textwidth]{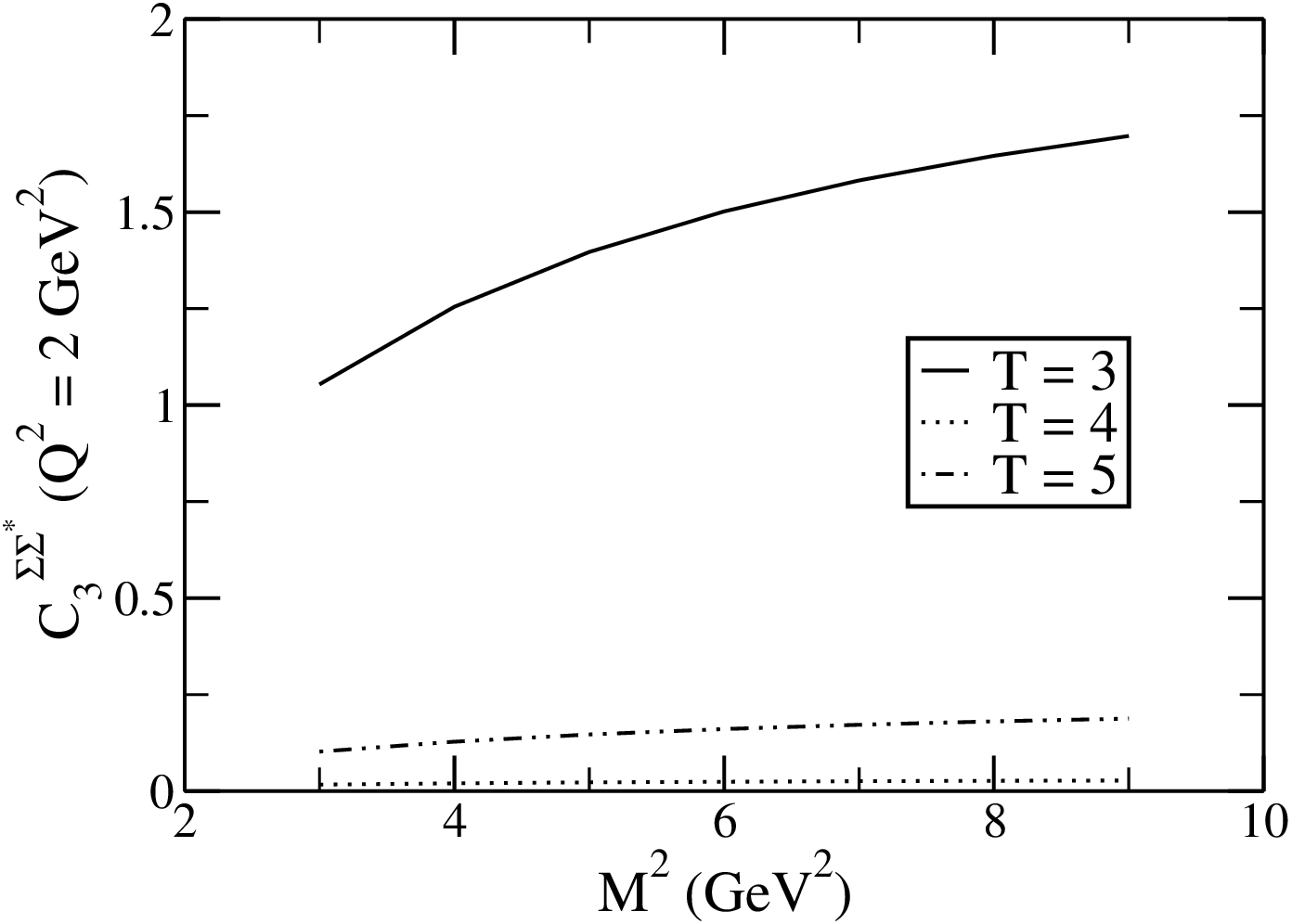}}
  \subfloat[]{\label{fig:SigC4con.eps}\includegraphics[width=0.4\textwidth]{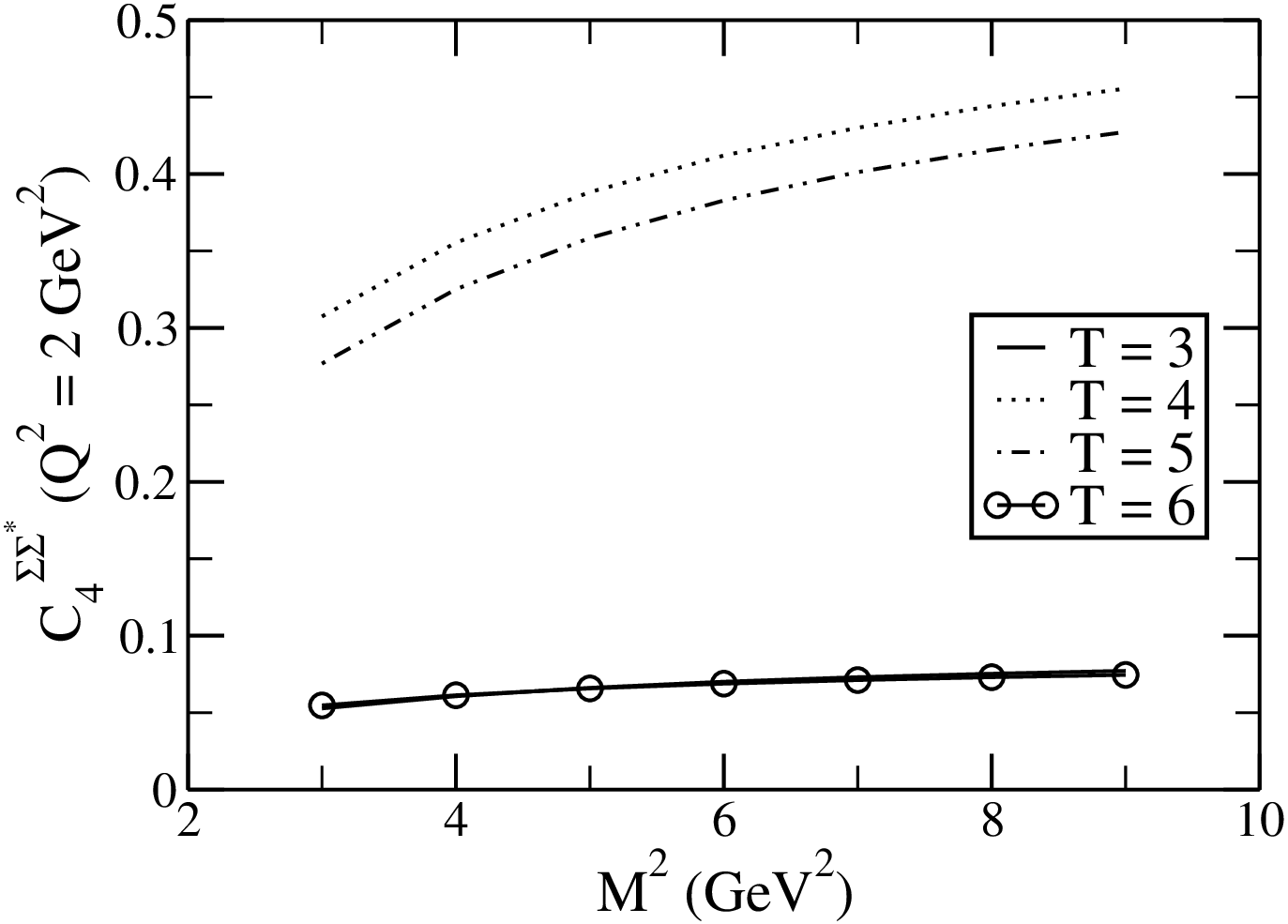}}\\
   \subfloat[]{\label{fig:SigC5con.eps}\includegraphics[width=0.4\textwidth]{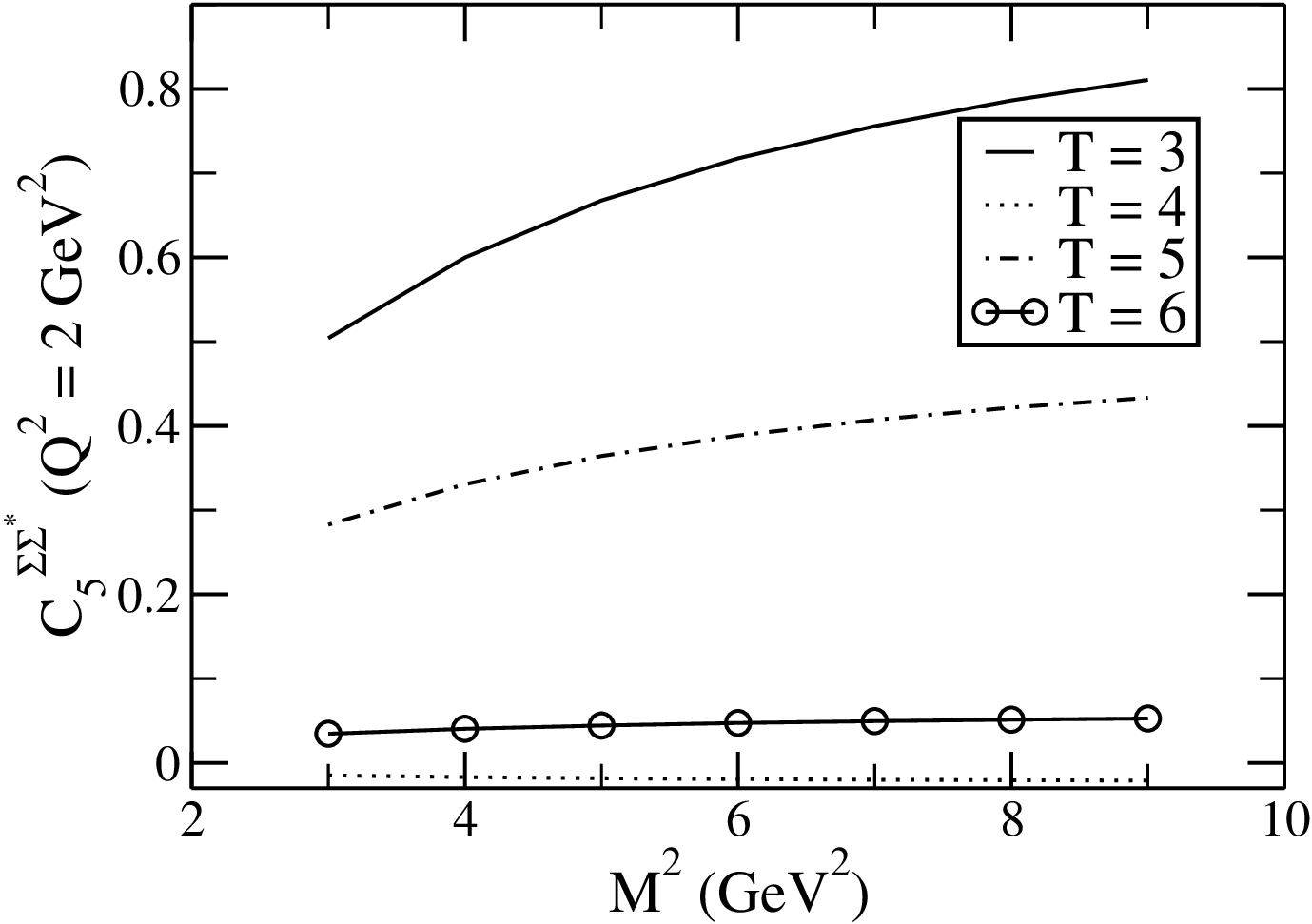}}
    \subfloat[]{\label{fig:SigC6con.eps}\includegraphics[width=0.4\textwidth]{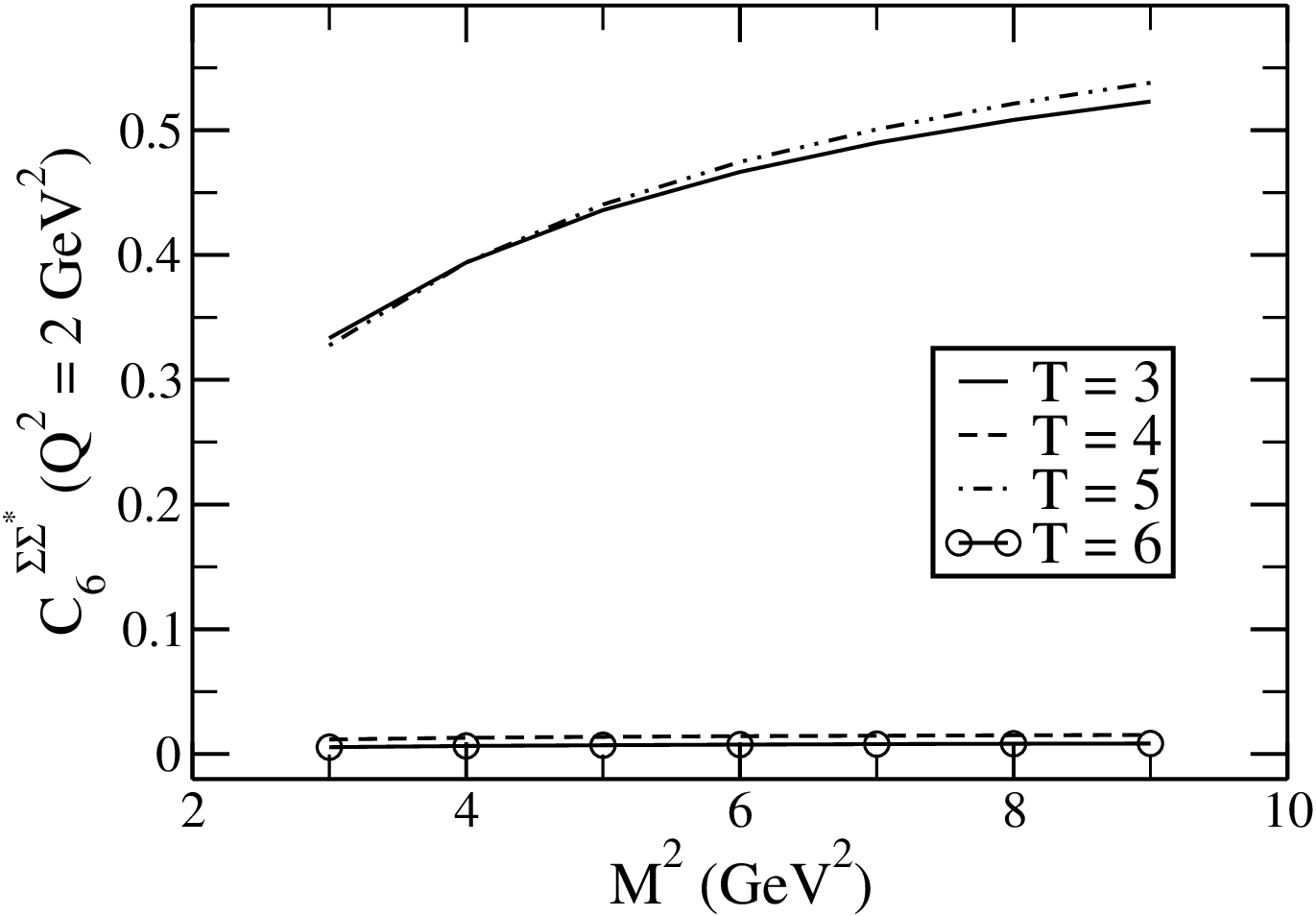}}
\caption{ The convergence of form factors at $Q^2 = 2~GeV^2$
(a) for $C_3^{\Sigma\Sigma^*}$ form factors,
(b) for $C_4^{\Sigma\Sigma^*}$ form factors,
(c) for $C_5^{\Sigma\Sigma^*}$ form factors,
(d) for $C_6^{\Sigma\Sigma^*}$ form factors.
The T = 3,4,5 and 6 are twist-3, twist-4, twist-5 and twist-6 contributions, respectively.}
\end{figure}

\newpage
\begin{figure}[htp]
\centering
 \subfloat[]{\label{fig:C3con.eps}\includegraphics[width=0.4\textwidth]{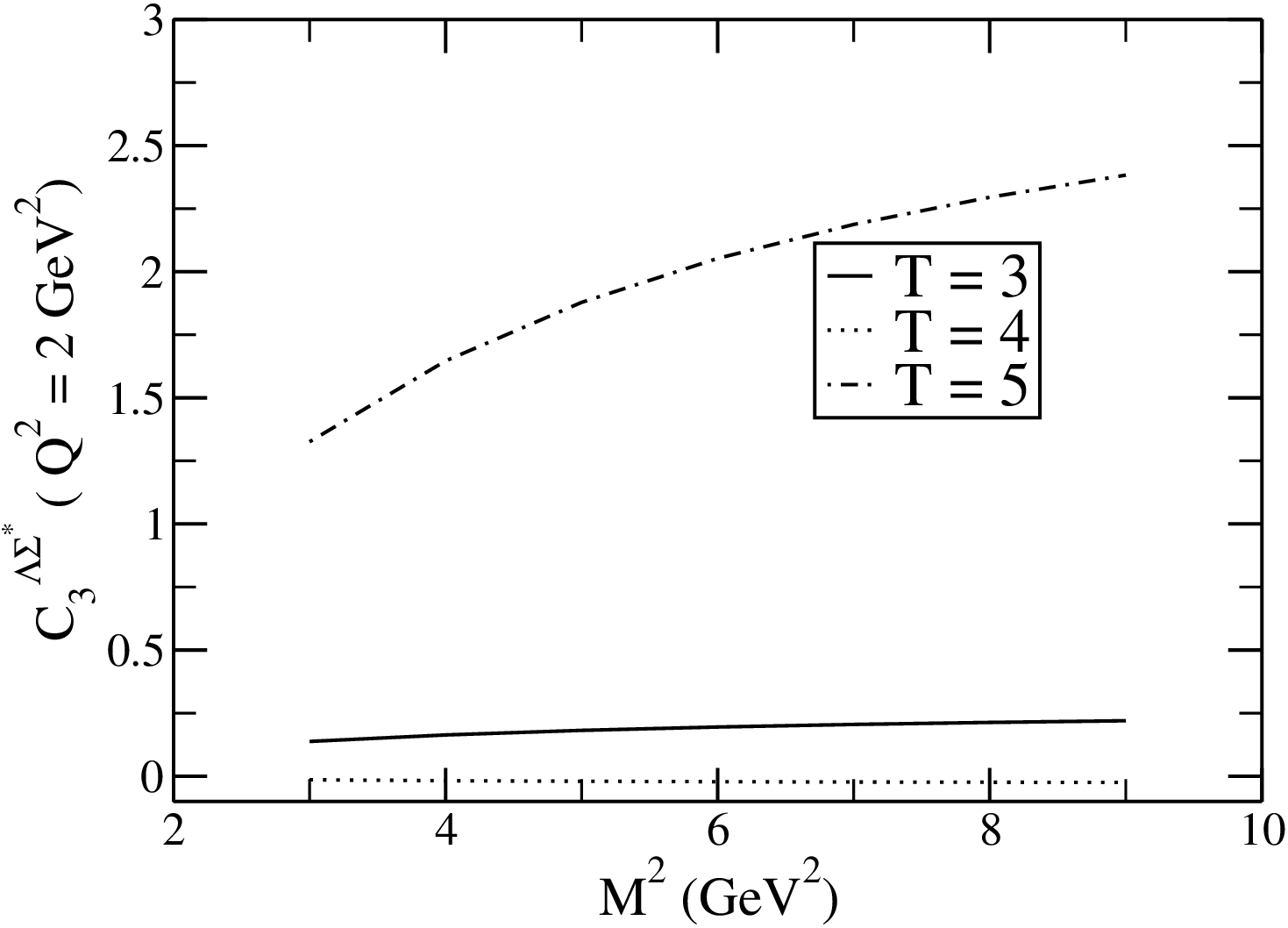}}
  \subfloat[]{\label{fig:C4con.eps}\includegraphics[width=0.4\textwidth]{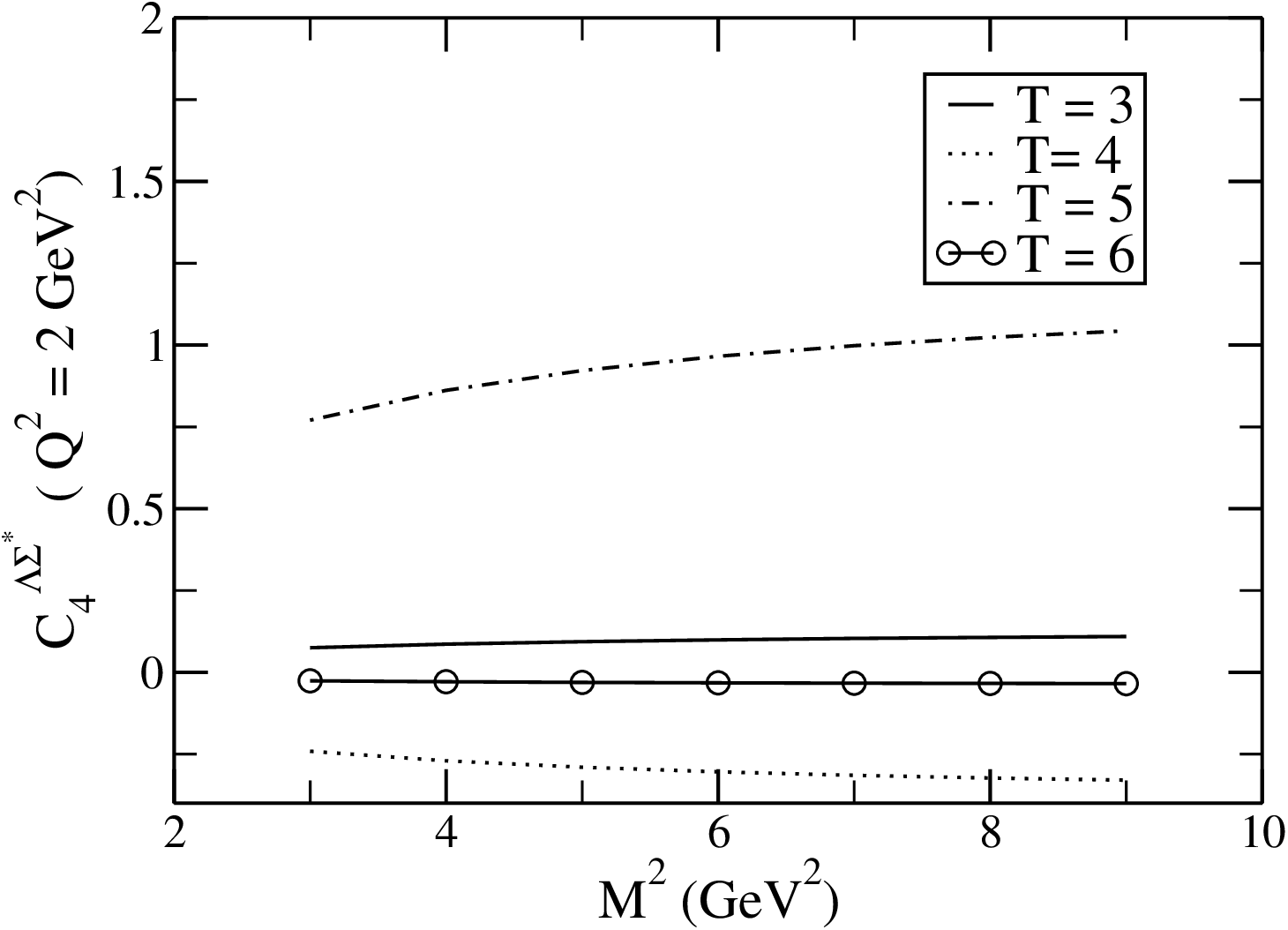}}\\
   \subfloat[]{\label{fig:C5con.eps}\includegraphics[width=0.4\textwidth]{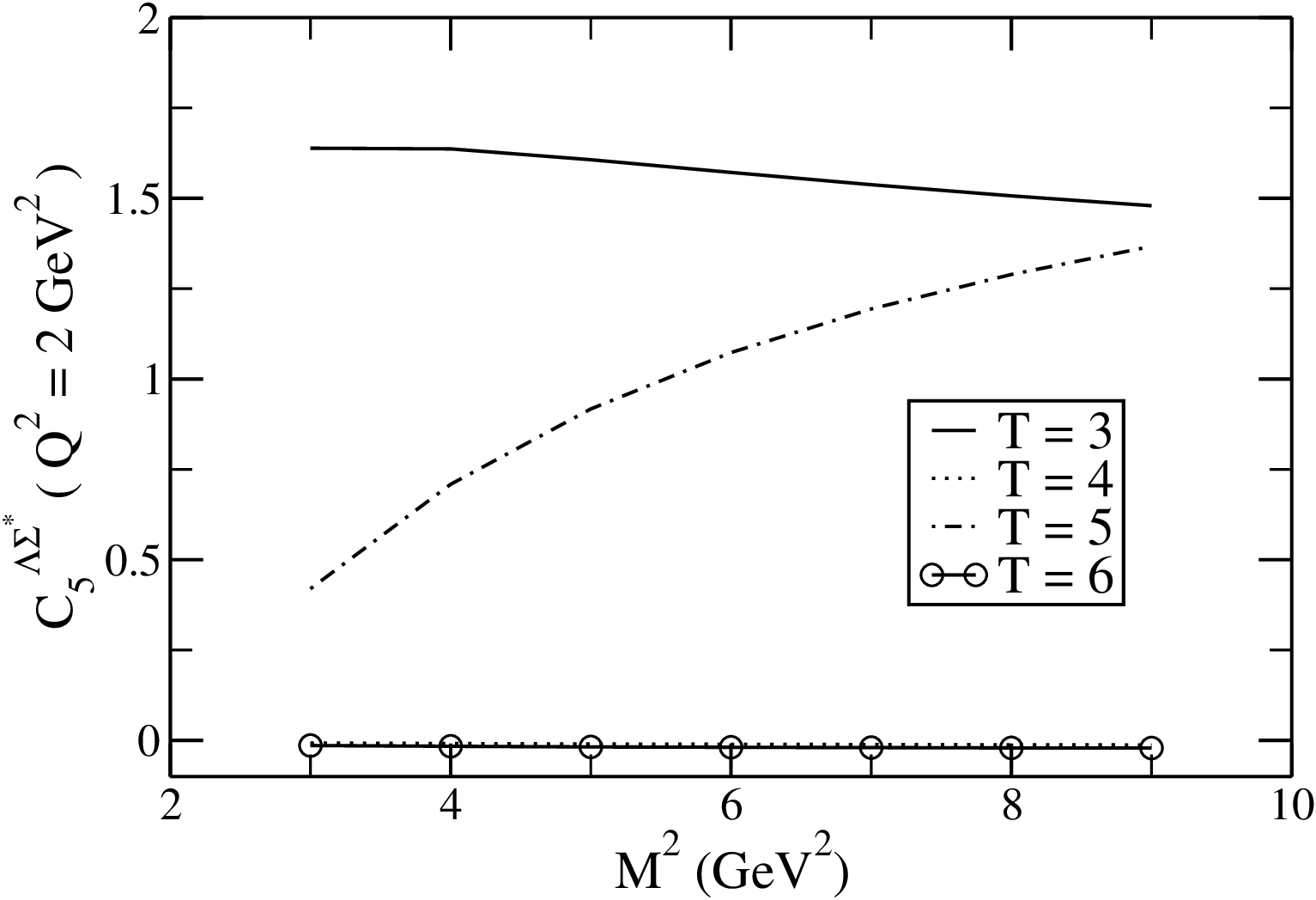}}
    \subfloat[]{\label{fig:C6con.eps}\includegraphics[width=0.4\textwidth]{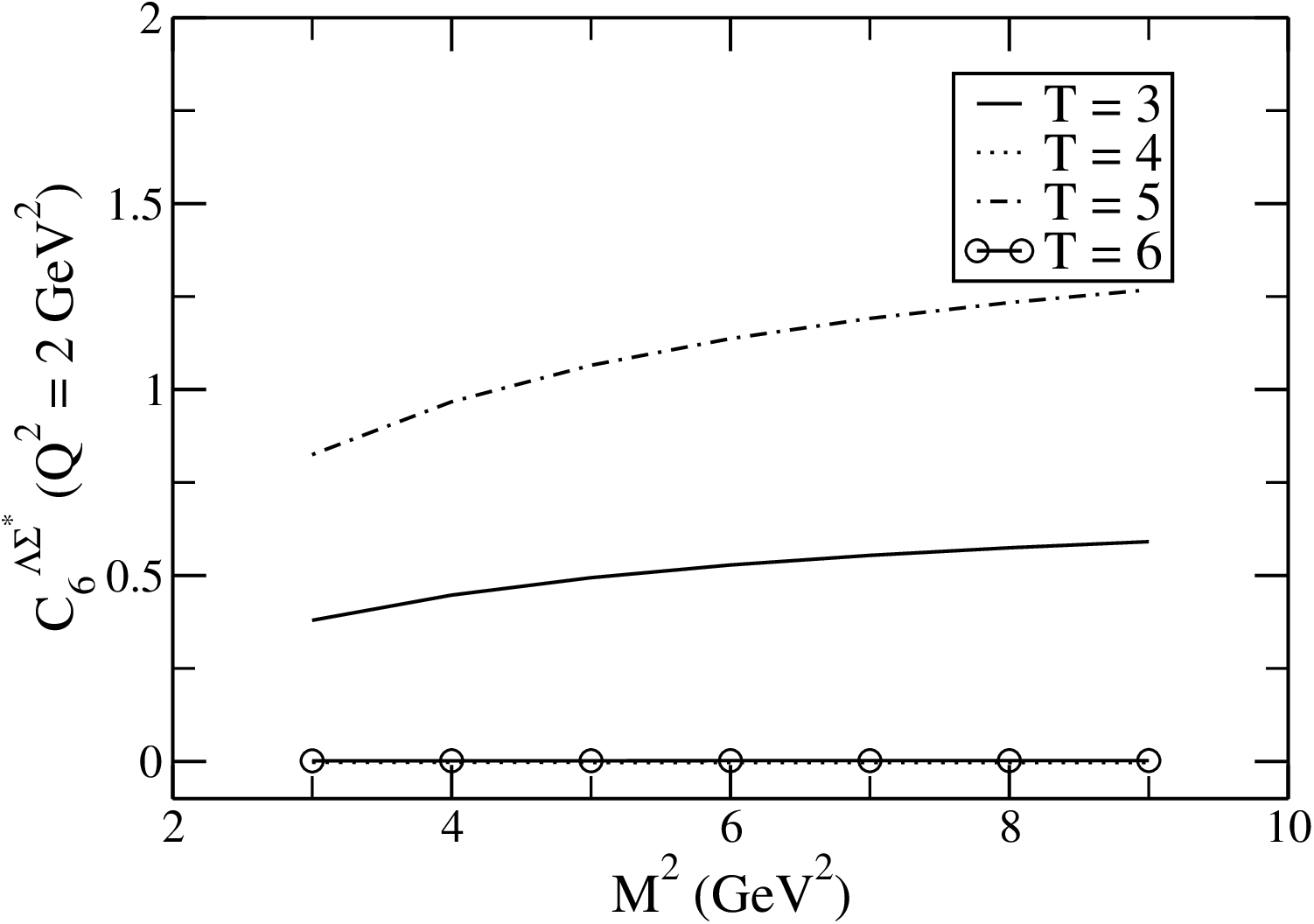}}
\caption{The convergence of form factors at $Q^2 = 2~GeV^2$
(a) for $C_3^{\Lambda\Sigma^*}$ form factor,
(b) for $C_4^{\Lambda\Sigma^*}$ form factor,
(c) for $C_5^{\Lambda\Sigma^*}$ form factor and
(d) for $C_6^{\Lambda\Sigma^*}$ form factor.
The T = 3,4,5 and 6 are twist-3, twist-4, twist-5 and twist-6 contributions, respectively.}
\end{figure}

\newpage

\begin{figure}[htp]
\centering
 \subfloat[]{\label{fig:ChiC3con.eps}\includegraphics[width=0.4\textwidth]{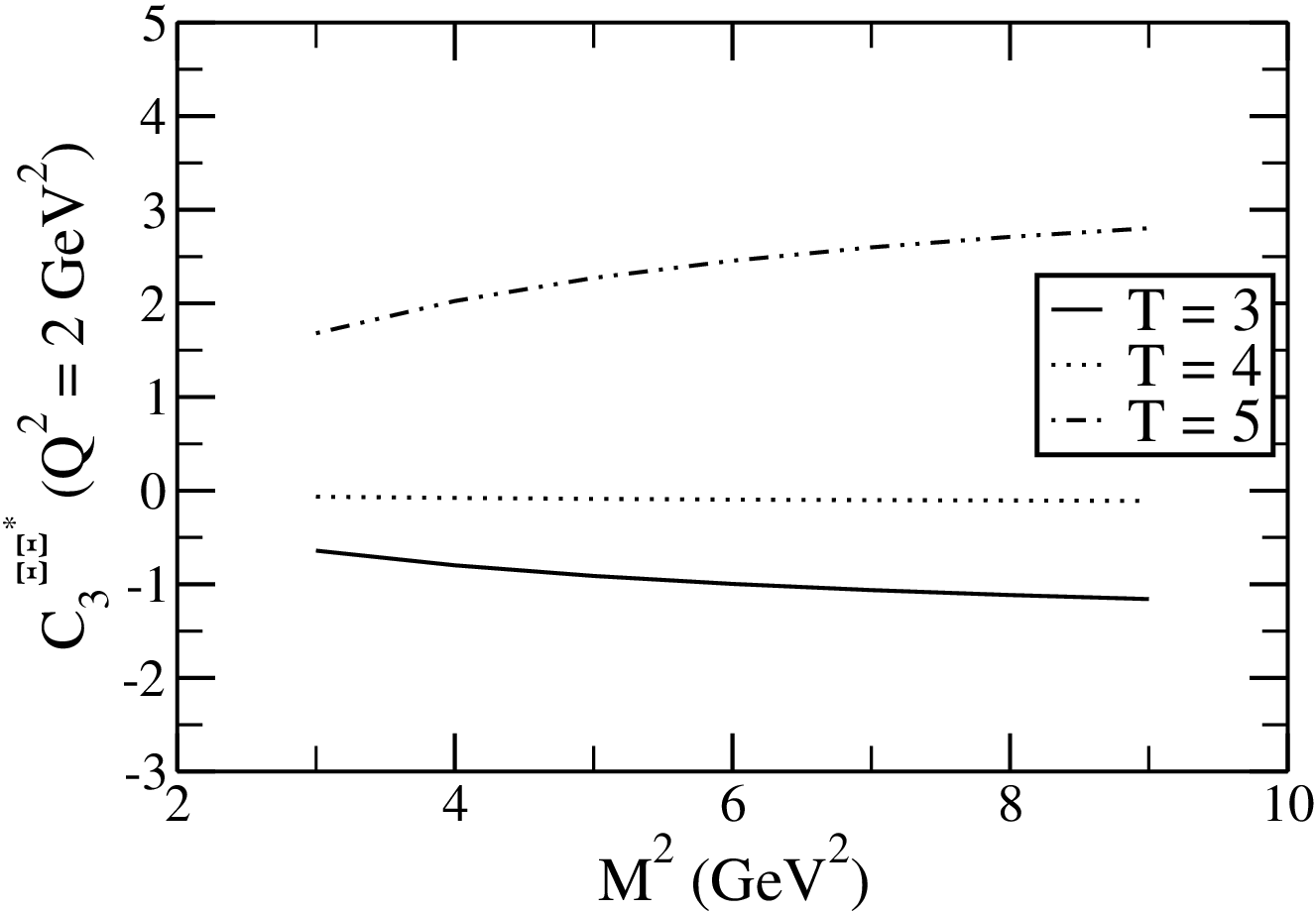}}
  \subfloat[]{\label{fig:ChiC4con.eps}\includegraphics[width=0.4\textwidth]{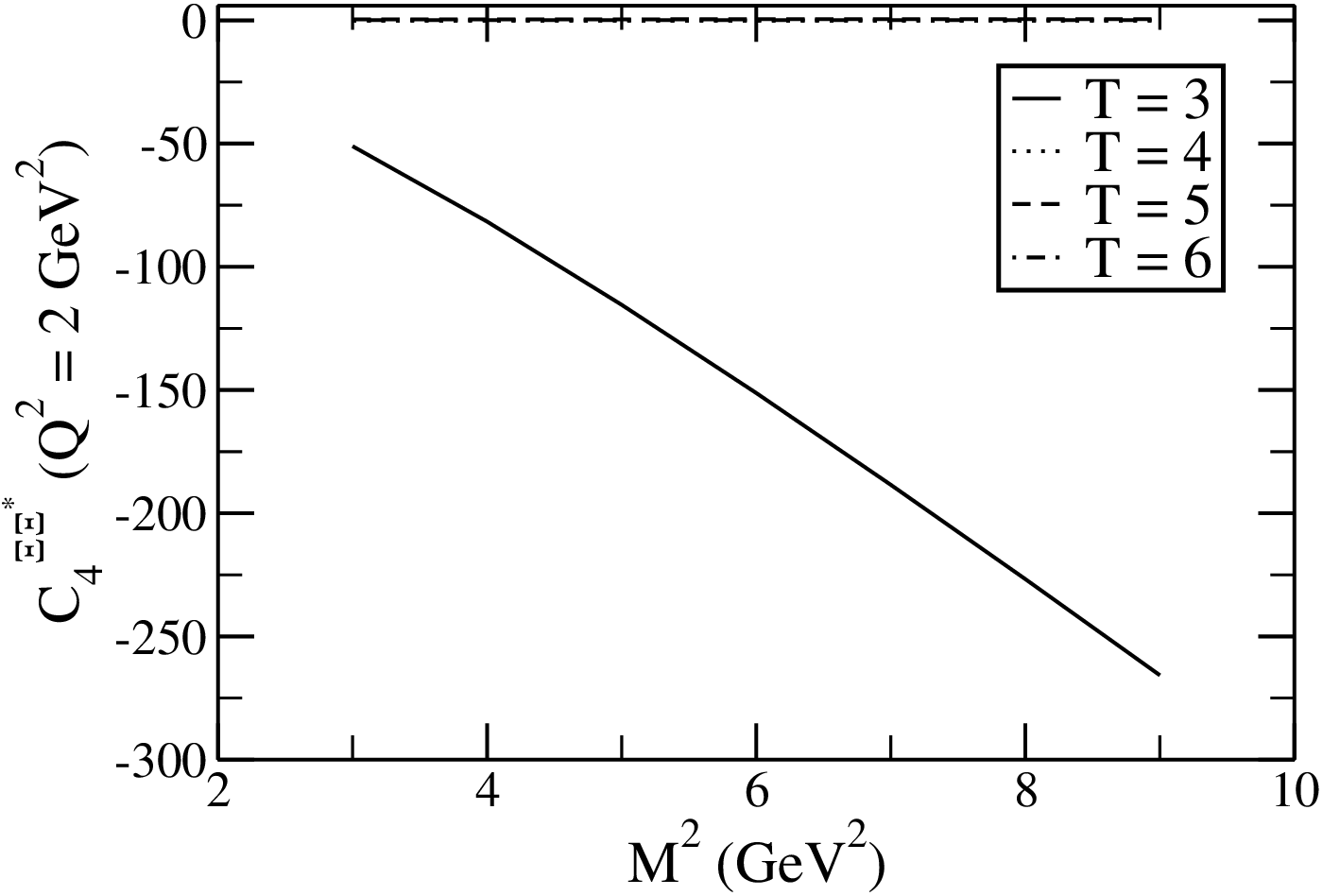}}\\
   \subfloat[]{\label{fig:ChiC5con.eps}\includegraphics[width=0.4\textwidth]{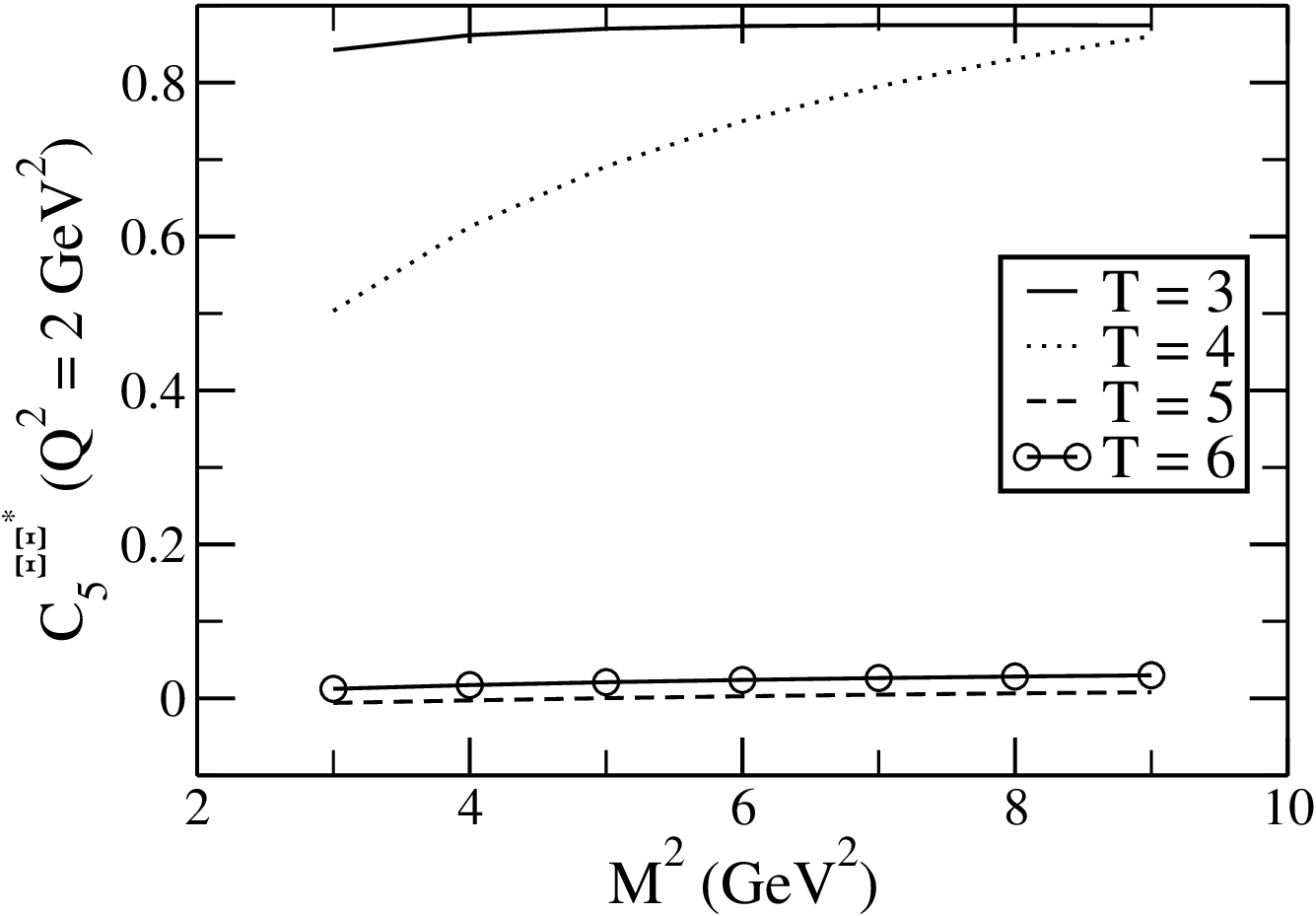}}
    \subfloat[]{\label{fig:ChiC6con.eps}\includegraphics[width=0.4\textwidth]{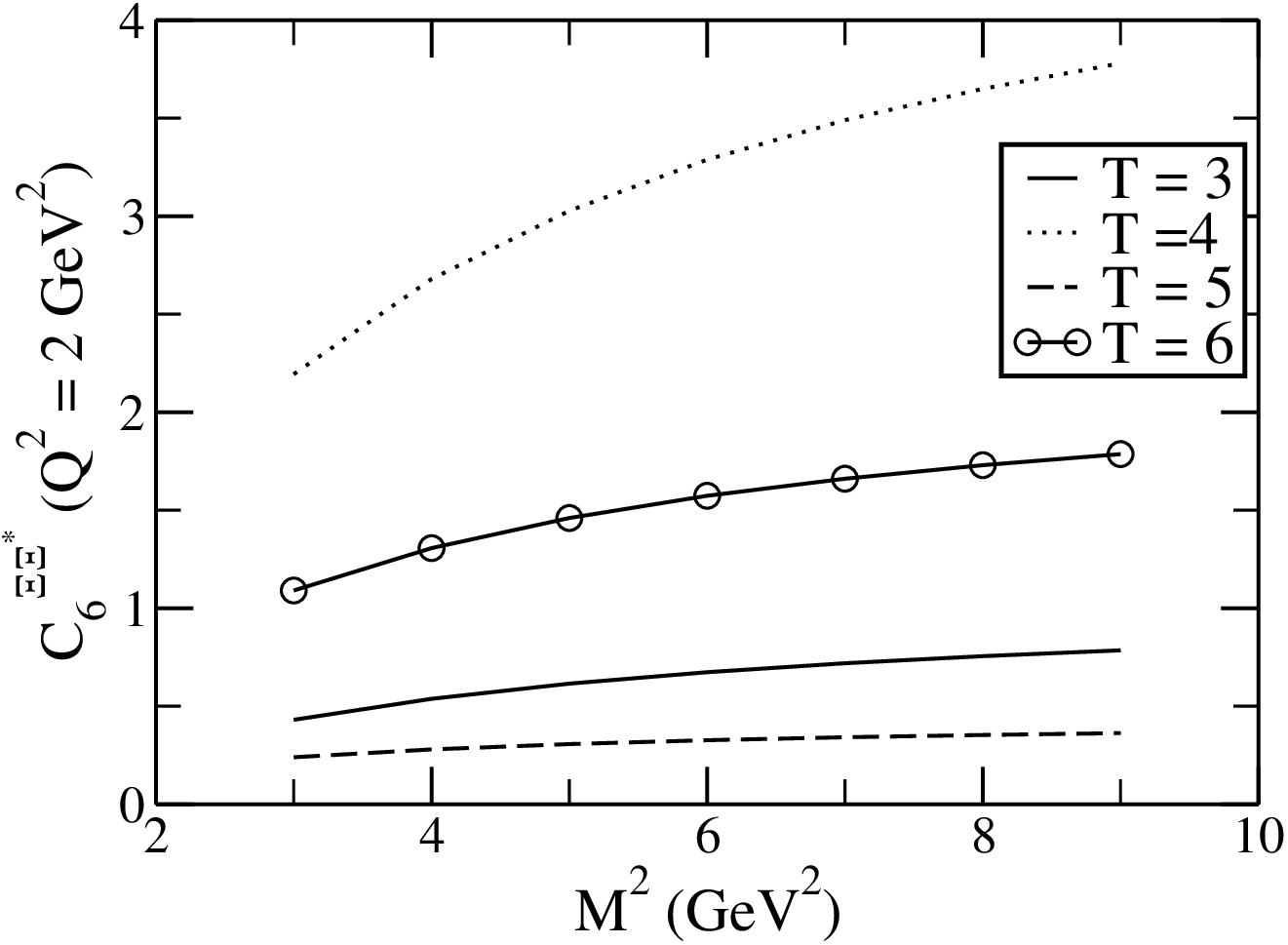}}
\caption{The convergence of form factors at $Q^2 = 2~GeV^2$
(a) for $C_3^{\Xi\Xi^*}$ form factor
(b) for $C_4^{\Xi\Xi^*}$ form factor,
(c) for $C_5^{\Xi\Xi^*}$ form facto and
(d) for $C_6^{\Xi\Xi^*}$ form factor.
The T = 3,4,5 and 6 are twist-3, twist-4, twist-5 and twist-6 contributions, respectively.}
\end{figure}

\newpage
\begin{figure}[htp]
\centering
 \subfloat[]{\label{fig:SigC3Qsq.eps}\includegraphics[width=0.4\textwidth]{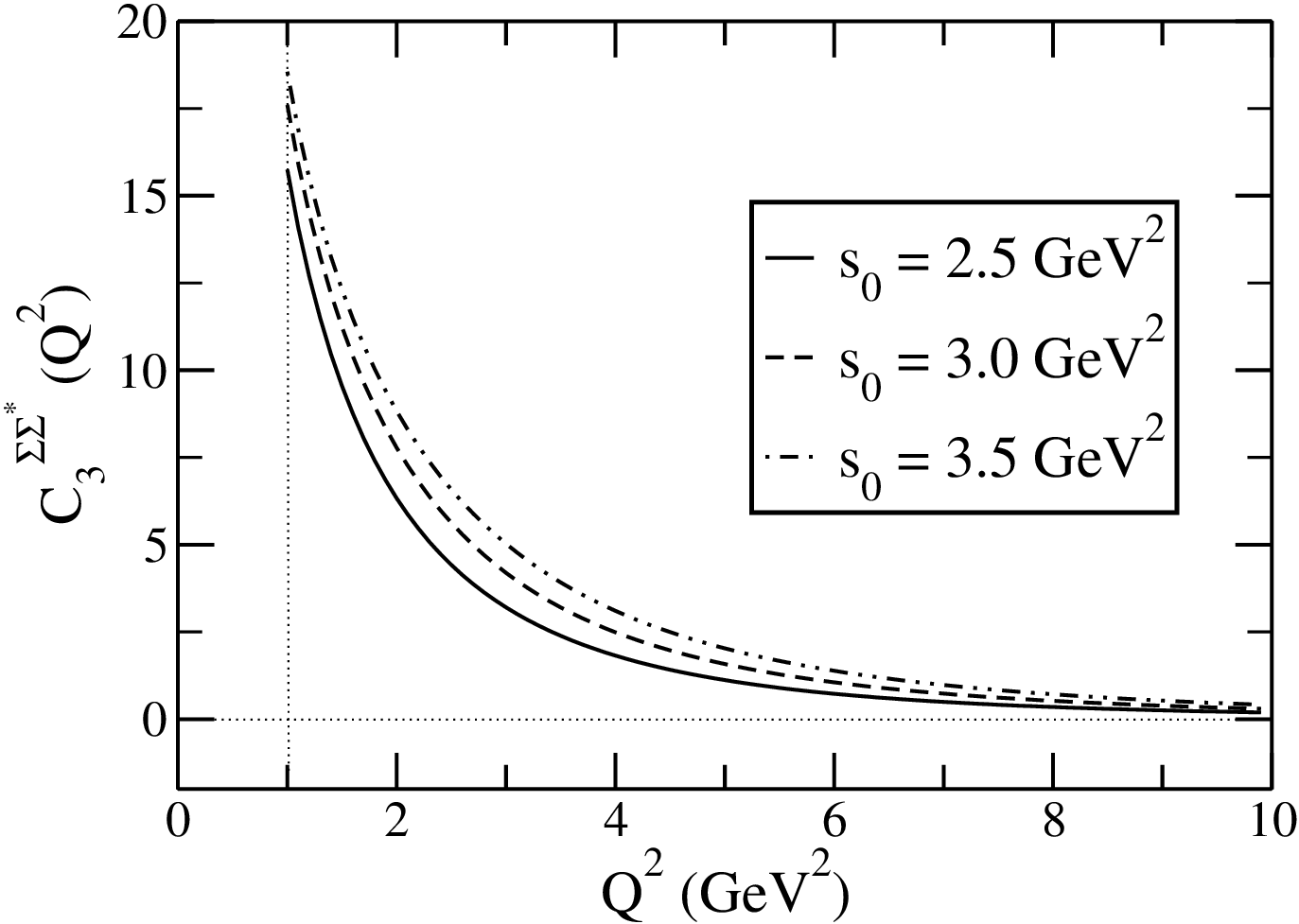}}
  \subfloat[]{\label{fig:SigC4Qsq.eps}\includegraphics[width=0.4\textwidth]{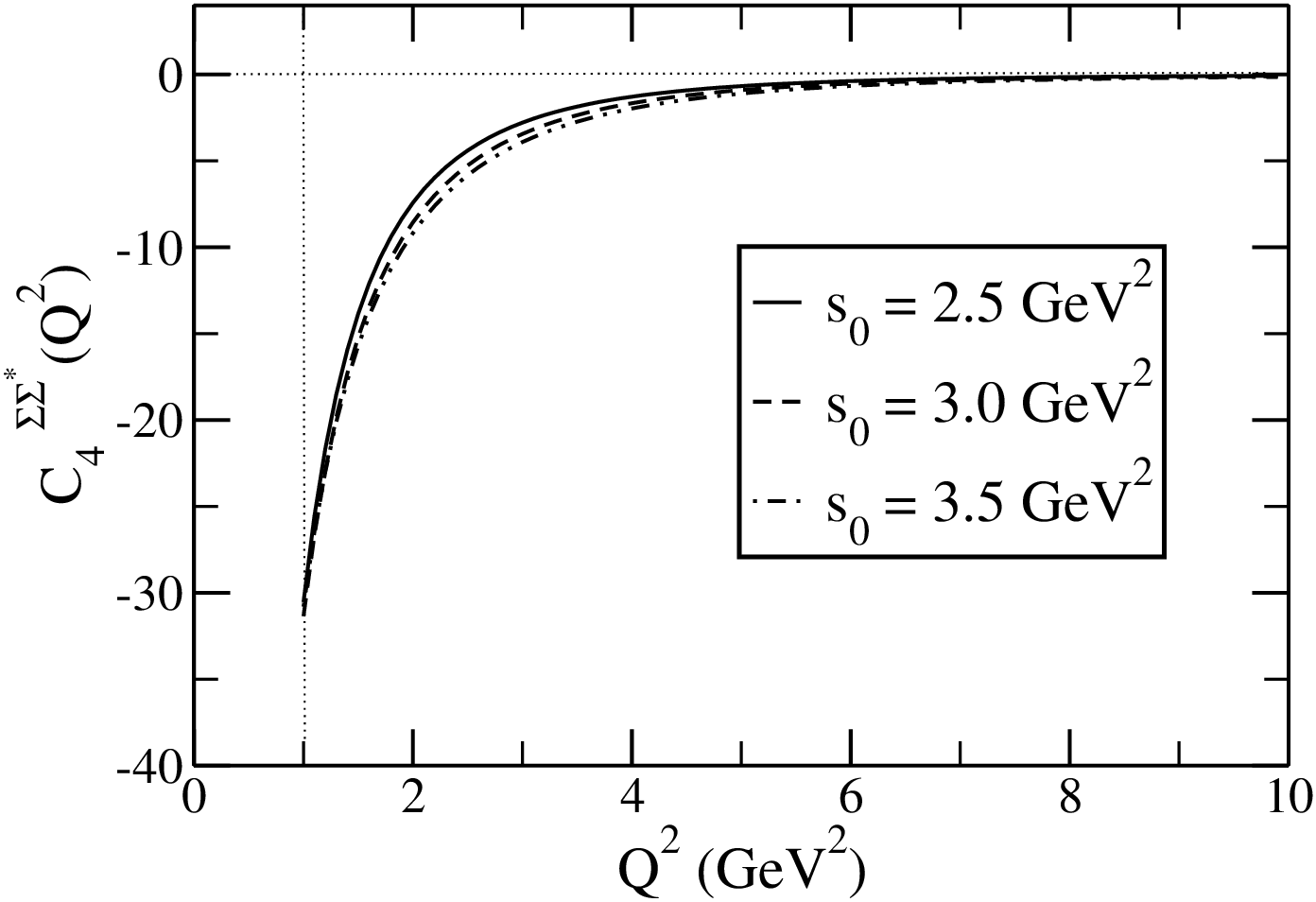}}\\
   \subfloat[]{\label{fig:SigC5Qsq.eps}\includegraphics[width=0.4\textwidth]{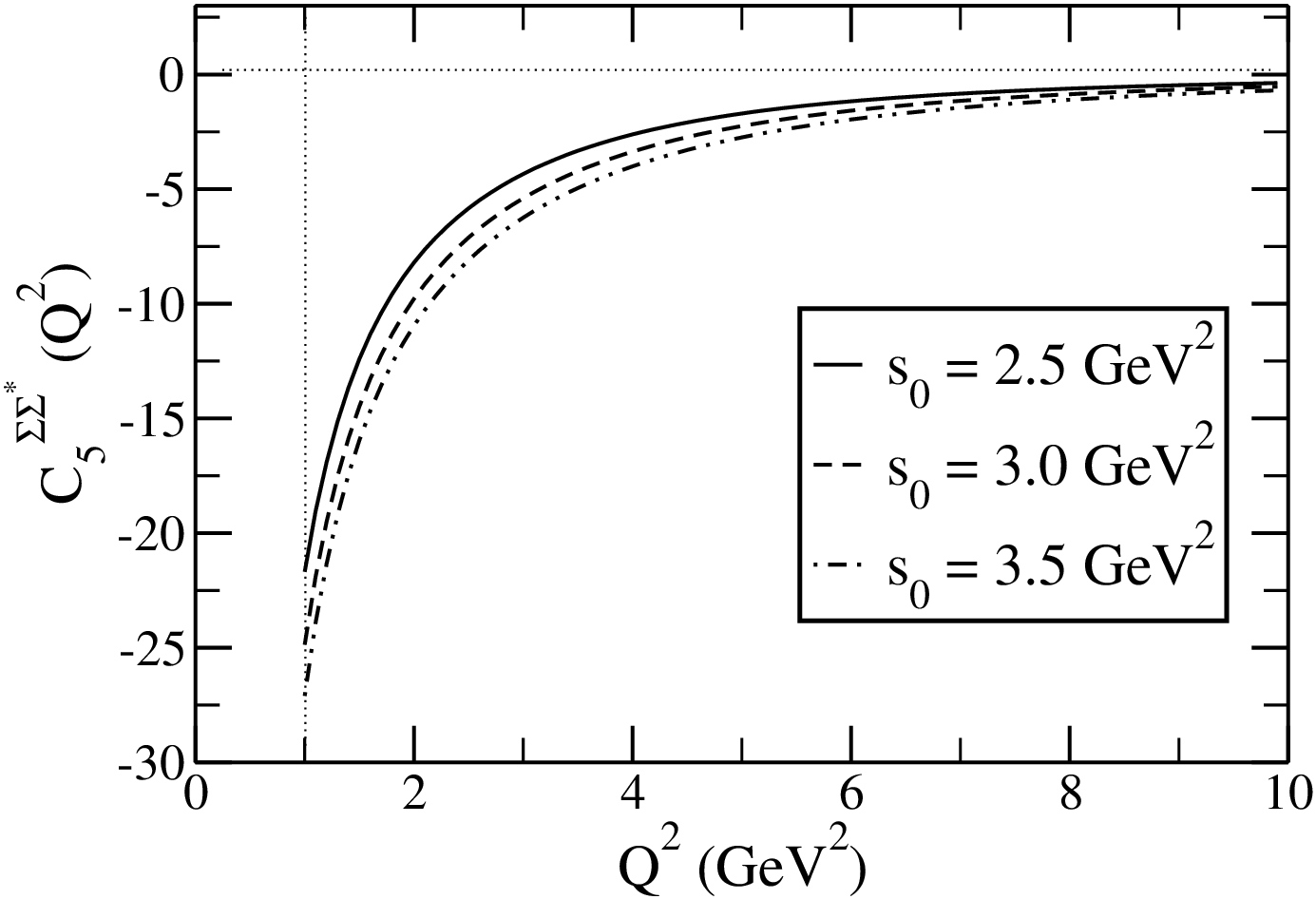}}
    \subfloat[]{\label{fig:SigC6Qsq.eps}\includegraphics[width=0.4\textwidth]{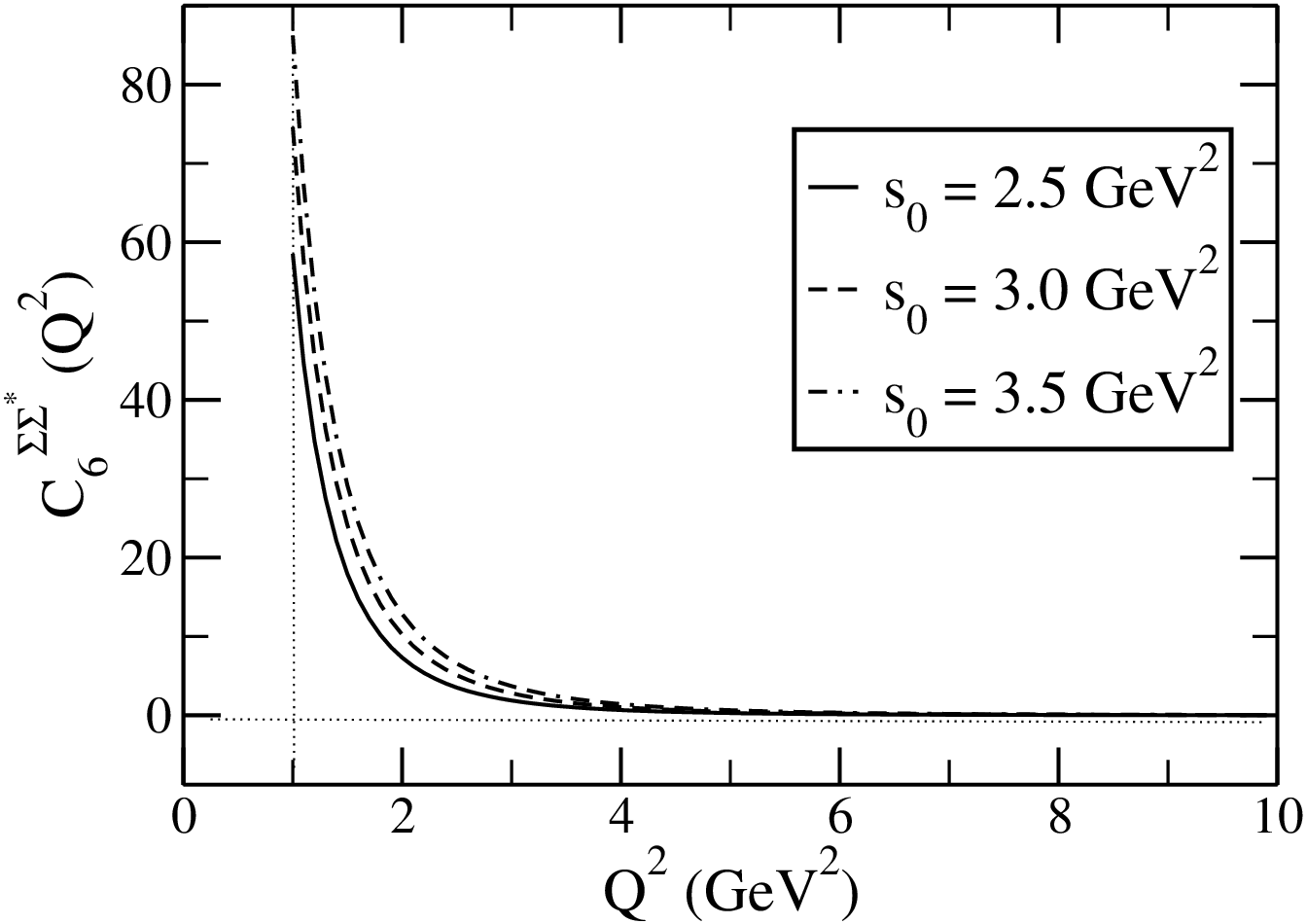}}
\caption{(a)The dependence of the  form factors  for the values of the continuum threshold
 $s_0 = 2.5 ~GeV^2$, $s_0 = 3.0~GeV^2$, $s_0 = 3.5~GeV^2$ and $M^{2}=3~GeV^2$
(a) for $C_3^{\Sigma\Sigma^*}$ form factor,
(b) for $C_4^{\Sigma\Sigma^*}$ form factor,
(c) for $C_5^{\Sigma\Sigma^*}$ form factor and
(d) for $C_6^{\Sigma\Sigma^*}$ form factor.}
\end{figure}

\newpage
\begin{figure}[htp]
\centering
 \subfloat[]{\label{fig:C3.eps}\includegraphics[width=0.4\textwidth]{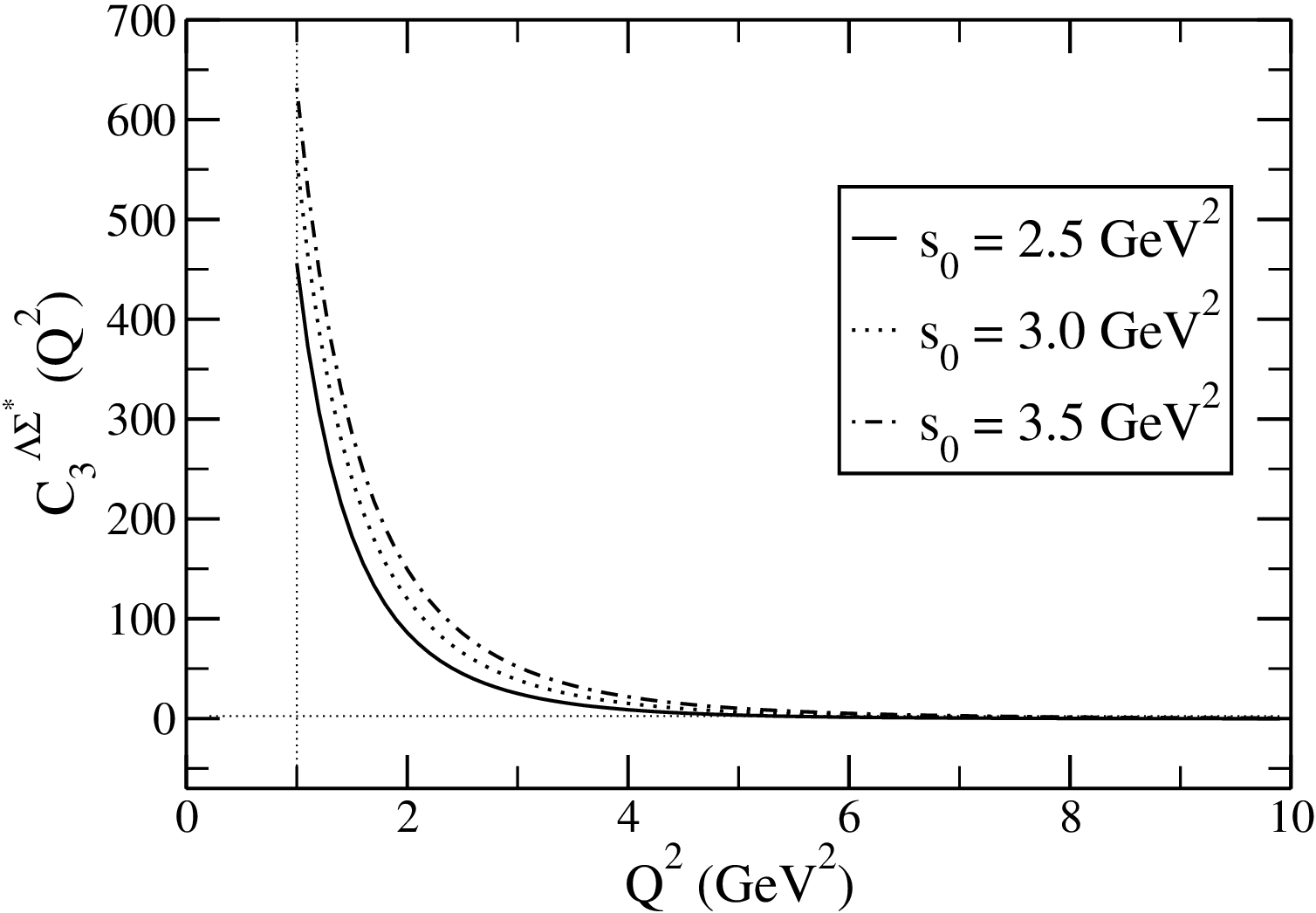}}
  \subfloat[]{\label{fig:C4.eps}\includegraphics[width=0.4\textwidth]{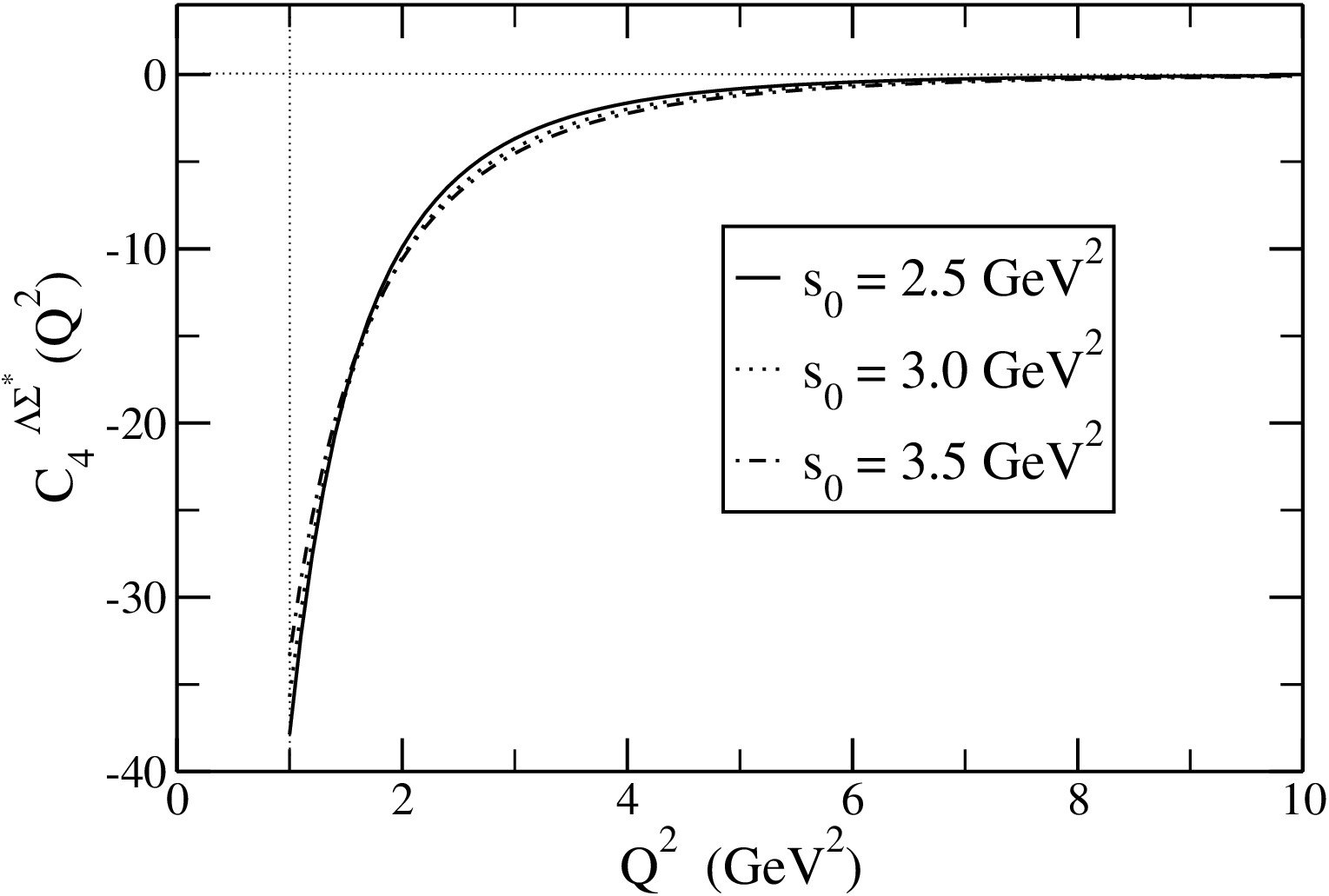}}\\
   \subfloat[]{\label{fig:C5.eps}\includegraphics[width=0.4\textwidth]{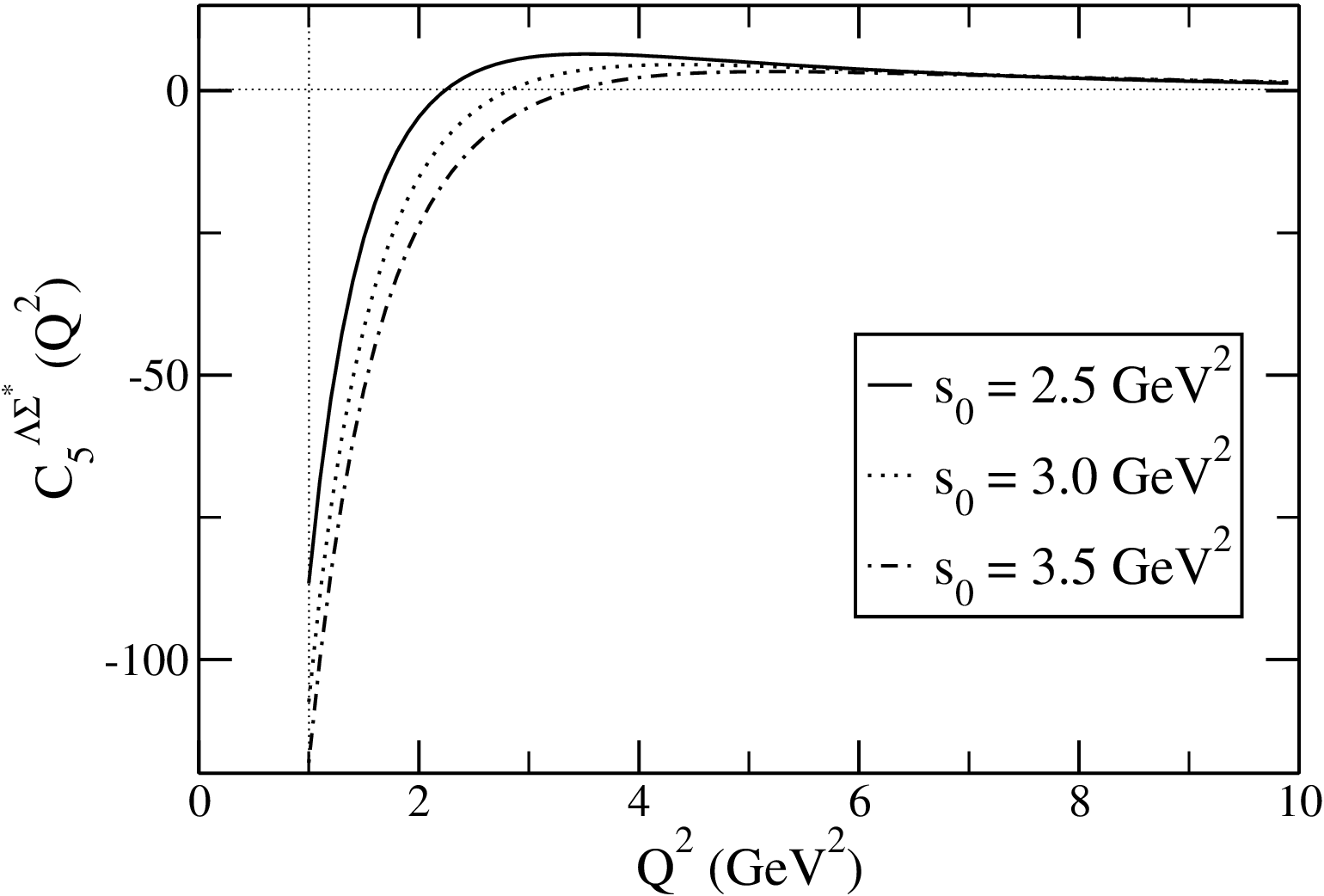}}
    \subfloat[]{\label{fig:C6.eps}\includegraphics[width=0.4\textwidth]{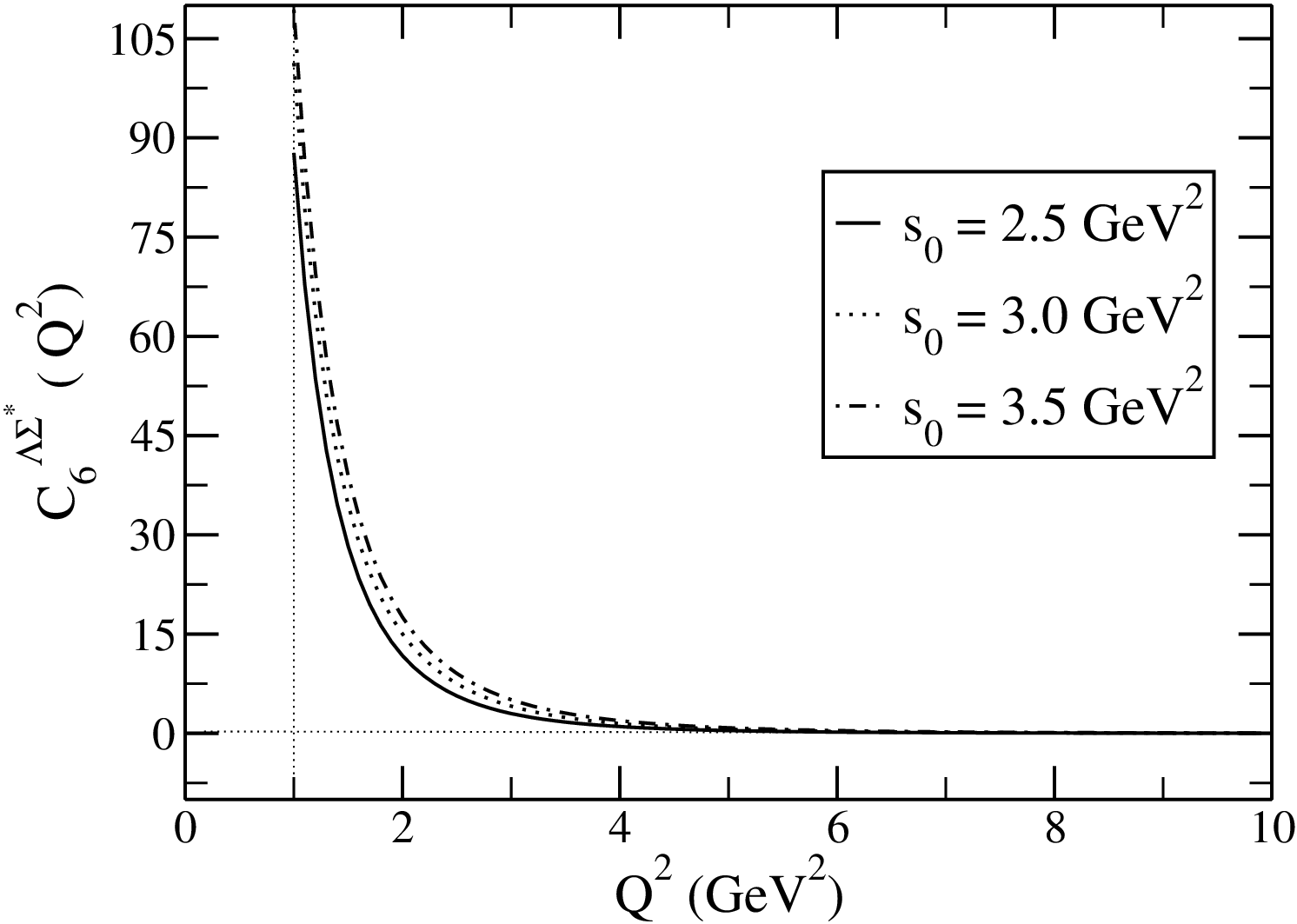}}
\caption{The dependence of the  form factors  for the values of the continuum threshold
 $s_0 = 2.5 ~GeV^2$, $s_0 = 3.0~GeV^2$, $s_0 = 3.5~GeV^2$ and $M^{2}=3~GeV^2$
(a) for $C_3^{\Lambda\Sigma^*}(Q^2)$ form factor,
(b) for $C_4^{\Lambda\Sigma^*}(Q^2)$ form factor,
(c) for $C_5^{\Lambda\Sigma^*}(Q^2)$ form factor and
(d) for $C_6^{\Lambda\Sigma^*}(Q^2)$ form factor.}
\end{figure}

\newpage
\begin{figure}[htp]
\centering
 \subfloat[]{\label{fig:ChiC3Qsq.eps}\includegraphics[width=0.4\textwidth]{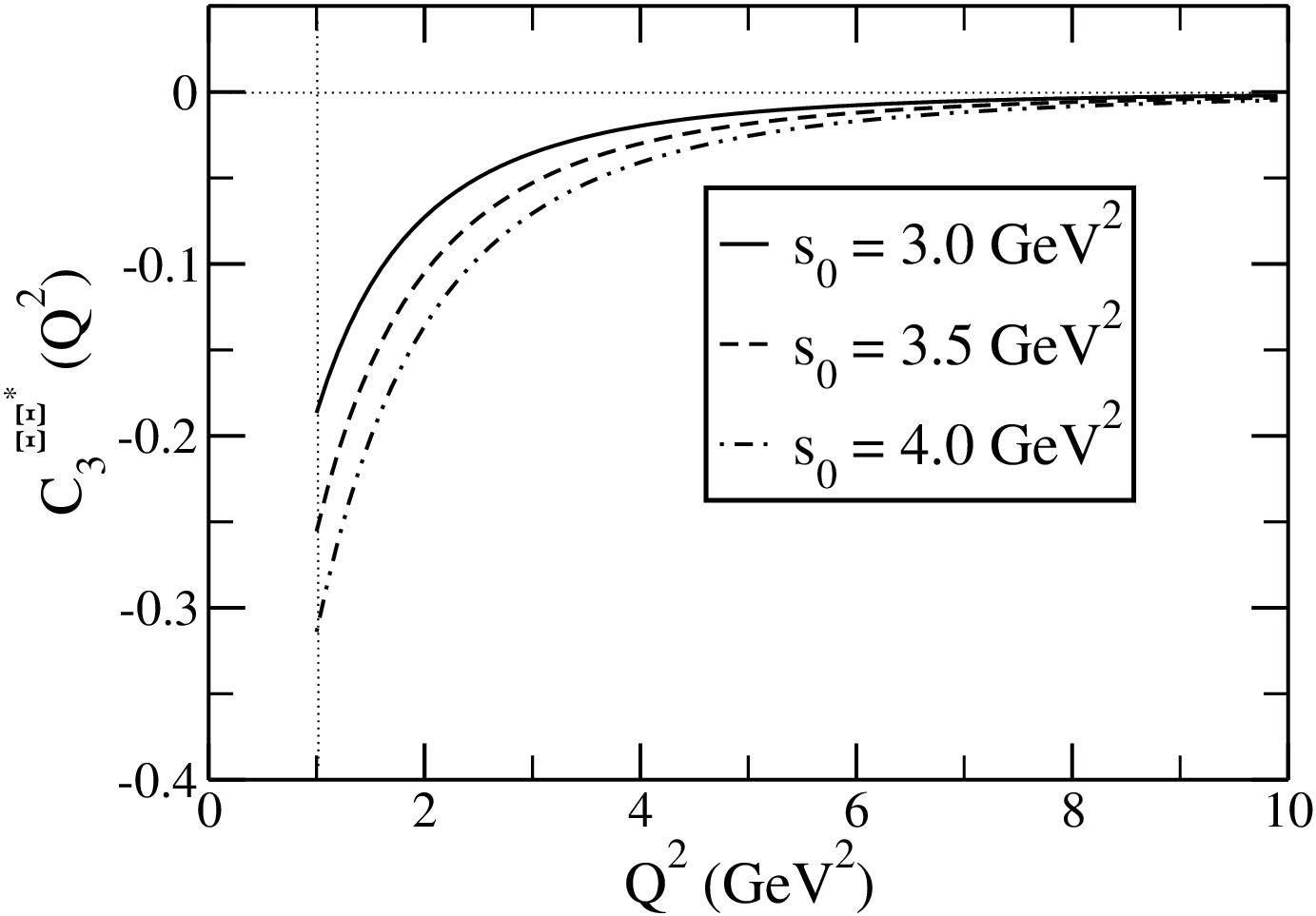}}
  \subfloat[]{\label{fig:ChiC4Qsq.eps}\includegraphics[width=0.4\textwidth]{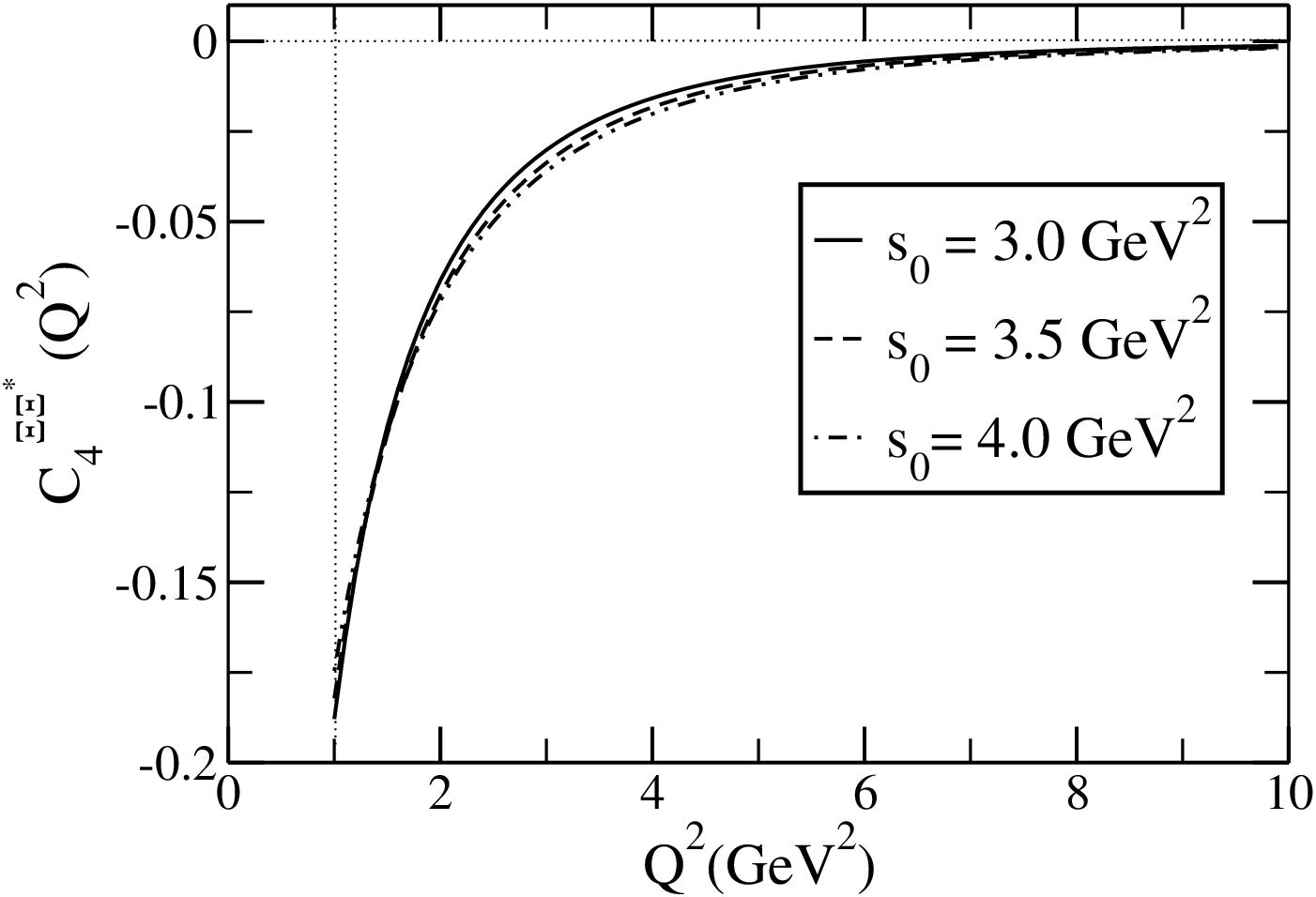}}\\
   \subfloat[]{\label{fig:ChiC5Qsq.eps}\includegraphics[width=0.4\textwidth]{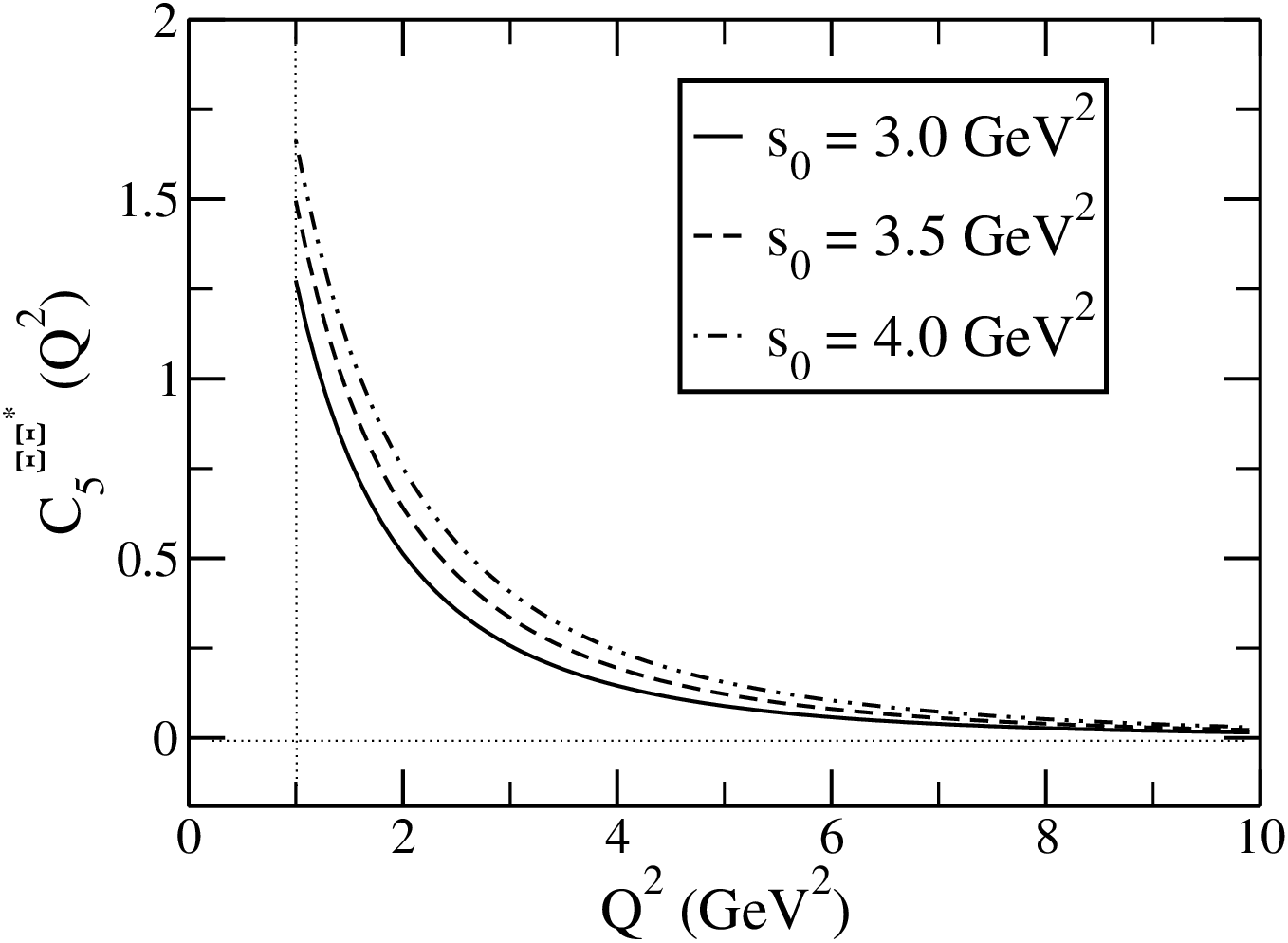}}
    \subfloat[]{\label{fig:ChiC6Qsq.eps}\includegraphics[width=0.4\textwidth]{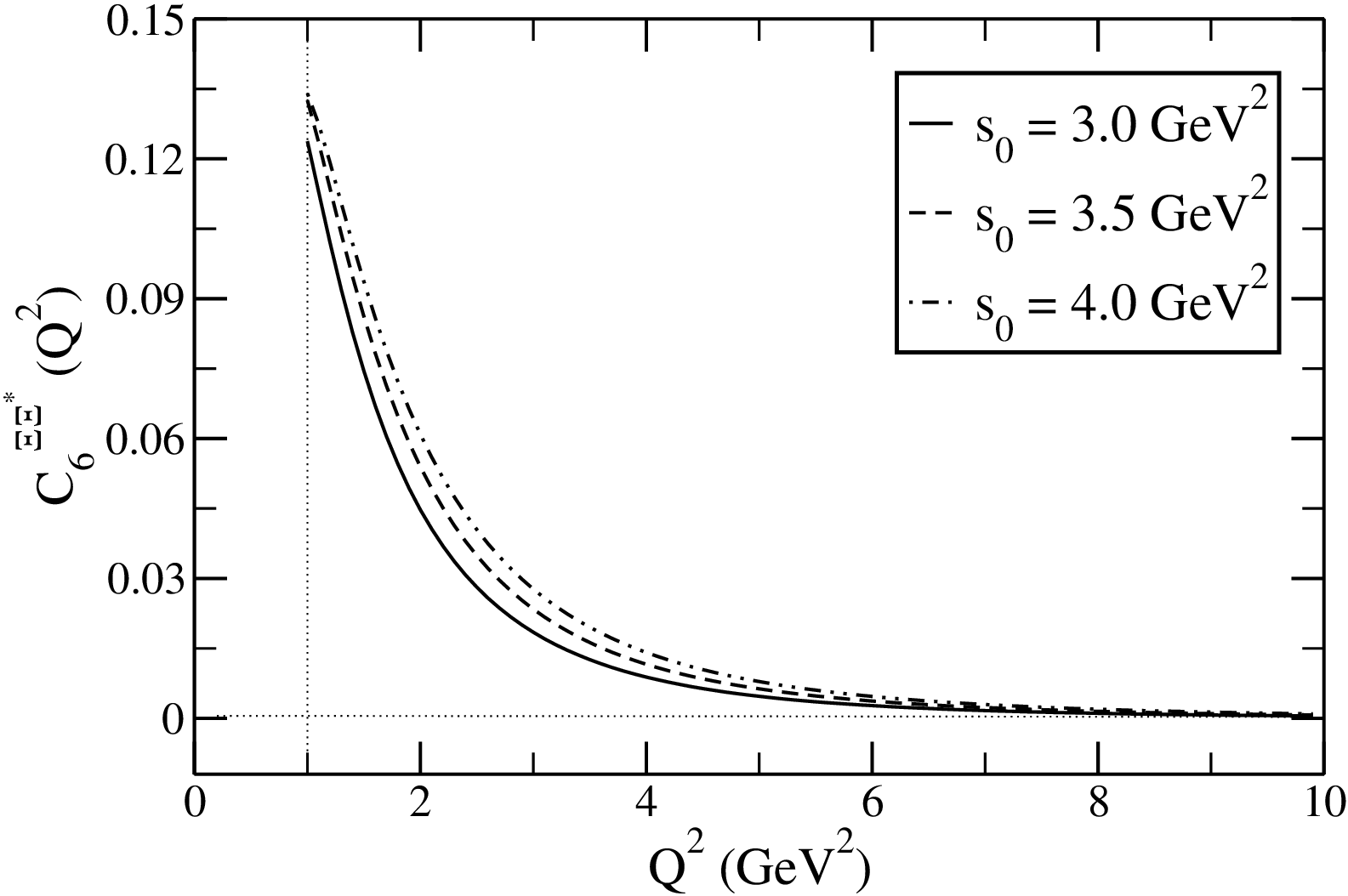}}
\caption{ The dependence of the  form factors  for the values of the continuum threshold
 $s_0 = 3.0 ~GeV^2$, $s_0 = 3.5~GeV^2$, $s_0 = 4.0~GeV^2$ and $M^{2}=3~GeV^2$
(a) for $C_3^{\Xi\Xi^*}$ form factor
(b) for $C_4^{\Xi\Xi^*}$ form factor,
(c) for $C_5^{\Xi\Xi^*}$ form facto and
(d) for $C_6^{\Xi\Xi^*}$ form factor.}
\end{figure}


\newpage
\appendix
\section{Explicit forms of the functions $F_i$ for the hyperon transitions}
\begin{eqnarray*}
F_1 &= \int_\alpha^{1}dx_2 \int_0^{1-x_2}dx_1 (2S_1-2S_2+2P_2-2P_1-10V_1+10V_2\\
&+4V_3+6V_4-2A_1+2A_2+4A_3-6A_4+4T_1-2T_2-2T_5\\
&-4T_7-4T_8)(x_1,x_2,1-x_1-x_2)\\
F_2 &= \int_0^{\beta}d\alpha\int_\alpha^{1}dx_2 \int_0^{1-x_2}dx_1 (T_2-T_3-T_4+T_5+T_7+T_8)\\
&(x_1,x_2,1-x_1-x_2)\\
F_3 &= \int_0^{\beta}d\alpha\int_\alpha^{1}dx_2 \int_0^{1-x_2}dx_1 (T_2-T_3-T_4+T_5+T_7+T_8)\\
&(x_1,x_2,1-x_1-x_2)\\
F_4 &= \int_0^{1-x_2}dx_1 (5V_1-T_1+A_1)(x_1,x_2,1-x_1-x_2)\\
F_5 &= \int_0^{\beta}d\alpha\int_\alpha^{1}dx_2 \int_0^{1-x_2}dx_1(-V_1+V_2+V_3+V_4+V_5-V_6\\
&+3A_1-3A_2+3A_3+3A_4-3A_5+3A_6+2T_2-2T_3-2T_4\\
    &+2T_5+2T_7+2T_8)(x_1,x_2,1-x_1-x_2)\\
F_6 &= \int_\alpha^{1}dx_2 \int_0^{1-x_2}dx_1(A_1-A_2+A_3-T_1-T_2+2T_3+3V_1\\
&-3V_2-3V_3)(x_1,x_2,1-x_1-x_2)\\
F_7 &= \int_\alpha^{1}dx_2 \int_0^{1-x_2}dx_1(A_1-A_2+A_3-V_1+V_2+V_3)\\
&(x_1,x_2,1-x_1-x_2)\\
F_8 &= \int_0^{\beta}d\alpha\int_\alpha^{1}dx_2 \int_0^{1-x_2}dx_1 (5A_1-5A_2+5A_3+5A_4-5_A5\\
            &+5A_6+V_1-V_2-V_3-V_4-V_5+V_6-T_1-T_2+2T_3+2T_4 \\
            &-T_5-T_6)(x_1,x_2,1-x_1-x_2)\\
F_9 &=  \int_0^{1-x_2}dx_1 (2P_1-2S_1-14V_1+6V_2+A_1+A_2+5A_3-2T_3\\
&-2T_7)(x_1,x_2,1-x_1-x_2)\\
F_{10} &= \int_0^{\beta}d\alpha\int_\alpha^{1}dx_2 \int_0^{1-x_2}dx_1(-V_1+V_2+V_3+V_4+V_5-V_6\\
&+3A_1-3A_2+3A_3 +3A_4-3A_5+3A_6+2T_2-2T_3-2T_4\\
&+2T_5+2T_7+2T_8)(x_1,x_2,1-x_1-x_2)\\
F_{11} &=  \int_0^{1-x_2}dx_1(-2P_1+2S_1+4V_1-6V_2-2A_2-4A_3+2T_3\\
&+2T_7)(x_1,x_2,1-x_1-x_2)\\
F_{12} &=  \int_0^{\beta}d\alpha\int_\alpha^{1}dx_2 \int_0^{1-x_2}dx_1 (T_2-T_3-T_4+T_5+T_7+T_8)\\
&(x_1,x_2,1-x_1-x_2)\\
F_{13} &=  \int_0^{\beta}d\alpha\int_\alpha^{1}dx_2 \int_0^{1-x_2}dx_1 (-T_1+T_2T_5-T_6+2T_7+2T_8)\\
&(x_1,x_2,1-x_1-x_2)\\
F_{14} &= \int_\alpha^{1}dx_2 \int_0^{1-x_2}dx_1 (2S_1-2S_2+2P_2-2P_1-10V_1+10V_2\\
      & 	+4V_3+6V_4-2A_1+2A_2+4A_3-6A_4+4T_1-2T_2-2T_5\\
      &-4T_7-4T_8)(x_1,x_2,1-x_1-x_2)\\
F_{15} &=  \int_0^{1-x_2}dx_1(5V_1-T_1+A_1)(x_1,x_2,1-x_1-x_2)\\
F_{16} &= \int_\alpha^{1}dx_2 \int_0^{1-x_2}dx_1(T_2-T_3-T_4+T_5+T_7+T_8)\\
&(x_1,x_2,1-x_1-x_2)\\
F_{17} &= \int_0^{\beta}d\alpha\int_\alpha^{1}dx_2 \int_0^{1-x_2}dx_1(-V_1+V_2+V_3+V_4+V_5-V_6\\
&+3A_1-3A_2+3A_3 +3A_4-3A_5+3A_6+2T_2-2T_3-2T_4\\
&+2T_5+2T_7+2T_8)(x_1,x_2,1-x_1-x_2)\\
F_{18} &= \int_\alpha^{1}dx_2 \int_0^{1-x_2}dx_1(-2V_1+2V_2+2V_3-T_1-3T_2+4T_3\\
&-2T_7)(x_1,x_2,1-x_1-x_2)\\
F_{19} &= \int_0^{\beta}d\alpha\int_\alpha^{1}dx_2 \int_0^{1-x_2}dx_1(-V_1+V_2+V_3+V_4+V_5-V_6\\
&+3A_1-3A_2+3A_3)(x_1,x_2,1-x_1-x_2)\\
F_{20} &= \int_\alpha^{1}dx_2 \int_0^{1-x_2}dx_1(3A_1-3A_2+3A_3-T_1-T_2+2T_3+V_1\\
&-V_2-V_3)(x_1,x_2,1-x_1-x_2)\\
F_{21} &= \int_\alpha^{1}dx_2 \int_0^{1-x_2}dx_1(2T_1-T_3-T_4-T_7-T_8+A_3-A_4+S_2\\
&-S_1+P_1-P_2-2V_1+2V_2+V_3+V_4)(x_1,x_2,1-x_1-x_2)\\
F_{22} &= \int_0^{1-x_2}dx_1(V_1-T_1)(x_1,x_2,1-x_1-x_2)\\
F_{23} &= \int_0^{\beta}d\alpha\int_\alpha^{1}dx_2 \int_0^{1-x_2}dx_1 (T_2-T_3-T_4+T_5+T_7+T_8)\\
&(x_1,x_2,1-x_1-x_2)\\
F_{24} &= \int_0^{\beta}d\alpha\int_\alpha^{1}dx_2 \int_0^{1-x_2}dx_1(T_1-T_3-T_4+T_6-T_7-T_8-A_1\\
&+A_2 -A_3-A_4+A_5-A_6)(x_1,x_2,1-x_1-x_2)\\
F_{25} &= \int_\alpha^{1}dx_2 \int_0^{1-x_2}dx_1 (2A_1-2A_2+2A_3-4T_1+4T_3+4T_7\\ &+2V_1-2V_2-2V_3)(x_1,x_2,1-x_1-x_2)\\
F_{26} &= \int_\alpha^{1}dx_2 \int_0^{1-x_2}dx_1(A_1-A_2+A_3-V_1+V_2+V_3)\\
&(x_1,x_2,1-x_1-x_2)\\
F_{27} &= \int_0^{\beta}d\alpha\int_\alpha^{1}dx_2 \int_0^{1-x_2}dx_1(-T_1-T_2+2T_3+T_4-T_5-T_6\\
&+2A_1+2A_3-2A_4-2A_5-2A_6)(x_1,x_2,1-x_1-x_2)\\
F_{28} &= \int_0^{1-x_2}dx_1(P_1-S_1+V_1+V_2-A_1+A_2-T_3-T_7)\\
&(x_1,x_2,1-x_1-x_2)\\
F_{29} &= \int_0^{\beta}d\alpha\int_\alpha^{1}dx_2 \int_0^{1-x_2}dx_1(-T_1+T_3+T_4-T_6+T_7+T_8\\
&+A_1-A_2+A_3+A_4 -A_5+A_6)(x_1,x_2,1-x_1-x_2)\\
F_{30} &= \int_0^{1-x_2}dx_1(-P_1+S_1-2V_2-V_3-A_3+T_3+T_7)\\
&(x_1,x_2,1-x_1-x_2)\\
F_{31} &= \int_0^{\beta}d\alpha\int_\alpha^{1}dx_2 \int_0^{1-x_2}dx_1 (-T_1+T_2+T_5-T_6+2T_7+2T_8)\\
&(x_1,x_2,1-x_1-x_2)\\
F_{32} &= \int_\alpha^{1}dx_2 \int_0^{1-x_2}dx_1(2T_1-T_3-T_4-T_7-T_8+A_3-A_4+S_2\\
&-S_1+P_1-P_2-2V_1+2V_2+V_3+V_4)(x_1,x_2,1-x_1-x_2)\\
F_{33} &= \int_0^{1-x_2}dx_1(V_1-T_1)(x_1,x_2,1-x_1-x_2)\\
F_{34} &= \int_0^{\beta}d\alpha\int_\alpha^{1}dx_2 \int_0^{1-x_2}dx_1(T_1-T_3-T_4+T_6-T_7-T_8-A_1\\
&+A_2-A_3-A_4+A_5-A_6)(x_1,x_2,1-x_1-x_2)\\
F_{35} &=  \int_0^{\beta}d\alpha\int_\alpha^{1}dx_2 \int_0^{1-x_2}dx_1 (T_2-T_3-T_4+T_5+T_7+T_8)\\
&(x_1,x_2,1-x_1-x_2)\\
F_{36} &= \int_\alpha^{1}dx_2 \int_0^{1-x_2}dx_1(A_1-A_2+A_3-V_1+V_2+V_3)\\
&(x_1,x_2,1-x_1-x_2)\\
F_{37} &= \int_0^{\beta}d\alpha\int_\alpha^{1}dx_2 \int_0^{1-x_2}dx_1(T_1-T_3-T_4+T_6-T_7-T_8\\
          &-A_1+A_2-A_3-A_4+A_5-A_6)(x_1,x_2,1-x_1-x_2)\\
F_{38} &= \int_\alpha^{1}dx_2 \int_0^{1-x_2}dx_1(A_1-A_2+A_3-T_1+T_3+T_7)\\
&(x_1,x_2,1-x_1-x_2)\\
\end{eqnarray*}
\end{document}